\documentclass[useAMS, usenatbib, usegraphicx]{mn2e} 
\usepackage{amssymb}

\newcommand{\si}{~\mbox{s}^{-1}}
\newcommand{\yr}{~\mbox{yr}}
\newcommand{\Myr}{~\mbox{Myr}}
\newcommand{\eV}{~\mbox{eV}}
\newcommand{\Mpc}{~\mbox{Mpc}}
\newcommand{\cMpch}{~h^{-1}~\mbox{comoving Mpc}}
\newcommand{\cmsq}{~\mbox{cm}^{2}}
\newcommand{\cmsqi}{~\mbox{cm}^{-2}}
\newcommand{\cmci}{~\mbox{cm}^{-3}}
\newcommand{\cmc}{~\mbox{cm}^{ 3}}
\newcommand{\kpc}{~\mbox{kpc}}
\newcommand{\K}{~\mbox{K}}
\newcommand{\vect}[1]{\mathbf{#1}}
\newcommand{\traphic}{{\sc traphic}}

\newcommand{\gadget}{{\sc gadget-2}}
\newcommand{\ctworay}{{\sc c}$^2${\sc -ray}}
\newcommand{\crash}{{\sc crash}}

\newcommand{\simplex}{{\sc simplex}}
\newcommand{\ftte}{{\sc ftte}}

\title[TRAPHIC - Radiative Transfer for SPH]{TRAPHIC - Radiative Transfer for
  Smoothed Particle Hydrodynamics Simulations} 

\author[Andreas H. Pawlik \& Joop Schaye] {Andreas H. Pawlik$^{1}$\thanks{E-mail: pawlik@strw.leidenuniv.nl} and 
  Joop Schaye$^{1}$\thanks{E-mail: schaye@strw.leidenuniv.nl} \\ 
  $^{1}$Leiden Observatory, Leiden University, P.O. Box 9513, 2300RA Leiden, The Netherlands} 
       
\begin{document} 
  
\date{} 
 
\pagerange{\pageref{firstpage}--\pageref{lastpage}} \pubyear{0000} 
 
\maketitle 
  
\label{firstpage} 
  
\begin{abstract} 
We present \traphic, a novel radiative transfer scheme for Smoothed Particle
Hydrodynamics (SPH) simulations. \traphic\ is designed for use in
simulations exhibiting a wide dynamic range in physical length scales and containing a
large number of light sources. It is adaptive both in
space and in angle and can be employed for application on distributed
memory machines. The commonly encountered computationally expensive scaling with the number of light sources
in the simulation is avoided by introducing a source merging procedure. 
The (time-dependent) radiative transfer equation is
solved by tracing individual photon packets in an explicitly
photon-conserving manner directly on the unstructured grid traced out by the set of
SPH particles. To accomplish directed transport of radiation despite the irregular spatial
distribution of the SPH particles, photons are guided inside cones. 
We present and test a parallel numerical implementation of \traphic\ in the
SPH code \gadget, specified for the transport of mono-chromatic hydrogen-ionizing radiation. The
results of the tests are in excellent agreement with both analytic solutions and results obtained
with other state-of-the-art radiative transfer codes. 
\end{abstract} 
 

\section{Introduction} 
\label{Sec:Introduction}
Radiation is one of the fundamental constituents of our Universe. 
Its interaction with baryons may lead to an energy exchange that can both heat
and cool the matter, initiating pressure forces that may strongly
influence the subsequent hydrodynamical evolution. Radiation may also exert a direct pressure force
upon the matter through the exchange of momentum. Radiative interactions are
furthermore often the dominating process in governing the excitation and
ionization state of atoms and molecules. The inclusion of the transport of radiation
into hydrodynamical simulations may therefore provide the key for interpreting
the outcome of physical experiments and observational campaigns.
 \par
To perform hydrodynamical simulations, the Lagrangian technique  
Smoothed Particle Hydrodynamics (SPH; \citealp{Gingold:1977};
\citealp{Lucy:1977}) is often employed. In SPH, the continuum fluid 
is discretized using a finite set of point particles,
each carrying its own collection of variables. The deformation of the fluid by
internal and external processes manifests itself in a steady redistribution of
the point particles in space. All it takes to determine the values 
of physical field variables at a given point in the simulation box is to perform a weighted
average over the values the relevant variables take on the particles
in its surrounding. This elegant simplicity of the SPH technique is one of the
many reasons for its success.
\par
Although the first radiative transfer calculation was already included in SPH 
at the very birth of this numerical technique more than thirty
years ago (\citealp{Lucy:1977}), the detailed treatment of radiation transport 
in SPH is still an enormous challenge. One of the main
reasons for this is certainly the high dimensionality of the problem. In fact, 
the radiative transfer equation depends on no less than seven variables (three
space coordinates, two angles, frequency and time). Moreover, existing numerical
schemes to solve the radiative transfer equation have most often 
only been formulated for use with uniform grids.  Despite these difficulties, 
there has been encouraging progress over the last years, with several
interesting and rather different approaches (see Section~\ref{Sec:RT}).
In this paper we present a new method to solve the radiative transfer equation
in SPH. We specifically designed our method for use in hydrodynamical simulations 
exhibiting a wide dynamic range in physical length scales and  
containing a large number of light sources.
\par
A prominent example for such simulations are cosmological simulations of large-scale
structure formation. Performing cosmological simulations is an
exceedingly demanding computational task. Difficulties in describing the 
growth of the initially tiny matter density perturbations produced before and during
the event of recombination and their metamorphosis into the rich 
structure observable today do not only arise because of our ignorance of how
to properly model the governing physical processes. Often, it is simply the
lack of computational power which prevents us from faithfully representing
the basic actions involved: Cosmological simulations are both time consuming 
and memory exhaustive. To overcome these computational challenges, one can make
use of advanced techniques and resort to clever approximations, reducing the computational effort.
\par
Consider the wide range of scales encountered in cosmological
hydrodynamical simulations. According to the hierarchical model of structure formation,
the first structures and building blocks of the evolving universe 
are expected to form at small scales. The non-linear evolution at these
scales shapes the distribution of the matter at all times, thus necessitating
simulations of high spatial resolution. On the other hand, we also
require sufficiently large simulation boxes
in order to properly account for the modulation of the small-scale nonlinear
evolution by the large-scale structure formation processes (e.g.~\citealp{Barkana:2004})
and not to be deceived by the cosmic scatter. 
\par
To accommodate these two antithetic demands while
keeping the number of particles representing the matter low enough to be 
computationally manageable, {\it spatially
adaptive} SPH simulations have been invoked. It is then immediately clear that
when solving the transport of radiation along with the gravito-hydrodynamics
of the matter, one requires the radiative transfer scheme to be adaptive, too.
Even spatially adaptive cosmological SPH simulations,
however, make use of hundreds of millions of particles and 
are therefore still extremely memory-consuming. It
is then indispensable to distribute the computational load over a large number of
machines. For this reason, we require a radiative transfer scheme that is 
{\it parallel on distributed memory} machines.
\par
When performing radiative transfer simulations, sources of
light are assigned a special importance. As a result, the computation time
of most of the available radiative transfer schemes scales linearly with the
number of sources in the simulation box. However, in cosmological simulations,
even at times as early as 1 billion years after the Big Bang, i.e. at a
redshift of $z \sim 6$, a non-negligible amount ($\sim 1$ per cent)
of baryonic matter has undergone star formation. In addition to these stellar
sources of light, the intergalactic gas emits photons too, producing a
radiation component often referred to as diffuse radiation. Hence, 
without {\it breaking the linear scaling} with the number of 
light sources, radiative transfer simulations at the resolution of
state-of-the-art cosmological hydrodynamic simulations need to be dispatched to the realm of
the future.
\par
In this paper we present a
radiative transfer scheme for use in SPH simulations that is
adaptive, parallel on distributed memory and that avoids the linear scaling of the computation time
with the number of sources, making it ideal for application in large
simulations covering a wide range of length scales and containing many sources. 
In our scheme we follow the propagation of individual photon packets. Hence, it is   
{\it explicitly photon-conserving} (\citealp{Abel:1999}) and can be applied to solve the
{\it time-dependent} radiative transfer equation. Because the photon packets
are traced in cones, we refer to our scheme as \traphic\ - TRAnsport of
PHotons In Cones. The introduction of cones is required in order to perform
the transport of photon packets directly on the unstructured grid defined by the SPH
particles. Although we have designed our radiative transfer scheme to be readily coupled to the
hydrodynamic evolution of the matter in SPH simulations, here we limit
ourselves to the description and testing of the radiative transfer scheme itself. We will report
on fully coupled radiation-hydrodynamics SPH simulations employing \traphic\ in
future work.
\par
The outline for the rest of this paper is as follows. We start with a brief
review of the principles of SPH that we consider crucial for the understanding of the
present work. In Section~\ref{Sec:RT} we then recall the radiative transfer equation and place
our radiative transfer method in context by reviewing the approaches that have
been used to solve the radiative transfer problem in SPH so far. In
Section \ref{Sec:Method} we give a detailed description of the ideas behind our
method. Throughout we will emphasize  how we
satisfy the requirements set by our primary aim of performing radiative transfer in
simulations exhibiting a wide range in length scales and a large number of
light sources. Thereafter, in Section \ref{Sec:Application}, we apply our
method to the transport of hydrogen-ionizing radiation, describe its numerical
implementation in the parallel state-of-the-art SPH code \gadget\ and report its behaviour in test problems. Finally, we summarise
our approach and conclude with an outlook.

\section{Smoothed Particle Hydrodynamics}
\label{Sec:SPH}
Excellent reviews of SPH exist (see e.g.~\citealp{Monaghan:2005}; \citealp{Monaghan:1992};
\citealp{Price:2005}). Here, we just briefly outline the basic principles we
consider critical for the understanding of our radiative transfer method.
\par
At the basis of SPH lies the representation of a field $A(\vect{r})$ by its 
integral interpolant $A_I(\vect{r})$,
\begin{equation}
A_I(\vect{r}) = \int d^3 r^{\prime} \ A(\vect{r}^{\prime}) W(\vect{r} -
\vect{r}^{\prime}, h),
\label{Eq:SPH:IntegralInterpolant}
\end{equation}
where the smoothing length $h$ determines the spatial resolution.
The interpolation kernel $W$ satisfies
\begin{eqnarray}
 \int d^3r^{\prime}\ W(\vect{r} - \vect{r}^{\prime}, h) &=& 1\\
 \lim_{h\to 0} W(\vect{r} - \vect{r}^{\prime}, h) &=& \delta_D(\vect{r} - \vect{r}^{\prime}),
\end{eqnarray}
where $\delta_D$ is the Dirac delta function, such that $A_I(\vect{r})$ coincides with
$A(\vect{r})$ in the limit $h\to 0$. The last two conditions do not fix the functional form for
$W$. An often adopted choice is the spherically symmetric compact spline 
\begin{eqnarray}  
  W(\vect{r} - \vect{r}^{\prime}, h)=\frac{8}{\pi h^3}\left\{\begin{array}{ll}   
  1-6\left(\frac{r}{h}\right)^2+6\left(\frac{r}{h}\right)^3, 
  &  0\le\frac{r}{h}\le\frac{1}{2}\\  
  2\left(1-\frac{r}{h}\right)^3,& \frac{1}{2}<\frac{r}{h}\le 1\\  
  0,& \frac{r}{h}>1,\\  
\end{array}\right.  
\label{Eq:Kernel:Spline}
\end{eqnarray}  
where $r = |\vect{r} - \vect{r}^{\prime}|$. From here on, when referring to the
interpolation kernel $W$, we assume that it is of this form. The discrete representation of a fluid by SPH
particles  is achieved by approximating the integral interpolant
(Eq.~\ref{Eq:SPH:IntegralInterpolant}) by the
summation interpolant, 
\begin{equation}
A_I(\vect{r}) \approx A_S(\vect{r}) = \sum_j \frac{m_j} {\rho_j}\ A(\vect{r}_j) W(\vect{r} - \vect{r}_j, h),
\end{equation}
where the summation is over the particles of mass $m$ and mass density $\rho$.
For a self-contained numerical treatment, any field needs to be discretized only at the
positions of the particles $i$ representing the fluid,  
\begin{equation}
A_i \equiv A_S(\vect{r}_i) = \sum_j \frac{m_j} {\rho_j}\ A(\vect{r}_j) W(\vect{r}_i - \vect{r}_j, h).
\label{Eq:SPH:Sum}
\end{equation}
Since the kernel $W$ is compact, only local
information needs to be accessed for the evaluation of the last sum, which can be carried out in a computationally
efficient manner in parallel on distributed memory machines.
\par 
For the interpretation of Eq.~\ref{Eq:SPH:Sum}, two main approaches can be taken,
depending on the choice for $h$ (\citealp{Hernquist:1989}). In the scattering
approach each particle $j$ is considered as a cloud
of radius $h_j$ and contributes to the field at the position of particle $i$  with the weight
$W(\vect{r}_i - \vect{r}_j, h_j)$. In the gathering approach, each particle $i$
searches the sphere of radius $h_i$ (in the following referred to as the sphere of
influence, or {\it neighbourhood}) for particles (its {\it neighbours}) and obtains an estimate of
the field value at its position by summing the field values at the positions
of the neighbours $j$, each weighted by $W(\vect{r}_i - \vect{r}_j, h_i)$.
\par
The two interpretations are identical only if the spatial resolution is fixed
throughout the fluid, i.e. if $h_i = h_j = h$. However, the SPH formalism only unfolds 
its true strength by allowing the resolution to vary in space, according to
the local density field. This is usually achieved by either directly fixing the number of
neighbours $N_{ngb}$ for all particles, or by requiring that the kernel volume
contains a constant mass (\citealp{Springel:2002}).

\section{Radiative Transfer in SPH - Previous Work}
\label{Sec:RT}
Before describing \traphic, our method to solve the radiative transfer equation in SPH, 
we briefly recall the radiative transfer equation and give a short overview
of the main numerical methods that have been employed so far to obtain its
solution in SPH simulations. 
\par
The classical equation of radiative transfer reads (see e.g. \citealp{Mihalas:2000})
\begin{equation}
\frac{1}{c}\frac{\partial I_\nu}{\partial t} 
+ \vect{n} \cdot \nabla I_\nu = \epsilon_\nu
- \kappa_\nu \rho I_\nu.
\label{Eq:RTE}
\end{equation} 
In Eq.~\ref{Eq:RTE}, $I_\nu\equiv I_\nu(\vect{r}, \vect{n}, t)$ is the monochromatic intensity (units
$~\mbox{ergs}~\mbox{cm}^{-2}~\mbox{s}^{-1}~\mbox{Hz}^{-1}~\mbox{sr}^{-1}$) of
frequency $\nu$ at position $\vect{r}$,
$\vect{n}$ is a unit vector along the direction of light
propagation and $c$ is the speed of light. Sources and sinks of radiation are described by the emissivity $\epsilon_\nu$ 
(units $~\mbox{ergs}~\mbox{cm}^{-3}~\mbox{s}^{-1}~\mbox{Hz}^{-1}~\mbox{sr}^{-1}$)
and the mass absorption coefficient $\kappa_\nu$ (units $~\mbox{cm}^{2}~\mbox{g}^{-1}$),
respectively. In general, both $\epsilon_\nu$ and $\kappa_\nu$ are functions of 
$\vect{r}, \vect{n}$ and time $t$. If we define the photon
number density $\psi_\nu$ such that $\psi_\nu d\Omega d\nu$ is the number of photons
per unit volume with frequencies $(\nu, \nu + d\nu)$ travelling with
velocity $c$ into a solid angle $d\Omega$ around $\vect{n}$, the intensity is
given by $I_\nu = c h_p \nu \psi_\nu$, where $h_p$ is the Planck
constant. Eq.~\ref{Eq:RTE} is a partial differential equation depending on seven
variables, which in general is hard to solve. Different approximations 
have been employed to obtain its solution, giving rise to different numerical approaches. Below we discuss the main approaches
that have been taken so far to accomplish the transport of radiation in SPH simulations. 
\par
In his study of optically thick proto-stars, \cite{Lucy:1977} modelled the 
transport of radiation as a diffusion process by including a corresponding
term in the SPH formulation of the energy equation. \cite{Brookshaw:1994}
pointed out drawbacks of the particular formulation used in \cite{Lucy:1977},
and re-phrased the diffusion equation in SPH such as to reduce its sensitivity to
particle disorder. This new formulation of the diffusion equation was employed by  \cite{Viau:2006} to perform
collapse simulations of centrally condensed clouds. 
\par
Treating the transport of radiation in the diffusion limit is, however, only a
good approximation in optically thick media. In the optically thin regime,
infinite signal propagation speeds result, since the diffusion
equation is a partial differential equation of parabolic type. This defect can be
remedied by supplementing the diffusion equation with a so-called
flux-limiter. SPH simulations with flux-limited diffusion have been carried
out by e.g. \cite{Herant:1994},  \cite{Whitehouse:2006}, \cite{Fryer:2006} and
\cite{Mayer:2007}, covering a wide range of physical applications, from
supernovae explosions and proto-stellar collapse to the study of proto-planetary
disks. The (flux-limited) diffusion approach to solve the radiative transfer equation
fails in complex geometries. In particular, opaque obstacles illuminated by a point source do
not cast sharp shadows, because the shadow region is filled in by the diffusion.
\par
To solve the radiative transfer equation without being restricted to the optically thick
regime or simple geometries, Monte Carlo techniques can be employed (\citealp{Oxley:2003}, \citealp{Stamatellos:2005},
\citealp{Croft:2007}, \citealp{Semelin:2007}). In Monte Carlo radiative 
transfer, individual photon packets emitted by each source are directly 
followed as they pass through the matter, thus simulating the physical process 
of radiation transport as encountered in nature. While Monte Carlo simulations
allow realistic shadows to be cast behind opaque obstacles, they are
computationally very demanding, since the Poisson noise inherent to the statistical description of the
radiation field leads to a signal-to-noise ratio that grows only with the
square-root of the number of photon packets emitted. 
\par
Ray-tracing schemes keep the advantages of Monte Carlo radiative transfer
simulations, whilst not being affected by the statistical noise. In short,
in ray-tracing schemes rays are cast from each source throughout the
simulation box. The optical depth to each point in space is
calculated along theses rays, and the attenuation of the flux emitted by the sources is
obtained. Variants of ray-tracing working with photons instead of
fluxes blur the differences with Monte-Carlo schemes.
\par
Ray-tracing schemes are most easily implemented on a regular grid superimposed
on the SPH density field. However, this implies that all
information about substructure on scales smaller than the size
of the grid cells is ignored. As a cure, adaptive grids have been invoked
(e.g.~\citealp{Razoumov:2006}, \citealp{Alvarez:2006}). 
In SPH, {\it grid-less} ray-tracing has been introduced by \cite{Kessel-Deynet:2000} 
to investigate the effects of ionizing UV radiation by massive stars on the surrounding
interstellar medium. The Str\"omgren volume method
(\citealp{Dale:2007}, compare also \citealp{Johnson:2007}) and the ionization front
tracking technique employed by \cite{Yoshida:2007} are related
approaches. Another grid-less radiation tracing scheme that has been applied
to SPH simulations is presented in \cite{Ritzerveld:2006}.
Inspired by the \cite{Kessel-Deynet:2000}  approach, \cite{Susa:2006} describes a
radiation-hydrodynamics scheme for the transport of ionizing radiation in
cosmological simulations of structure formation. 
\par
One major drawback most ray-tracing schemes share with the Monte-Carlo technique
is that the computational effort to solve the radiative transfer equation 
scales linearly with the number of sources in the simulation box (but see
Section~\ref{Sec:TRAPHIC:Merging} later on). For this reason,
SPH radiation-hydrodynamical simulations employing the ray-tracing approach 
have mostly been restricted to the study of problems involving only a few
sources. 
\par
In the following sections we will describe \traphic, a novel method to solve the radiative
transfer equation in SPH. \traphic\ employs a photon tracing technique that
works directly on the discrete set of irregularly distributed SPH particles. It is 
designed to overcome the challenges set by our goal of carrying out large radiation-hydrodynamics 
SPH simulations exhibiting a wide dynamic range in length
scales and containing a large number of light sources.

\section{TRAPHIC - TRAnsport of PHotons in Cones}
\label{Sec:Method}
In this section we give a detailed description of \traphic, our method to solve the 
radiative transfer equation in SPH.  We start the presentation 
of our radiative transfer scheme by sketching the main
idea, in order to give an overview and to introduce the reader to the underlying
concepts, which will be illustrated and
explained in more detail in the subsequent sections. 
\traphic\ can be applied to solve the radiative transfer
equation in both two-dimensional and three-dimensional SPH simulations. 
When illustrating concepts with schematic figures  we
will, however,  restrict ourselves to two dimensions for the sake of clarity.
\par
Although we ultimately aim at performing
simulations in which the radiation transport is fully coupled to the hydrodynamics, here we assume
that the SPH density field is static and that the radiation does not
exert any thermodynamic or hydrodynamic feedback on the matter. We will report on the coupling of
radiation and hydrodynamics in future work.
\subsection{Overview}

We solve the radiative transfer equation
(Eq.~\ref{Eq:RTE}) in finite steps $(t_r, t_r +  \Delta
t_r)$, where $t_r$ is the current simulation time,  
by tracing photon packets radially away from their location of emission until a
certain stopping criterion is fulfilled. We do not introduce a grid on which the photons are
propagated. Instead, we propagate the photons directly on the discrete set of particles in the
simulation. For the transport of photons we employ the same particle-to-neighbour communication scheme 
that is already used in the SPH simulation to solve the equations of hydrodynamics.
That is, we accomplish the global transport of photons by propagating them
only locally, between a particle and its $\tilde{N}_{ngb}$ neighbours. 
We allow $\tilde{N}_{ngb}$ to be different from $N_{ngb}$, the number of
neighbours used during the SPH calculations. 
In SPH, $N_{ngb}$  determines the adaptive spatial resolution of the hydrodynamical simulation.
Similarly, in our scheme, $\tilde{N}_{ngb}$ sets the adaptive spatial resolution of the
radiation transport. It is the first of two numerical parameters in our method
(see also the left-hand panel of Fig.~\ref{Fig:Emission}).
\par
Working directly on the set of SPH particles, our scheme
exploits the full spatial resolution of the hydrodynamical simulation.
A further advantage of our approach is that the radiation transport is automatically parallel on
distributed memory, as long as the SPH simulation itself is parallel on
distributed memory. This is in contrast to radiation-hydrodynamical schemes
that employ individual communication schemes for the transport of radiation
and the SPH computations. Our radiative transfer scheme can hence be coupled
to the SPH simulation without introducing any additional computational
structures. 
\par
The main challenge we face when performing radiation transport by
propagating photons only from a particle to its neighbours is achieving 
a transport that is directed: Photons travel along straight  
rays, while the spatial distribution of the SPH particles is generally highly 
irregular. In the following we give an overview of the concepts employed in our scheme
and describe how we overcome this and other challenges to accomplish the
transport of radiation in SPH simulations. 
\par
We distinguish two types of particles present in 
SPH simulations that are relevant for the transport of photons: {\it Star} particles
and  SPH, or {\it gas}, particles. Only gas particles can interact with
photons, via absorptions and scatterings. We assume, however, that both star 
and gas particles can be sources of photons and will therefore refer to them as
source particles, or simply sources.
\par
The transport of radiation at simulation time $t_r$ over a single radiative
transfer time step $\Delta t_r$ starts with the emission of photons by
the source particles. Subsequently, these photons are propagated downstream, radially away from
the source, simultaneously for all sources. In our particle-to-neighbour
transport scheme photons are propagated from gas particle to gas
particle. On their way, photons may interact with the matter field discretized by the gas particles.
Each gas particle can simultaneously receive photons, possibly emitted at
different times, from multiple (in principle all) sources in the
simulation. Their further propagation would require a loop over all
propagation directions. We explicitly avoid the resulting linear scaling of the
computation effort with the total number of sources by introducing a source merging
procedure.
\par
The distribution of photons amongst the neighbours of the source particles must  
respect the angular dependence of the source emissivity. We achieve this by
introducing for each source a set of randomly oriented, tessellating 
{\it emission cones}\footnote{We use the word cone in a general sense. It does not necessarily
describe, but includes in its meaning, a regular cone, which is defined as a pyramid with a circular cross
section.}(Section~\ref{Section:EmissionBySourceParticles}), as illustrated in the middle panel of Fig.~\ref{Fig:Emission}. The fraction of the
total number of photons emitted into each cone is proportional to its solid angle. The
emission direction (which we will equivalently refer to as the propagation direction) 
associated with each emitted photon packet is determined by the
central axis of the emission cone. The number of cones used in the
emission cone tessellation, $N_c$, is the second numerical
parameter in our method. It sets the angular resolution. The random
orientation given to the cones ensures that photons are not restricted to
propagate along a fixed number of directions and prevents artefacts
introduced by the shape of the emission cones.
\par
Some of the emission cones may not contain any neighbours, as is the case for
the bottom-right cone of the tessellation  shown in the middle panel of Fig.~\ref{Fig:Emission}. To nevertheless
emit photons into the corresponding directions, we create additional neighbours
to which the photons can then be transferred (Fig.~\ref{Fig:Emission},
right-hand panel). We refer 
to these neighbours as virtual particles (ViPs), since they do not take part in
the SPH simulation. The properties of the ViPs are obtained from those of
the neighbouring SPH particles using SPH interpolation. ViPs compensate for the lack of neighbours in certain solid angles around the
emitting source. Hence, by employing ViPs we introduce the freedom of choosing emission
directions independently of the geometry of the SPH particle distribution. 
As we will argue in Section \ref{Section:EmissionBySourceParticles} and
investigate in more detail in Appendix \ref{Appendix:A}, this freedom
is a necessary requirement for achieving a desired angular dependence
of the transport process, e.g. the isotropic emission of photons by star
particles, in any particle-to-neighbour transport scheme.  
\par
The photon packets distributed amongst the
neighbours of the sources are further transferred downstream until they
experience an interaction. In contrast to the emission of photons by source particles, gas particles distribute photons
only amongst the subset of neighbours located in the downstream direction (see
Fig.~\ref{Fig:Transmission}). 
We distinguish this directed particle-to-neighbour transport 
from the emission process by referring to it as {\it transmission} (Section~\ref{Section:Transmission}).  In more
detail, each gas particle transmits photons downstream by distributing photon packets
amongst the subset of neighbours located in certain regular cones only.
These so-called transmission cones are  centred on the emission (propagation)
direction associated with each photon packet and have an opening angle determined by the
angular resolution, i.e. by $N_c$. They confine the photon packets to the cone
into which they were emitted by the corresponding source particle. As a
result, the photon packets are propagated radially away from the sources,
in a manner that is independent of the geometry of the SPH particle distribution. 
Like emission cones, transmission cones might not contain any
neighbours. Again we employ ViPs to accomplish the photon transport into the
corresponding directions.  
\par
By keeping the opening angle of the cones fixed, the
solid angle subtended by a transmission cone as viewed from the original source
decreases with the distance from the source. As a result, the photon transport becomes
adaptive in angle, as in the ray-splitting
technique employed in ray-tracing schemes (e.g.~\citealp{Abel:2002}). We will demonstrate later on,
when testing a numerical implementation of our scheme
(Section~\ref{Section:Test2}), that the use of implicitly adaptive transmission cones 
confining the photon propagation yields sharp shadows behind opaque obstacles. 
\par 
Gas particles receive photon packets from other particles through the processes of emission
and transmission described above. A given gas particle can
simultaneously receive multiple photon packets, which may each be associated with
a different emission direction. In fact, our assumption that all star and gas particles 
in the simulation box can be source particles would imply that the
number of directions from which photon packets can be simultaneously received
by a given gas particle would scale linearly with the total number of
particles in the simulation.
The computational effort to
accomplish the subsequent photon transmission along the associated directions
would then also scale linearly with the total number
of particles. Since this consideration applies to the transmission of photons
by any SPH particle in the simulation, the total computational effort to solve the radiative transfer
equation would scale no less than quadratically with the total number of SPH particles.
\par
To avoid this computationally expensive scaling, we introduce a source {\it merging}
procedure (cp. Fig.~\ref{Fig:Merging}, which will be explained in more
detail in Section~\ref{Sec:TRAPHIC:Merging}). We make use of the fact that source
particles that are seen as close (in angle) to each other from a given gas particle  
can be approximated by, or merged into, a single point source. We
average the emission directions associated with the photon packets received
from sources in solid angle bins
defined by a so-called reception cone tessellation which, except for a
random rotation, is identical to the cone tessellation employed for the emission
process. Consequently, photon packets need to be transmitted into at most $N_c$ directions.
\par
We distinguish two types of interactions with the matter field sampled by
gas particles and ViPs that photons may experience. Absorption interactions 
change the number of photons contained in each packet. Scattering interactions change the propagation direction
(and possibly the frequency) of the photons in a packet. All interactions are taken into account
in an explicitly  photon-conserving manner. We give a detailed description of interactions in
Section~\ref{Sec:Method:Interaction}.
\par

\subsection{Transport of photons}
We now expand on the description of our radiative transfer scheme given
above. We comment in detail on the main concepts underlying \traphic, i.e. the 
emission of photons by source particles,
the transmission of photons by gas particles and the source merging
procedure.  The description of how these
concepts are employed to advance the solution of the radiative transfer equation
in time is deferred to Section~\ref{Sec:Evolution}.

\subsubsection{Photon emission by source particles}
\label{Section:EmissionBySourceParticles}
\begin{figure*} 
 
  \includegraphics[width=55mm]{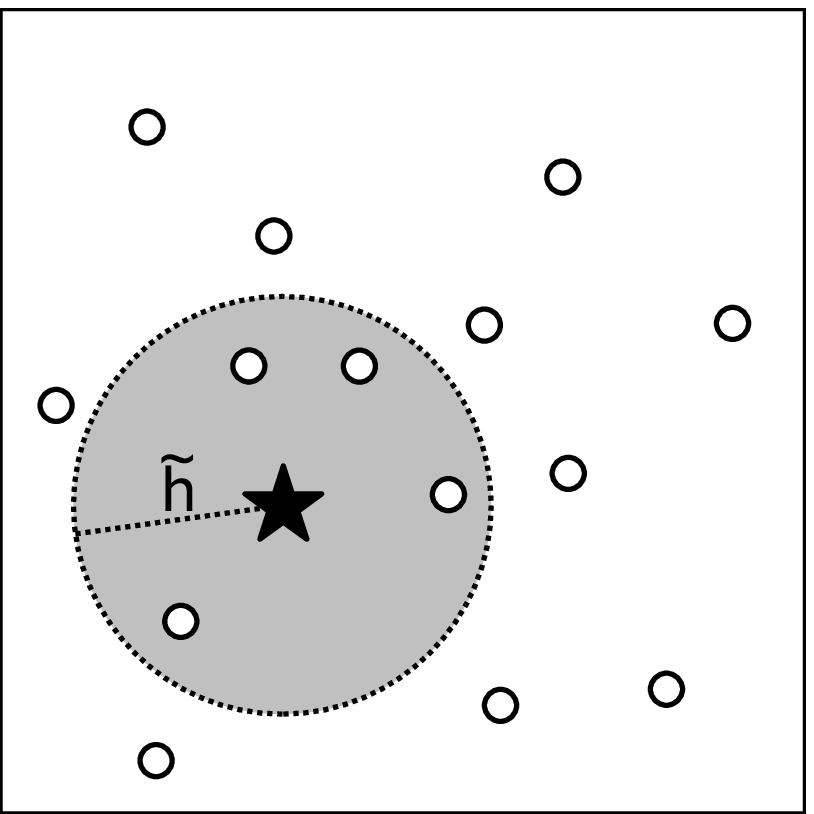} 
  \includegraphics[width=55mm]{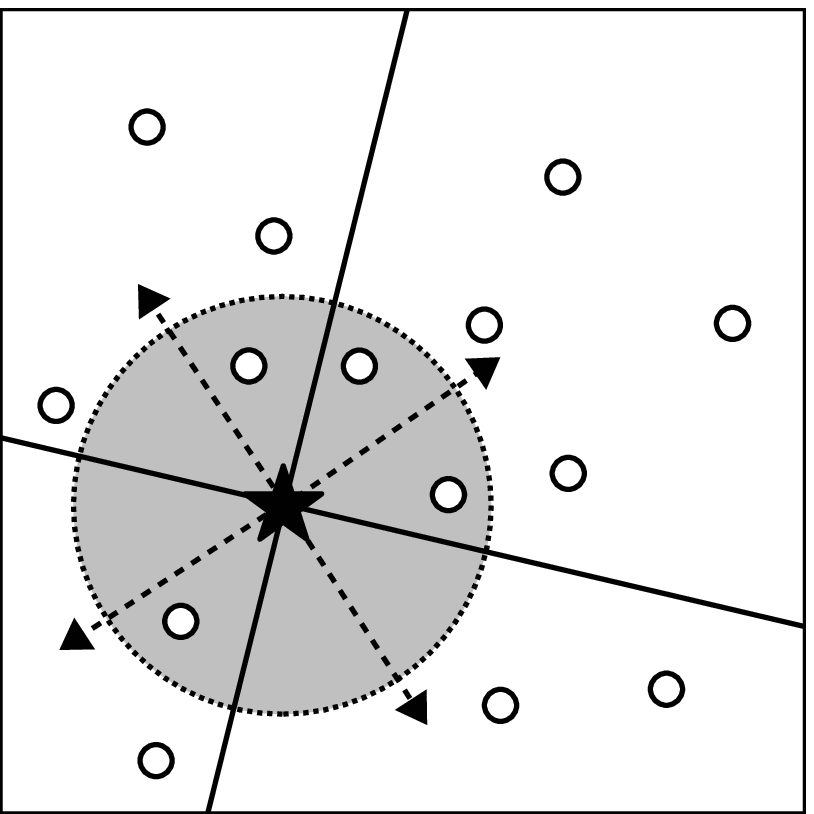} 
  \includegraphics[width=55mm]{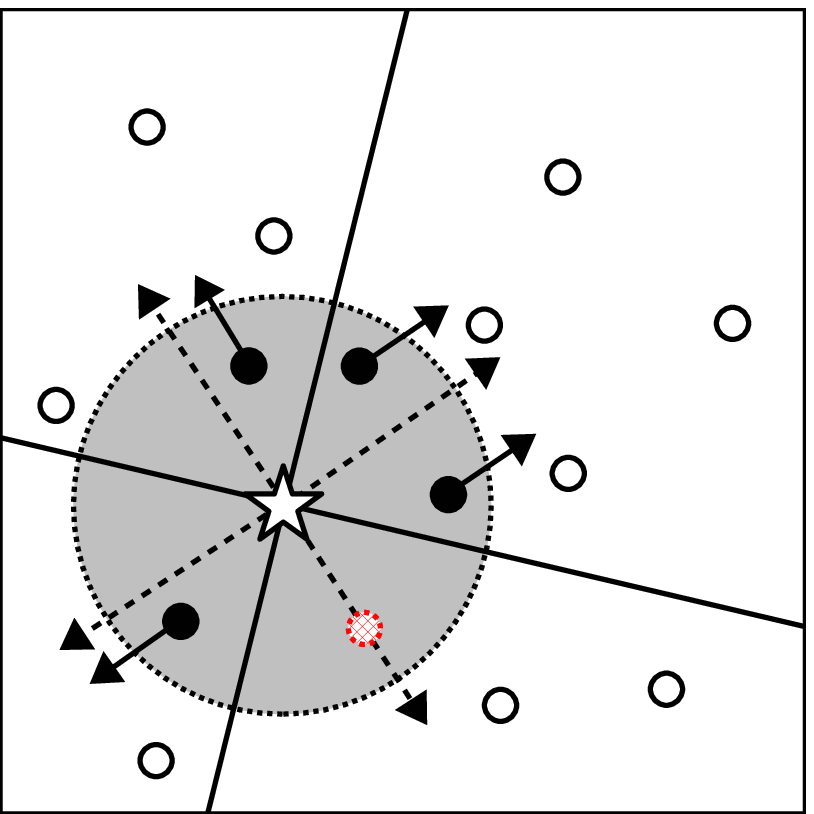} 

  \caption{Emission of photon packets by source particles. Left-hand panel: A source particle (black
  star) and its neighbourhood (big grey disc). In this example the radius $\tilde{h}$ of the
  neighbourhood is defined such that there are $\tilde{N}_{ngb} = 4$ neighbouring gas
  particles (white discs). Note that SPH simulations typically use $N_{ngb}
  \approx 48$. Middle panel: The randomly oriented emission cone tessellation of the source
  particle (solid lines) is shown for the case $N_c = 4$. The dashed arrows
  indicate the central axes of the cones. Note that the bottom-right cone does
  not contain a neighbouring gas particle. Right-hand panel: The source particle has
  transferred its photon packets to its neighbours (black discs). The emission
  direction (small solid arrows) associated with each packet points in the
  direction of the axis of the cone in which the neighbour resides. For the
  empty cone, a virtual particle is created (red hatched disc), placed randomly
  along the central axis of the cone.}
  \label{Fig:Emission}
\end{figure*} 
In this section we describe the emission of photons by source
particles. Star particles emit photons according to their intrinsic
luminosity, which can for instance be obtained from population synthesis models.
An example of the emission of photons by gas particles is the
emission of recombination radiation by a recombining ion. 
\par
Let us  consider the emission of photons by source
particle $i$, located at $\vect{r}_i$. We denote the number of photons emitted
per unit time  per unit frequency $\nu$ per unit solid angle $\Omega$ around the unit
direction vector $\vect{n}$ by
$\dot{\mathcal{N}}_{\nu,i}(\vect{n},\vect{r},t)$, 
or simply $\dot{\mathcal{N}}_{\nu,\vect{n},i}$.
With this notation, the number of photons $\dot{\mathcal{N}}_{\nu,i}$ 
emitted per unit time at frequency $\nu$ is 
$\dot{\mathcal{N}}_{\nu,i} \equiv \int d\Omega\ \dot{\mathcal{N}}_{\nu,\vect{n},i}$,
and the total number of photons emitted per unit time is 
$\dot{\mathcal{N}}_{\gamma,i} \equiv \int d\nu\ \int d\Omega\ \dot{\mathcal{N}}_{\nu,\vect{n},i}.$
\par
Source particle $i$ emits $\int_t^{t + \Delta t_r}\dot{\mathcal{N}}_{\gamma,i}$ photons
over the radiative transfer time step $\Delta t_r$. In agreement with our
particle-to-neighbour based transport approach, the emission process is modelled
by distributing these photons amongst the $\tilde{N}_{ngb}$ nearest gas
particles, residing in a sphere of radius $\tilde{h}_i$
centred on $\vect{r}_i$. This sphere is schematically depicted in the left-hand panel of
Fig.~\ref{Fig:Emission}.
\par 
Source particle $i$ distributes its photons amongst its neighbours with the help of
a set of space-filling emission cones, defined as follows (middle
panel of Fig.~\ref{Fig:Emission}). We tessellate the
simulation box into $N_c$ cones of (generally not identical) solid angles $\Omega_i^k$, $k
= 1,2,...,N_c$, with the apexes located at the position $\vect{r}_i$ of
particle $i$.  
The number of neighbours $\tilde{N}^k_{ngb,i}$ in each cone $k$ is determined, taking
values in the range $0 \le \tilde{N}^k_{ngb,i} \le \tilde{N}_{ngb}$. 
\par
For what follows we consider the emission of photons for each frequency $\nu$ separately.
To each neighbour $j$ in cone $k$ a photon packet of characteristic frequency $\nu$ is emitted. The packet contains 
a fraction $w_i^{kj} > 0$  ($j = 1...\tilde{N}^k_{ngb,i}$)  of the total number
of photons $\dot{\mathcal{N}}_{\nu,i}$ to be emitted per unit time at
frequency $\nu$, with 
\begin{equation}
w_i^{kj} = \frac{\int_{\Omega_i^k} d\Omega\
  \dot{\mathcal{N}}_{\nu,\vect{n},i}}{\int d\Omega\ \dot{\mathcal{N}}_{\nu,\vect{n},i}} \times \frac{w_i^j}{\sum_{l=1}^{\tilde{N}_{ngb,i}^k} w_i^l},
\label{Eq:ConeWeights}
\end{equation}
where $w_i^j$ are weights attached to neighbour $j$ in cone $k$. The first
factor on the right-hand side of Eq.~\ref{Eq:ConeWeights} determines which fraction of the total
number of photons is emitted into cone $k$, whereas the second factor controls
the fraction of photons that is transferred to
neighbour $j$, that is, it controls the distribution of photons amongst the 
particles within cone $k$. Here we set $w_i^j = 1$, i.e. equal weights for all
neighbours in a given cone. The number of cones used in the tessellation, the parameter  $N_c$, determines
the angular resolution of the radiation transport. A cone tessellation 
may consist of cones with different solid
angles\footnote{We note that  in the numerical
  implementation of our radiative transfer scheme the tessellation  
is made of cones having equal solid angles, see Appendix~\ref{Appendix:B:Cones}.}. We
therefore define the angular resolution to be the size of the average solid
angle, $\langle\Omega\rangle = 4\pi / N_c$.
\par
For each emission cone $k$ the central axis is defined, characterised by the unit vector
$\vect{n}_k$ pointing away from the source (Fig.~\ref{Fig:Emission}, middle
panel). We refer to this vector as the
emission direction associated to photon packets emitted into cone $k$. 
When a photon packet is emitted to a neighbouring gas particle,
the emission direction is
transferred in addition to the number of photons it contains
(Fig.~\ref{Fig:Emission}, right-hand panel). 
Since the orientation of the emission cone tessellation is randomised by
applying a random rotation\footnote{For a numerical implementation of the
random rotations we refer to Appendix~\ref{Appendix:B:Rotation}.} at each emission, 
there is no a priori limit on the values the emission directions can take. 
Note that while we transfer the emission direction, we do
not transfer the position of the source. Photon packets are traced
further downstream based on their emission direction only, as we will
explain in Section~\ref{Section:Transmission}. Each photon packet has also
an associated {\it clock} $t^\star$. At emission the clock time is set to the 
time of the simulation, $t^\star = t_r$. In Section~\ref{Sec:Evolution} we will see that the
clock can be employed to propagate the photon packets at the speed of light.
\par
Some cones may not contain any neighbouring gas particles. This is for instance
the case for the bottom right cone in the middle panel of
Fig.~\ref{Fig:Emission}. For a fixed number of neighbours these empty cones will
occur more frequently if the angular resolution is higher (i.e. if the solid angles of the
cones are smaller). On the other hand, for a fixed angular resolution $N_c$, empty cones will
occur more frequently if the number of neighbours $\tilde{N}_{ngb}$ is smaller. Spatial
clustering of the neighbours also increases the probability of the occurrence of empty
emission cones. In the absence of a neighbouring gas particle photons cannot be
transported along the emission direction of the empty cone. We therefore
create a new neighbour, a so-called virtual particle (ViP), to which the photons
are transferred. The ViP is placed along the cone axis at a random distance $<
\tilde{h}_i$ from the source particle, such that the volume of the cone is
uniformly sampled (see the right-hand panel of Fig.~\ref{Fig:Emission}). The properties of
ViPs (e.g. density) are determined from their neighbouring gas particles
using SPH interpolation. ViPs are only employed for the
transport of photons. We stress that ViPs are not used by the SPH simulation.
\par
We emphasize that the introduction of cones is essential for accomplishing the emission process
in our particle-to-neighbour transport scheme. To see this, consider the alternative
of distributing the photons directly amongst the neighbours of gas particle
$i$. This amounts to setting $\Omega_i^1 = 4\pi$ and $\Omega_i^j = 0$ for $j
> 1$ in Eq.~\ref{Eq:ConeWeights}. Without any reference system, the emission directions associated to the
photon packets would then be given by the unit vectors pointing from
source $i$ to neighbour $j$. In Appendix \ref{Appendix:A} we show that in this
case the net emission direction in general would correlate with the direction towards the centre of mass of the
neighbours. As a result, the emission process would depend on the
clustering of the neighbours in space as set by the geometry of the SPH
simulation. The use of emission cones combined with ViPs gives us the freedom to choose
emission directions independently of the spatial clustering of the neighbours. 
This allows us to model any angular dependence of the source emissivity, within the bounds of
the chosen angular resolution.
\par
In summary, source particles transfer photon packets to their neighbouring gas
particles using a set of randomly oriented, tessellating cones. Each cone
defines a direction of transport, i.e. the emission direction, associated with the packets. The random orientation
applied to the cone tessellation ensures that photon packets are not
transferred along a fixed set of directions only. Virtual particles (ViPs) are placed in emission cones not containing any
neighbours, to which the photon packets are then transferred to. The
combination of emission cones and ViPs makes the radiation transport independent of the spatial distribution of the neighbours.

\subsubsection{Photon transmission by gas particles and ViPs}
\label{Section:Transmission}
\begin{figure*} 
      \includegraphics[width=55mm]{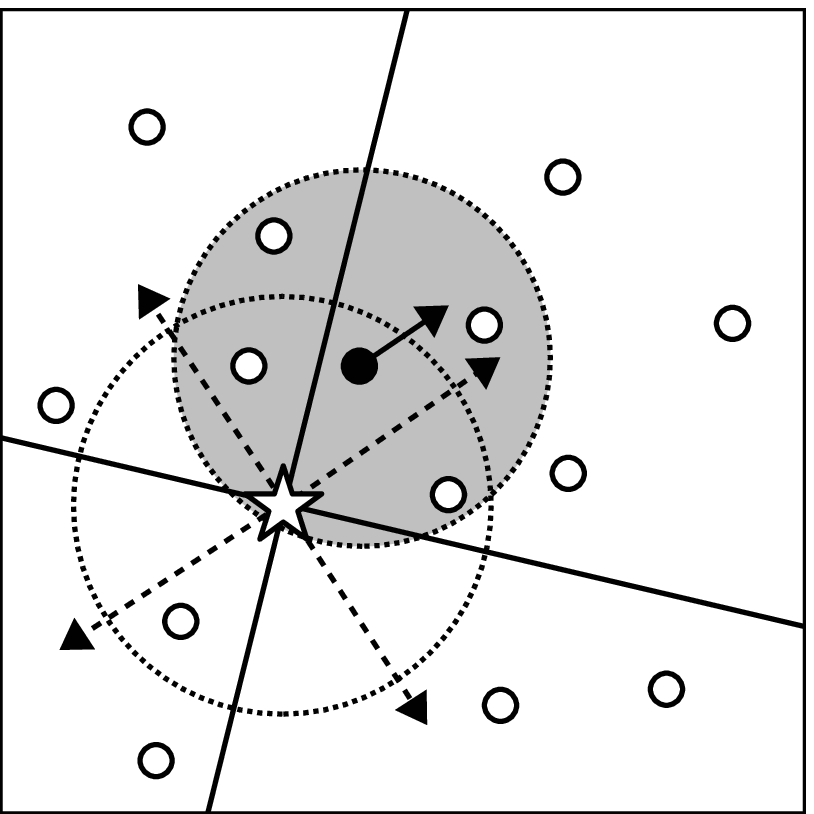} 
     \includegraphics[width=55mm]{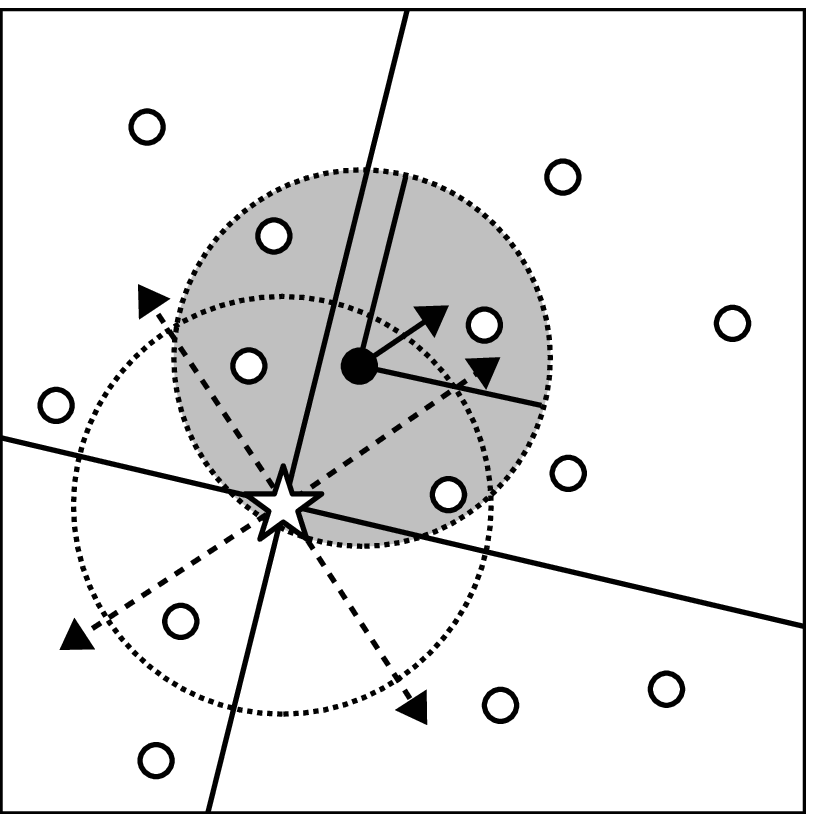} 
     \includegraphics[width=55mm]{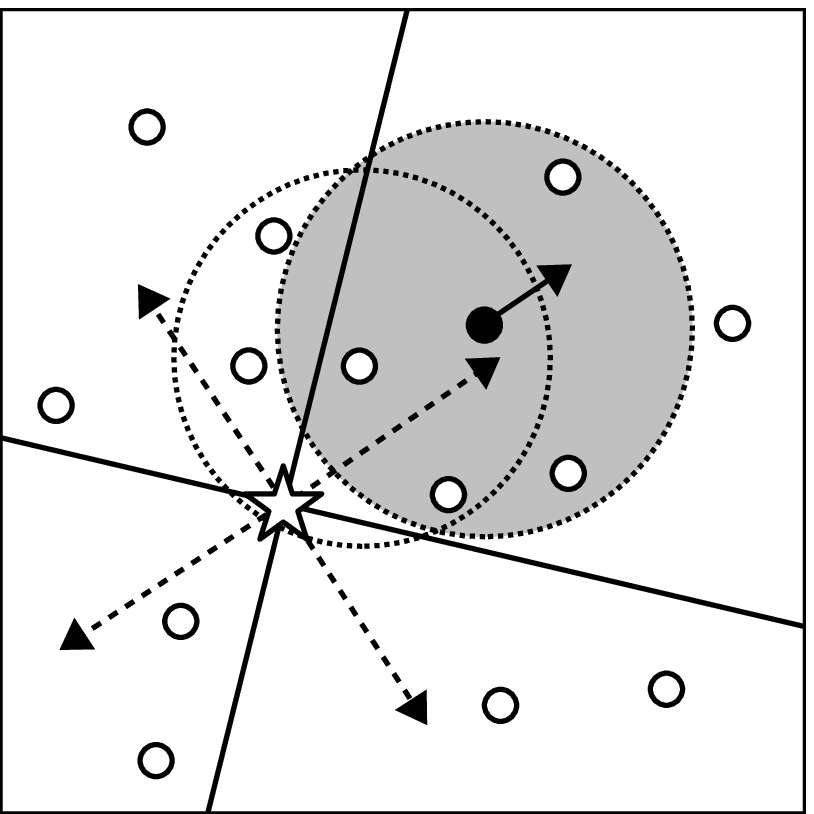} 

  \caption{Transmission of photon packets by gas and virtual particles. The
  particle positions shown are the same as in Fig.~\ref{Fig:Emission}, as
  are the neighbourhood (dashed circle) and emission cone tessellation of
  the source particle (white star). The transmission of photons is  
  illustrated for one of the neighbours that received radiation from the star
  in Fig.~\ref{Fig:Emission}. Left-hand
  panel: The neighbourhood (big grey disc) of the transmitting gas particle (black disc) is defined.
  As in Fig.~\ref{Fig:Emission}, $\tilde{N}_{ngb} = 4$. The emission
  direction associated with the photon packet to be transmitted is shown as
  the short solid arrow. Middle
  panel: The transmission cone is shown, centred on 
  the emission direction of the photon packet that is to be transmitted. In
  this cone one downstream neighbour is found. Right-hand panel: The photon packet is transmitted to the downstream
  neighbour, turning it into a transmitting particle. The cycle repeats with
  defining the neighbourhood for this particle. As a result, the photon packet is
  propagated downstream, radially away from the source particle.}
  \label{Fig:Transmission}
\end{figure*} 

In the last section we described how a source particle distributes photon packets amongst the gas particles 
in its neighbourhood. For each packet we defined an emission direction,
independently of the spatial distribution of the neighbouring gas particles
by employing a randomly oriented set of emission cones. In this section we
describe how the packets are propagated through the simulation box along their
emission directions, by employing directed particle-to-neighbour transport, a
process which we refer to as transmission.
\par
Consider a gas particle $i$, which receives a photon packet.
Recall that the neighbours of particle $i$ are the $\tilde{N}_{ngb}$ nearest gas
particles located in the sphere with radius $\tilde{h}_i$, centred on
particle $i$ (Fig.~\ref{Fig:Transmission}, left-hand panel).  
Analogous to the case of emission, particle $i$ re-distributes the photons
contained in the packet amongst its neighbours. In contrast to 
emission, photons are only distributed amongst the subset of neighbours
$\tilde{N}_{t,ngb} \le \tilde{N}_{ngb}$ located {\it downstream}. 
These are the neighbours residing in a {\it regular}\footnote{A regular cone is
  a pyramid with a circular cross-section.} cone centred on the direction of propagation of the
packet (see the middle panel of Fig.~\ref{Fig:Transmission}). We refer to this
cone as transmission cone. 
\par
\begin{figure} 
  \includegraphics[trim = 15mm 0mm 35mm 10mm, width=0.49\textwidth, clip=true]{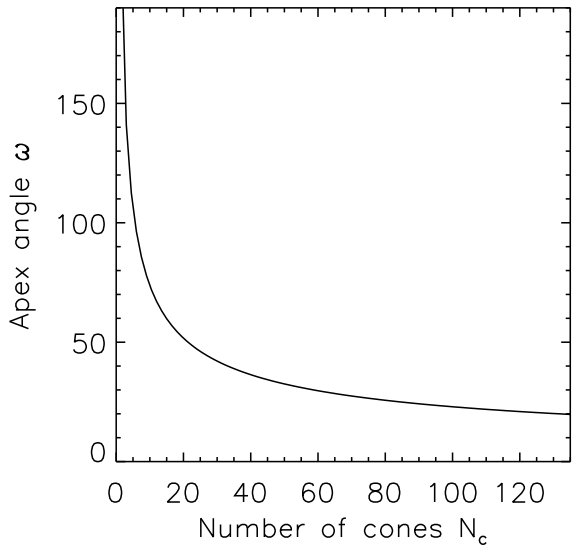} 
  \caption{The apex angle $\omega$ of the transmission cones as a function of the
  angular resolution $N_c$ (see Eq.~\ref{Eq:Apex}).} 
  \label{Fig:Apex} 
\end{figure}

 
The apex of the transmission cone is located at the position of gas particle
$i$. The solid angle of the cone is set by the
angular resolution, $\Omega_t = 4\pi / N_c \equiv \langle\Omega\rangle$, and
determines the apex angle $\omega$  through the standard relation
\begin{equation}
  \omega = 2 \arccos \left(1- \frac{\Omega_t}{2\pi}\right)\times \frac{180}{\pi}.
  \label{Eq:Apex}
\end{equation}
We show this relation in Fig.~\ref{Fig:Apex}. 
By definition, a neighbour $j$ with position $\vect{r_j}$ is interior to the transmission cone with apex at
the position $\vect{r}_i$ of the transmitting particle $i$ if the inner angle between
the transmission cone axis and the vector $\vect{r_j} - \vect{r}_i$ is
less than $\omega/2$.
\par
The received photon packet is split into $\tilde{N}_{t,ngb}$
packets, each of which is transferred to one of the downstream neighbours $j$
of particle $i$. The number of photons
contained in each packet is set by the weights $w^j_i$, such that each
neighbour $j$ receives a photon fraction $w^j_i / \sum_k w^k_i$  
of the parent photon packet, where the sum is over all downstream neighbours $k
= 1, ...,\tilde{N}_{t,ngb}$. Here we assume $w^j_i = 1$, giving equal weight
to each neighbour. If there are no
downstream neighbours, i.e. if the transmission cone is empty, we simply create a
neighbour as in the case of empty emission cones described in the previous
section. That is, we place a virtual particle (ViP) along the emission direction of the
photon packet, a random (in volume) radial distance $<\tilde{h}_i$ away from
the transmitting gas particle. The properties of the ViP are determined by SPH interpolation
from its neighbours, and the photon packet is transferred. 
\par
Each photon 
packet inherits the emission direction from its parent packet
(Fig.~\ref{Fig:Transmission}, right-hand panel). After all packets have been transferred to the downstream
neighbours, the transmission is finished. Subsequently, the transmission
procedure can be applied again. The
transmission process thus generally splits each photon packet into multiple packets, propagating with the same
emission direction. With each subsequent transmission individual photon packets are
confined to a smaller fraction of the solid angle traced out by the 
emission cone they were emitted into (see Fig.~\ref{Fig:Transmission}, middle panel). This is similar to the technique of ray
splitting employed in ray-tracing codes.
Hence, in \traphic\ the photon transport takes place adaptively in angle. 
\par
The transmission procedure specified above for gas particles is also applied for ViPs. Again,
the transmission cone of the ViP can either contain gas neighbours or be
empty. If it is empty, the ViP creates another ViP, as in
the case of transmission by gas particles. After the ViP has performed the
transmission, it is deleted. ViPs are therefore temporary constructs. For 
$N_c \gg \tilde{N}_{ngb}$ the total number of ViPs in the simulation is
proportional to $N_c$, whereas for $\tilde{N}_{ngb} \gg N_c$ the total number
of ViPs approaches zero.
\par
The main purpose of the transmission cones is to confine the downstream propagation of photons to
the solid angles into which they were emitted by the source particles, which is 
the main challenge for schemes using  particle-to-neighbour transport. Just as
emission cones were introduced to make the emission of photons by
source particles independent of the geometry of the SPH particle
distribution, transmission cones are introduced to further propagate the
photons downstream, independently of the geometry of the SPH particle
distribution. As a consequence, shadows will be thrown behind opaque
obstacles, and their sharpness will be in agreement with the chosen angular
resolution. In fact, as we will see in Section~\ref{Section:Test2}, the shadows
are much sharper than implied by the formal angular resolution.
\par
The cones making up the emission tessellation will generally be
irregular - in three dimensions there exists no tessellation of space with regular
cones. The use of regular cones for the downstream propagation of photon packets emitted
into a certain irregular emission cone is therefore an approximation. 
In principle, each photon packet could be transmitted into a cone having the
shape of the emission cone it originates from. However, tracing photons within
irregular cones is computationally more demanding. Furthermore, in the next section we will see
the necessity for combining multiple photon packets propagating along
different directions into a single photon packet that propagates along a
correspondingly averaged direction. Since there is no unique way of defining the
shape of a transmission cone associated with this average direction, 
we use a regular shape. 
\par

\subsubsection{Photon merging by gas particles}
\label{Sec:TRAPHIC:Merging}
\begin{figure*} 
  \includegraphics[width=55mm]{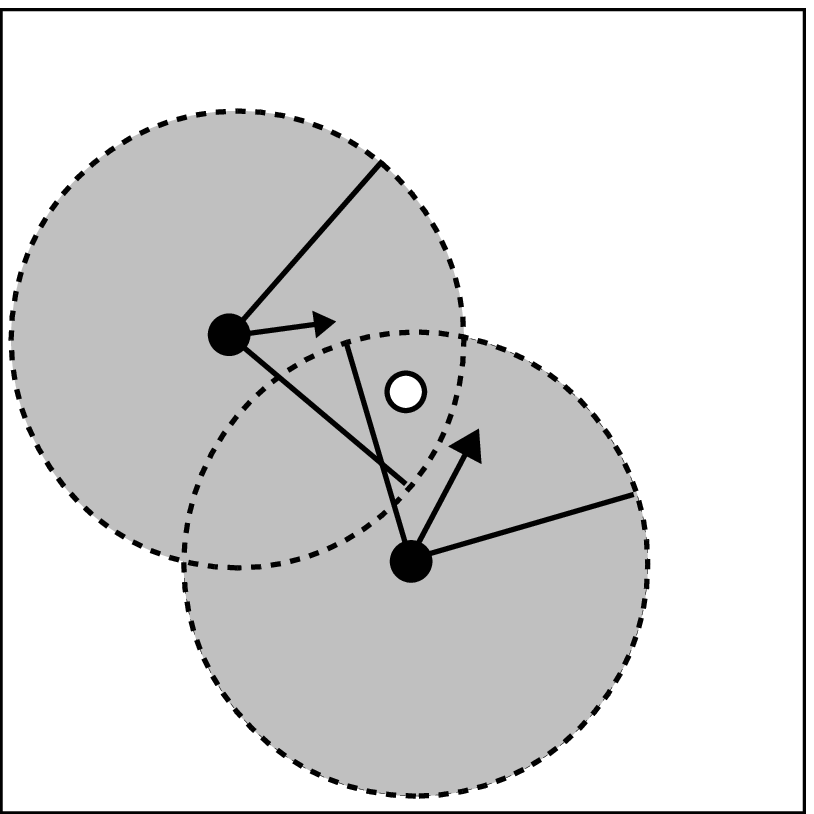} 
  \includegraphics[width=55mm]{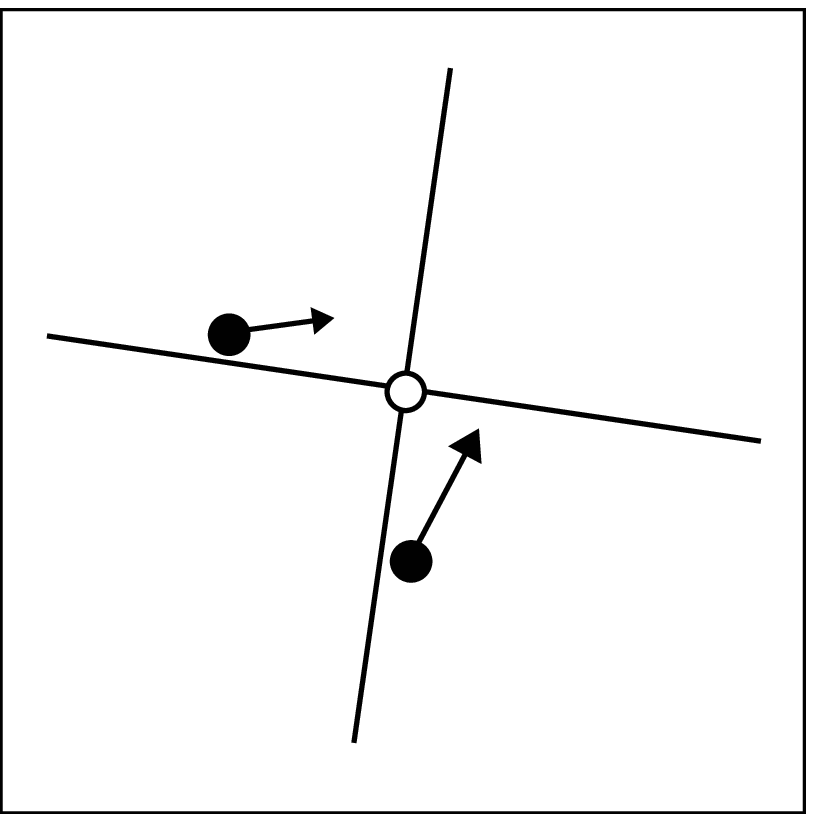} 
  \includegraphics[width=55mm]{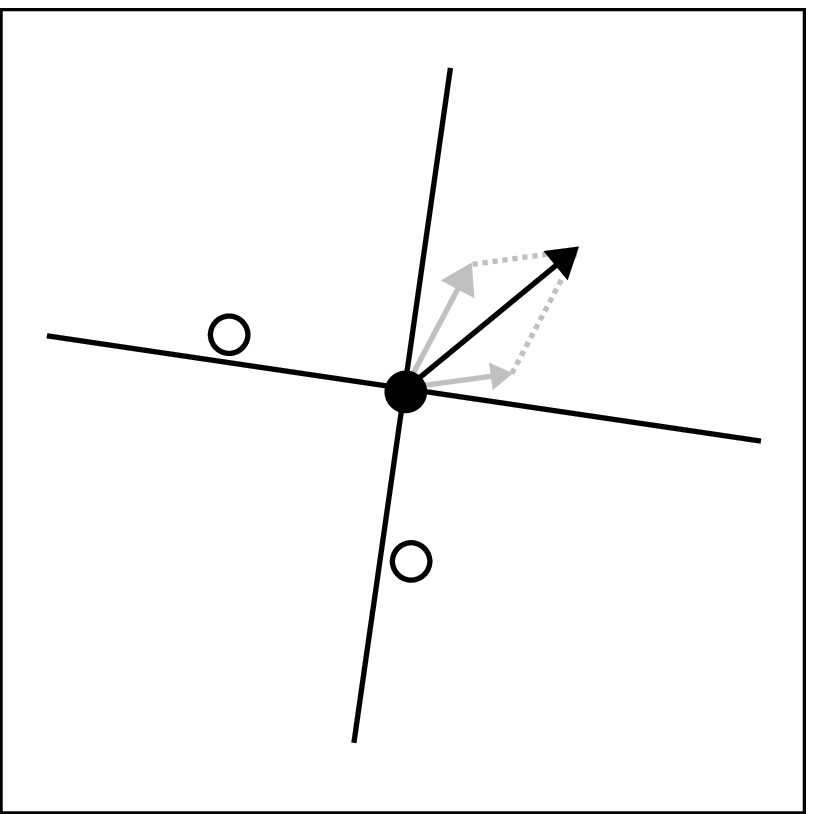} 

  \caption{Merging of photon packets by gas particles. Left-hand panel: A gas particle
  (white disc) is receiving photon packets simultaneously from two transmitting
  gas particles (black discs). The emission direction associated with each
  photon packet is indicated by an arrow, whose length is given by the number of
  photons contained in the packet. The big grey discs indicate the
  neighbourhoods of the transmitting gas particles, the solid lines are the
  transmission cones (we assume $N_c = 4$). For clarity, we do not show
  particles and photon packets that are not directly involved in the merging event. Middle panel: A randomly oriented reception cone tessellation is defined
  for the receiving gas particle. We omitted the
  transmission cones and the neighbourhoods of the transmitting particles. Right-hand
  panel: The photon packets have been transferred to the receiving gas
  particle, turning it into a transmitter. Their emission vectors (grey arrows), after translation to the position of the
  receiver, are both found to fall into the upper-right reception
  cone. Note that the particles itself fall into different
  reception cones. Their positions are irrelevant for the merging, since transmitted
  photon packets rather than transmitting particles are merged. The emission vectors are merged through a weighted average as described in Section
  \ref{Sec:TRAPHIC:Merging}, resulting in a single photon packet (black arrow).}
  \label{Fig:Merging}

\end{figure*} 

In the previous section we described the transmission of a {\it single} photon
packet arriving at a gas particle. In principle, each gas\footnote{We do not consider
  ViPs here, since by
  construction each ViP can receive only a single photon packet.} particle
can simultaneously receive {\it multiple} photon packets, possibly
emitted by different sources at different times, propagating along different emission
directions, a situation which is depicted schematically  in the left-hand panel of 
Fig.~\ref{Fig:Merging}. 
Accomplishing the transmission would then require executing a loop over all
received packets for each transmitting gas particle. Clearly, for each
gas particle this would imply 
a linear scaling of the computation effort with the number of packets that need to be transmitted, or in
other words, with the total number of source particles present in the
simulation. 
\par
In cosmological simulations of structure formation, for instance, where a significant number of
star particles can usually be found already at high redshift, the simultaneous
reception of multiple photon packets will be the rule rather than the
exception. Even without taking into account the diffuse radiation of the intergalactic gas
resulting from radiative de-excitations, recombinations and scatterings, 
a linear scaling with the number of sources would quickly turn the radiation transport into a computational
task that would be too expensive to be carried out even with the most advanced
supercomputer available today. 
\par
A few radiation tracing schemes have attempted to tackle this
scaling problem by replacing sources that are close to each
other with a single point source (e.g.~\citealp{Cen:2002};
\citealp{Razoumov:2006}; \citealp{Gnedin:2001}). Here we introduce a photon
packet merging procedure that strictly limits the number of packets that need to be 
transmitted per particle to at most $N_c$, without restricting the number of directions along which each individual
packet can propagate.
\par
For each gas particle $i$ receiving a photon packet we define a so-called reception
cone tessellation, as shown in the middle panel of Fig.~\ref{Fig:Merging}. 
A reception cone tessellation is a tessellating set of $N_c$
cones attached to a gas particle. It is identical to the
set of $N_c$ cones employed for the emission of photons by source particles, but
generally has a different orientation. As their name indicates, and in contrast to the emission cones, reception cones are used to
collect photons. Whenever a photon packet $p$ is transferred to gas particle $i$, the
packet is binned into one of the reception cones by examining into which
reception cone
its emission direction $\vect{n}_p$ falls, after translation to the location of the gas
particle. 
Photon packets whose  emission directions 
fall in one and the same cone are merged, leaving only a single packet for
transmission (Fig.~\ref{Fig:Merging}, right-hand panel). 
The reception cone tessellation is given a random orientation to
prevent artefacts.
\par
A merged photon packet created in cone $k$ is assigned the
new, merged emission direction $\vect{n}_{m,k}$. 
This new emission direction is defined as the weighted average  
$\vect{n}_{m,k} = \sum w_p \vect{n}_p/ \sum w_p$, where the sum is over all photon
packets received in emission cone $k$ ($k = 1, 2,..., N_c$), and the weights $w_p$ are
the number of photons per packet $p$ that are to be transmitted into direction $\vect{n}_p$ to the
downstream neighbours of particle $i$ (Fig.~\ref{Fig:Merging}, right-hand
panel). Similarly, the merged photon packet is associated a merged clock
$t^\star_{m,k}$ by averaging over the individual clock times  $t^\star_p$ in
reception cone $k$, i.e. $t^\star_{m,k} = \sum w_p t^\star_p/ \sum w_p$.
\par
As a result of the photon packet (resp. source) merging at most $N_c$ packets have
to be transmitted per gas particle, fully consistent with the angular resolution of our
transport scheme. The computation effort required to perform radiative
transfer simulations with \traphic\ can therefore be readily controlled: In
the most demanding situation, i.e. in the optically thin limit, when the whole box is filled
with radiation, it merely scales with the product of spatial and angular resolution, $N \times
\tilde{N}_{ngb} \times N_c$, where $N$ is the total number of particles (SPH +
stars). The merged photon packets are not associated with any existing source particle in the
simulation. They should be thought of as emerging from an imaginary source, whose
properties are implicitly defined by our merging prescription. The merged
photon packets are traced further downstream, radially away from this imaginary source, into the 
merged emission direction, exactly as was described for non-merged emission directions 
in the previous section.
\par

\subsection{Photon interactions with gas particles}
\label{Sec:Method:Interaction}
Photons are propagated radially away from each source until they interact with
the matter represented by the SPH particles. Here we distinguish between
absorptions and scattering interactions and describe how they are accounted
for in \traphic. We remind the reader that in this work we do not discuss the effect of
interactions on the thermodynamical and hydrodynamical evolution of the SPH particles.
\subsubsection{Absorption}

Absorptions remove photons from the beam emitted by the source. 
This process is described in terms of  
the mass absorption coefficient $\kappa_\nu$. From the formal solution of the
radiative transfer equation it can easily be seen (e.g. \citealp{Mihalas:2000}) that the fraction of photons of 
frequency $\nu$ that is absorbed while travelling along the straight line connecting the positions
$\vect{r}_1$ and $\vect{r}_2$ is given by $1-\exp(-\tau)$, where
\begin{equation}
\tau = \int_{\vect{r}_1}^{\vect{r}_2} dr\ \kappa_\nu(r) \rho(r),
\label{Method:OpticalDepth}
\end{equation}
is the optical depth over the distance $d = |\vect{r}_1 - \vect{r}_2|$.
\par
In \traphic, absorptions are accounted for by removing the photon fraction
$1-\exp(-\tau_{ij})$ from photon packets propagating between 
particle $i$ and its neighbouring particles $j$. For the calculation of the
optical depth $\tau_{ij}$ between these two particles we need to 
numerically evaluate the integral Eq.~\ref{Method:OpticalDepth}. This is 
computationally expensive, since it involves
the calculation of the density $\rho$ (and the absorption coefficient
$\kappa_\nu$) at a large number of points along the photon path using the SPH
formulation. Similar to the approach of \cite{Kessel-Deynet:2000}, 
we therefore approximate the density field near a gas or virtual particle $i$ 
by the SPH density $\rho_i$ evaluated at its position.
\par
Since photon packets are propagated between 
particle $i$, located at $\vect{r}_i$, and particle $j$, located at
$\vect{r}_j$, along their emission direction, i.e. in the direction of the unit vector $\vect{n}$, the
distance $d$ they cover is generally not equal to the distance between 
the particles. Instead, we set $d = |\vect{n} \cdot (\vect{r}_i - \vect{r}_j)|$,
i.e., we employ the projection of the particle distance along the emission
direction. This is a valid approximation far away
from the source from which the photon packets originate, but it generally fails close
to the source. Therefore, we only employ the projected distance when
propagating photons between a transmitting particle and its neighbours. On the
other hand, when propagating photons from a source particle to its neighbours, we use the
particle distance, i.e. we set $d = |\vect{r}_j - \vect{r}_i|$, corresponding
to radially outward propagation. 
\par
The distance $d$ is used to obtain the optical depth $\tau_{ij}$ between particle $i$ and particle $j$,
$\tau_{ij} = \kappa_\nu \rho_j d$. This approximation neglects
the existence of any substructure 
between particle $i$ and its neighbouring particle $j$. However, this does not result in
a loss of information if the radius $\tilde{h}_i$ of the
neighbourhood of particle $i$ is chosen such that $\tilde{h}_i \lesssim h_i$ 
(resp. $\tilde{N}_{ngb} \lesssim N_{ngb}$).
\par
The distance $d$ is furthermore used to advance the clock $t^\star$ of each photon packet
according to the rule 
\begin{equation}
t^\star \to t^\star + d / c.
\label{Eq:clock}
\end{equation}
By synchronising the clock $t^\star$ with the simulation time $t_r$, the clock can be employed to advance
photon packets at the speed of light, as we discuss in Sec.~\ref{Sec:Evolution}.
We note in passing that the clock $t^\star$ can also be used to implement the cosmological redshifting of photons. 
\par
The number of absorbed and transmitted photons is
now calculated as follows. From the photon packet emitted or transmitted by particle $i$ to
particle $j$, a fraction $1-\exp (-\tau_{ij})$ is absorbed by particle $j$. 
The photon packet is subsequently transmitted by 
particle $j$ containing a fraction $\exp(-\tau_{ij})$ of the original number
of photons. By construction, this procedure explicitly conserves photons,
since 
\begin{equation}
1-e^{-\tau_{ij}} + e^{-\tau_{ij}} = 1.
\end{equation}
Photons absorbed by ViPs are distributed amongst all 
neighbours of the ViP using (photon-conserving) SPH interpolation. This is
necessary, because ViPs are a temporary construct employed for
the transport of photons; permanent information is only stored at the
particles in the SPH simulation. For further reference, 
we denote the total number of photons impinging on and the total number of
photons absorbed by particle $i$  over a single time step $\Delta
t_r$ with  $\Delta \mathcal{N}_{in,i}$  and $\Delta \mathcal{N}_{abs,i}$, resp.
\subsubsection{Scattering}
In contrast to absorptions, scatterings change the direction and possibly the frequency of the interacting
photons. A scattering event can be thought
of as an absorption event followed by an immediate re-emission. 
Scatterings can hence be described by Eq.~\ref{Eq:RTE} after adapting the absorption coefficient $\kappa_\nu$  and the
emissivity $\epsilon_\nu$. The re-emission re-distributes the photons in angle (and possibly frequency). For
this reason the scattered photons are sometimes referred to as diffuse. 
Adapting the emissivity for scatterings requires
evaluating an integral over the intensity. Therefore, scatterings turn the
radiative transfer equation (Eq.~\ref{Eq:RTE}) into an integro-differential
equation. 
\par
In \traphic, scatterings, that is the transport of diffuse photons, are
modelled by a combination of absorption and re-emission events. Photons to be scattered
travelling between two particles are first removed from the photon packet as
described in the last section. For a true absorption event, the photon energy would be lost
to the thermal bath provided by the matter. In contrast, in case of scattering, a gas particle
that absorbed photons re-emits them, closely following the description for source
particles in Section~\ref{Section:EmissionBySourceParticles}. 
\par
In particular, we employ randomly oriented
emission cones in combination with Eq.~\ref{Eq:ConeWeights}  to re-emit the
absorbed photons.  Since modelling the
scattering process means following photons emitted earlier by a source, the clock of the photon packets that are
to be re-emitted is not set to the simulation time $t_r$, as was the case
for the intrinsic emission of photons by source particles. Instead, it is determined by the clock of the photon packets
that are scattered and the details of the scattering interactions (e.g. there
could be a time delay between absorption and re-emission, or photons could
have travelled a distance greater than the (projected) particle distance $d$
when inferring the properties of the scattering process from a sub-resolution model).
If the scattering details are taken into account by correspondingly adjusting $d$, Eq.~\ref{Eq:clock} can be employed
to find the new clock time.  Note that it is crucial for the modelling of
scatterings to employ a radiative transfer scheme which does not scale with the
number of sources, since every gas particle will soon re-emit photons. 
\par

\subsection{Solving the radiative transfer equation}
\label{Sec:Evolution}
\traphic\ solves the radiative transfer equation (Eq.~\ref{Eq:RTE}) in discrete time steps $\Delta t_r$, the
size of which depends on the details of the problem under
study and has to be decided on a case by case basis. Hence, we defer its discussion
to the second part of this work, where we present an implementation of our
method to solve specific problems. In this section we briefly review the
concepts described in the previous sections used to obtain the solution to the radiative transfer equation over a single
time step before discussing how this solution is advanced in time.
\par
Starting at simulation time $t_r$, the size of the radiative transfer time step
$\Delta t_r$ is determined. Source particles emit 
photon packets to their $\tilde{N}_{ngb}$ neighbouring gas
particles employing $N_c$ cones. The total number of photons
contained in the packets is determined by the
number of photons $\dot{\mathcal{N}}_{\nu,i}$ to be emitted per unit time  in
frequency bin $\nu$  and by the size of the time step  $\Delta t_r$. 
For each neighbouring gas particle the optical depth to the
source particle is evaluated and the number of photons interacting with the
matter on their way from the source particle is inferred. Some photons
may be absorbed, some may be scattered; the remaining photons are left
unimpeded for transmission. 
\par 
For the subsequent transmission of photons by the gas particles, photons associated to 
sources that are seen as close (in angle) to each other are
merged, limiting the number of directions into which photons have to be
transmitted to at most $N_c$. 
Next, for each transmitting and scattering gas particle the $\tilde{N}_{ngb}$ neighbours
are found. Photons to be transmitted are distributed amongst the subset
$\tilde{N}_{t,ngb}$ of neighbours
that are located downstream,  as determined by the corresponding (merged)
emission directions. Photons to be scattered are distributed amongst all
$\tilde{N}_{ngb}$ neighbours following the prescription in
Section~\ref{Sec:Method:Interaction}. Virtual particles (ViPs), which are placed in
cones that do not contain any neighbours, are deleted after
having transmitted or scattered their photons. 
Photons that were absorbed by a ViP are distributed amongst the
neighbours of the ViP using (photon-conserving) SPH interpolation.
\par
The cycle described in the previous paragraph repeats until a stopping
criterion is satisfied.
The form of the stopping criterion depends on whether one solves the
time-{\it independent} or the time-{\it dependent} radiative transfer equation.
When solving the time-independent radiative transfer equation,
i.e. when  neglecting the first term on the left hand side of
Eq.~\ref{Eq:RTE}, one formally assumes $c\to\infty$. 
Accordingly, the stopping criterion becomes independent of the speed of light. The
cycle can then for example be repeated until all photons have been absorbed or have left the simulation box.
In contrast, when solving the time-dependent radiative transfer equation,
the stopping criterion is directly tied to the speed of light $c$: the cycle is
repeated until all photons have propagated a distance $c \Delta t_r$. 
\par
To decide whether or not a photon packet has travelled over the distance $c \Delta t_r$,
we employ its clock $t^\star$. For further illustration it is convenient to explicitly follow the clock of a photon
packet as it is propagated through the simulation box. Upon emission, the clock of the photon packet is initialized with the
simulation time $t_r$, as already mentioned in Sec.~\ref{Section:EmissionBySourceParticles}. 
After the emission, when the photon packet has been transferred from the source to one of the neighbouring SPH particles (or to a ViP),
its clock is advanced according to Eq.~\ref{Eq:clock}. It is then checked whether the clock has reached or crossed the threshold value
\begin{equation}
t^\star_{th} = t_r + \Delta t_r.
\label{Eq:PropagationCondition} 
\end{equation}
If not, the photon packet is transmitted further downstream. As before, its clock is advanced according to Eq.~\ref{Eq:clock}.
The photon packet is repeatedly transmitted and its clock is advanced until it has reached or crossed the threshold value 
defined in Eq.~\ref{Eq:PropagationCondition}.
At this point, the value displayed by its clock can in 
full generality be expressed as $t^\star = t^\star_{th} + \epsilon_0$, with $\epsilon_0 \ge 0$. 
\par
The propagation error, $\epsilon_0$, is typically larger than zero because the set of particles on which the 
photon packets are propagated is discrete, with the particle-to-neighbour distances 
generally not related to $c\Delta t_r$. Employing Eq.~\ref{Eq:PropagationCondition} to stop photon packets 
therefore results in the photon packets typically being propagated too far.
Due to the Lagrangian nature of the SPH simulation, $\epsilon_0$ 
is individual for each photon packet. If $d$ is the particle-to-neighbour distance corresponding to the emission or 
transmission event after which the photon packet was stopped, then $\epsilon_0 < d / c$. The propagation error $\epsilon_0$ is therefore strictly 
limited, $\epsilon_0 < \max_i \tilde{h}_i / c$, where the maximum is over all emitting and transmitting particles. Note
that $\epsilon_0$ becomes smaller with higher spatial resolution.
\par
Once stopped, a photon packet is held (and hence its clock stands still) until the simulation time
progresses to a value $t_r + \Delta t_r > t^\star$, at which point the packet is transmitted and its clock is advanced again. 
The clocks of photon packets are thus kept in synchronization with the current simulation time, which effectively
matches their propagation
speed to become the speed of light\footnote{We note that the clocks can also be employed to propagate photon packets 
at speeds $\tilde{c}$ different from the speed of light (cp. the reduced speed of light approximation suggested in \citealp{Gnedin:2001}),
by simply making the replacement $c \to \tilde{c}$.}. 
Photon packets may be held over several radiative transfer time steps, as $\epsilon_0$ may be (several times) larger than 
$\Delta t_r$. While the photon packet is held, its propagation error with respect to the current simulation time 
steadily decreases with each passing radiative transfer time step.
Since $t^\star$ is fixed while $t_r$ is increasing in steps of $\Delta t_r$, the current propagation error decreases 
according to $\epsilon_n = \epsilon_{n-1} - \Delta t_r$, where $n$ 
indicates the number of radiative transfer time steps that have passed since the photon packet was stopped.
Immediately before the photon packet is propagated again, its propagation error is 
strictly limited by the size of the time step,  $-\Delta t_r \le \epsilon_n < 0$. The photon 
packet is then propagated such as to again bring the clock into synchronization with the simulation time, as described above.
\par
When multiple photon packets are merged
into a single packet, the clock of the merged packet is determined by the
averaging procedure described in Section~\ref{Sec:TRAPHIC:Merging}. As a
result, photons may be propagated over distances that differ
somewhat from the case without merging. We have seen above that, due to the synchronization of the photon packet propagation
with the current simulation time $t_r$, clocks display only values 
in the interval $t_r \le t^\star \le t_r + \Delta t_r + \epsilon_0$. 
The difference in the clocks of different photon packets and hence the photon propagation error introduced by the merging procedure
can therefore be controlled to become arbitrarily small by increasing the 
temporal (i.e. decreasing $\Delta t_r$) and spatial (leading to smaller $\epsilon_0$) resolution.
\par
After the stopping criterion (time-independent or time-dependent)
has been fulfilled for all photon packets, the properties of gas particles are updated according to the
radiative interactions that occurred.
We emphasize that the properties of the gas particles (e.g. the ionization state) 
are therefore only updated at the end of each time step. Within
each time step the order in which photon packets from different sources arrive at gas particles
is therefore irrelevant. After updating the particle properties, 
the simulation time is advanced, $t_r \to t_r + \Delta t_r$. Finally, the size of the
next time step is determined and the radiation is transported as described above.
\par
In summary, we have presented a
radiative transfer scheme for use in SPH simulations that works directly on
the unstructured grid formed by the 
discrete set of irregularly distributed SPH particles. \traphic\ thus employs the full spatial resolution
of the SPH simulation. It achieves directed transport of radiation by
adaptively tracing photon packets in cones. \traphic\ can be used to solve both the
time-independent and the time-dependent radiative transfer equation in an explicitly photon-conserving way. 
Our scheme is by construction parallel on distributed memory machines if
the SPH simulation itself is parallel on distributed memory machines. 
Furthermore, the computation time necessary to accomplish the radiation
transport does not scale linearly with the number of sources. Instead, it
merely scales with the product of spatial and angular resolution, making our scheme
suitable for simulations containing a large number of sources as well as for
taking into account a diffuse radiation component.
\par

\subsection{Reduction of particle noise}
\label{Sec:Method:Regularization}
The radiative transfer equation  describes the propagation of 
photons within a continuum medium. In our scheme, however, photons are 
propagated on the discrete set of SPH particles, localised at irregular positions 
dictated by the SPH simulation. Consequently, numerical noise may arise from
the discreteness and irregularity of the spatial distribution of SPH particles
and this could  influence the numerical solution of the radiative transfer equation obtained with \traphic.
\par
To see how this {\it particle noise} can be reduced, it is helpful to employ a formal analogy (e.g. \citealp{Monaghan:2005})
between estimating the density in SPH simulations and estimating a 
probability density from sample points: We can consider the positions 
of the SPH particles as a random\footnote{We stress that SPH itself is not 
a Monte Carlo method, as first noted in \cite{Gingold:1978}. 
We only appeal to the Monte Carlo picture for use with the radiative transfer, not 
for the hydrodynamic evolution of the particles in the SPH simulation.} 
sample drawn from a probability density function proportional 
to the mass density. We can therefore Monte Carlo resample\footnote{We explicitly note that
the resampling does not affect the hydrodynamical simulation, for which the original positions are used.} the density field
without sacrificing its information content. Periodically assigning new positions to the
SPH particles during the radiation transport by resampling the density field
should lead to a better representation of the continuum physics by the discrete
set of SPH particles and hence lead to a reduction of particle noise.
\par
To employ the resampling, we think of particle $i$ at position $\vect{r_i}$ in the 
simulation box as being de-localised within its sphere of influence of radius 
$h_i$ centred on $\vect{r_i}$ and assume that the probability of finding it in the volume 
$d^3r$ around a particular point $\vect{r}$ in that sphere is given by the value 
of the interpolation kernel at that point (cp. the scattering approach of
Section~\ref{Sec:SPH}). For the radiative transfer on a 
static set of particles with positions $\vect{r}_i$
we then periodically Monte Carlo re-distribute the particles within their sphere of influence according 
to this probability.  We note
that while we change the positions of the SPH particles, we keep all their
other properties (e.g. density) fixed in order to avoid introducing
scatter in the physical variables. We describe the numerical implementation of the
resampling in Appendix~\ref{Sec:Resampling:Implementation}.
In Sections~\ref{Section:Test1} and \ref{Section:Test2} we will demonstrate
that the application of this recipe leads to a significant reduction of
particle noise. We will, however, also see that for most applications the noise is
small. We therefore do not employ the resampling technique in our simulations,
unless explicitly stated.
\par
Since the resampling randomly offsets the SPH particles from their 
positions provided by the hydrodynamical simulation, the apexes of the transmission cones 
attached to them are randomly offset, too. The resampling procedure may therefore lead to a slight diffusion of photons out
of the emission cone they were emitted into, effectively decreasing the angular
resolution. Note that a similar diffusion of photons could
occur when solving the radiative transfer equation coupled to the
hydrodynamics (on which we will report in a future publication), 
because of the physical particle motion. We will study this diffusion in our numerical implementation of
\traphic\ in Section~\ref{Section:Test2}.

\section{Application - Transport of ionizing radiation}
\label{Sec:Application}
In this section we apply the radiative transfer scheme presented in
Section~\ref{Sec:Method} to the transport of ionizing radiation. 
Ionizing radiation is thought to play a key role in determining the
ionization state and shaping the spatial
distribution of the baryonic matter in the universe on both small and large
scales. Examples include the triggering and quenching of star formation through radiative
feedback from nearby ionizing stellar sources both in the early
(e.g.~\citealp{Yoshida:2007}; \citealp{Wise:2007}; \citealp{Johnson:2007};
\citealp{Alvarez:2006}; \citealp{Susa:2006b}) and
present-day universe (e.g.~\citealp{Gritschneder:2007}; \citealp{Dale:2007b}), 
the thin shell instability (for a recent simulation see \citealp{Whalen:2007}) and the growth and
percolation of ionized regions generated by the first stars and quasars during the
so-called Epoch of Reionization (for recent simulations see
e.g. \citealp{Iliev:2006a}; \citealp{Trac:2006}; \citealp{Kohler:2007};
\citealp{Paschos:2007}).
\par
In the following we describe a numerical implementation of \traphic, our radiative
transfer scheme, specified for the transport of mono-chromatic
hydrogen-ionizing radiation, in the state-of-the-art SPH code
\gadget\ (\citealp{Springel:2005}).
We start in Section~\ref{Sec:Photoionization} with
a short review of the photo-ionization process. In
Section~\ref{Sec:NumericalImplementation} we present the details of 
our numerical implementation, establishing the connection 
to the general description given in Section~\ref{Sec:Method}. Finally, in
Section~\ref{Sec:Application:Tests}, we perform several radiative transfer
simulations, comparing the performance of our numerical implementation of
\traphic\ in three different test problems to both analytic solutions and numerical simulations carried out
with other radiative transfer codes as reported in the literature.

\subsection{Photo-ionization rate equation}
\label{Sec:Photoionization}
Here we briefly recall the principles of the photo-ionization and
recombination process occurring for a hydrogen-only gas of mass density $\rho$ 
exposed to hydrogen-ionizing radiation. We will employ the equations derived in this section
in the description of the numerical implementation of \traphic.
\par
Hydrogen-ionizing photons can be absorbed by neutral hydrogen. 
We approximate the frequency-dependence of the mass absorption 
coefficient $\kappa_\nu$  for hydrogen-ionizing 
radiation (e.g.~\citealp{Osterbrock:1989}) by 
\begin{eqnarray}
  \kappa_\nu  &\equiv& \frac{\sigma_\nu n_{HI}}{\rho}\label{Photoionzation:Coefficient}\\
  \sigma_\nu &=& \sigma_0 \left(\frac{\nu}{\nu_0}\right)^{-3} \Theta(\nu - \nu_0),
  \label{Photoionzation:Cross-section}
\end{eqnarray}
with $n_{HI} = (1 - \chi) \rho / m_H$ the neutral hydrogen number density, 
$\nu_0$ the Lyman-limit frequency of energy $h_p\nu_0 = 13.6 \eV$, 
$\sigma_0 = 6.3 \times 10^{-18} \cmsq$ the absorption cross-section for
photons at the Lyman-limit, $m_H$ the mass of a hydrogen atom and $\Theta(x)$
the Heaviside step function; the ionized fraction is $\chi \equiv n_{HII}/n_H$. 
The number of photo-ionizations per unit time per neutral hydrogen atom at a certain 
point in space is determined by the photo-ionization rate $\Gamma$,
\begin{equation}
  \Gamma = \int_0^\infty d\nu\ \frac{4\pi J_\nu(\nu)}{h_p \nu} \sigma_\nu,
  \label{Photoionzation:Rate}
\end{equation}
where $J_\nu \equiv \int d \Omega\ I_\nu / (4\pi)$ is the mean ionizing intensity.
The rate of change of the neutral fraction $\eta \equiv 1-\chi$ at this point
is then
\begin{equation}
  \frac{d}{dt}\eta = \alpha(T) n_e \chi - \Gamma \eta \equiv
  \frac{\chi}{\tau_{rec}} - \frac{\eta}{\tau_{ion}}.
  \label{Photoionization:RateEquation}
\end{equation}
In the last equation,  $\alpha(T) n_e$ is the number of recombinations
occurring per unit time per ionized hydrogen atom, $\tau_{rec}$ is the
recombination time-scale and $\tau_{ion}$ is the  photo-ionization time-scale.\footnote{Note that
  in Eq.~\ref{Photoionization:RateEquation} collisional ionizations can
  easily be taken into account by replacing $\Gamma$  with $(\Gamma + C(T) n_e)$, where
  $C(T) n_e$ describes the number of collisional ionizations per unit time per neutral
  hydrogen atom.  In this work, however, we assume that
  collisional ionizations are unimportant, setting $C \equiv 0$ throughout.} 
\par
With the definition $\tilde{\chi}\equiv \tau_{rec}/ (\tau_{ion} +
\tau_{rec})$ we can rewrite Eq.~\ref{Photoionization:RateEquation} to read
\begin{equation}
  \frac{d\chi}{dt} = 
  -\frac{\chi-\tilde{\chi}}{\tau_{ion}\tilde{\chi}}.
  \label{Photoionization:RateEquation:Reformulation}
\end{equation}
Setting  $d\chi/dt  = 0$ yields the the equilibrium
ionized fraction  $\chi_{eq}= \tau_{rec,eq}/ (\tau_{ion} +
\tau_{rec, eq})$. Over time-scales that are short compared with 
$\tau_{rec}/|d\tau_{rec}/dt|$ and $n_e/|dn_e/dt|$, Eq.~\ref{Photoionization:RateEquation:Reformulation} 
constitutes a first order linear homogeneous differential equation
in $\chi-\tilde{\chi}$ with constant coefficients, whose   
solution reads
\begin{eqnarray}
  \chi(t)-\chi_{eq} & = & (\chi(t_0)-\chi_{eq})e^{-\frac{t -
      t_0}{\tau_{eq}}} 
  \label{Photoionization:RateEquation:Reformulation:Solution}\\
  \tau_{eq} & \equiv & \frac{\tau_{ion}\tau_{rec}}
      {\tau_{ion}+\tau_{rec}}.
\end{eqnarray}
From Eq.~\ref{Photoionization:RateEquation:Reformulation:Solution} we see
that the equilibrium ionized fraction is exponentially 
approached on the instantaneous ionization equilibrium time-scale $\tau_{eq}$. 
We will employ this time-scale later on for the numerical
integration of the rate equation.
\par

\subsection[]{Numerical implementation}
\label{Sec:NumericalImplementation}
We have adapted \traphic\ for the transport of hydrogen-ionizing radiation
according to the physics of photo-ionization presented in
Section~\ref{Sec:Photoionization} and implemented it using a single frequency
bin in the parallel N-body-Tree-SPH code 
\gadget\ (\citealp{Springel:2005}). The description of this
implementation is the subject of this section. 

\subsubsection{Transport of ionizing photons}
The transport of ionizing photons is performed in finite
radiative transfer time steps of size $\Delta t_r$, employing the 
absorption coefficient $\kappa_\nu$  given by
Eqs.~\ref{Photoionzation:Coefficient} and
\ref{Photoionzation:Cross-section}. Photon packets are transported in cones
according to the description given in Section~\ref{Sec:Method}.  
At the end of each radiative
transfer time step, i.e. at time $t_r+ \Delta t_r$, where $t_r$ is the current
simulation time, we know the number of ionizing photons
$\Delta\mathcal{N}_{in,i}$ impinging on and the number of ionizing photons $\Delta\mathcal{N}_{abs,i}$ absorbed by
particle $i$  over the time interval $\Delta t_r$ (cp. Section~\ref{Sec:Method:Interaction}). 
The photo-ionization rate $\Gamma_i$
is obtained directly, that is without explicitly referring to the mean intensity $J_{\nu}$, using
\begin{equation}
\eta_i^{t_r} \frac{m_i^{t_r} X_i^{t_r} }{m_H} \Gamma_i^{t_r} \Delta t_r=\Delta\mathcal{N}_{in,i} \left[1 -
\exp(-\tau_i^{t_r}) \right ],
\label{Eq:Photoionizationrate}
\end{equation}
where $X_i$ is the hydrogen mass fraction and
\begin{equation}
\tau_i^{t_r} \equiv - \ln\left(1- \frac{\Delta\mathcal{N}_{abs,i}} {\Delta\mathcal{N}_{in,i}}\right)
\label{Eq:AposterioriOpticalDepth}
\end{equation}
is the a posteriori optical depth and we use superscripts to indicate the time at which quantities
are evaluated. In the next section we describe how the photo-ionization rate is used to
update the neutral fraction of particle $i$.

\subsubsection{Solving the rate equation}
\label{Sec:Subcycling}
In the simulation the neutral fraction associated with any gas particle $i$ 
is assumed to be constant over the time step $\Delta t_r$ (see
Section~\ref{Sec:Evolution}). The correspondingly
discretized rate equation would read (cp. Eq.~\ref{Photoionization:RateEquation}),
\begin{equation}
\eta_i^{t_r+\Delta t_r} - \eta_i^{t_r}  =  
  \alpha_i^{t_r}  n_{e,i}^{t_r} \chi_i^{t_r}\Delta t_r - \Gamma_i^{t_r}
  \eta_i^{t_r} \Delta t_r
\label{Eq:DiscreteRateEquation}
\end{equation}
According to Eq.~\ref{Photoionization:RateEquation:Reformulation:Solution}, 
however, the neutral fraction evolves on the time-scale
$\tau_{eq}$ during the photo-ionization process. 
In order to accurately follow the evolution of the neutral fraction using
Eq.~\ref{Eq:DiscreteRateEquation} one would have to choose time steps $\Delta
t_r \ll \tau_{eq}$ limited by the ionization equilibrium time-scale, over
which our assumption of a constant neutral fraction is a good
approximation. 
\par
In our implementation we choose the radiative transfer time step $\Delta t_r$
independently of the time-scale $\tau_{eq}$, which can be prohibitively small, 
by employing the following sub-cycling strategy at the end of each radiative
transfer time step. We only assume 
that the flux $d\mathcal{N}_{in, i} / dt$ of ionizing photons impinging on particle $i$ is constant over
the time step $\Delta t_r$, i.e. $d\mathcal{N}_{in, i} / dt = \Delta
\mathcal{N}_{in, i}/\Delta t_r$, in agreement with the temporal discretisation of
the radiation transport in \traphic. We then a posteriori follow the
evolution of the neutral fraction in time $t_i \in (t_r, t_r+\Delta t_r)$ 
on sub-cycles $\delta t_i \le \Delta t_r$, 
\begin{equation} 
  \eta_i^{t_i+\delta t_{i}} - \eta_i^{t_i}  =  
  \alpha_i^{t_i}  n_{e,i}^{t_i} \chi_i^{t_i} \delta t_i - \Gamma_i^{t_i} \eta_i^{t_i} \delta
  t_i,
  \label{Photoionization:RateEquation:Discretization}
\end{equation}
Our assumption of a constant ionizing flux implies that the photo-ionization
rate $\Gamma \propto
(1-e^{-\tau})\eta^{-1}$ (see Eq.~\ref{Eq:Photoionizationrate}). Hence, a
change in the neutral fraction implies a change in the photo-ionization rate 
$ \Gamma_i^{t_i}$, 
\begin{equation}
\Gamma_i^{t_i} =  \Gamma_i^{t_r}
\left[\frac{1-\exp(-\tau_i^{t_i})}{1-\exp(-\tau_i^{t_r})}\right]
\frac{\eta_i^{t_r}}{\eta_i^{t_i}},
\label{Subcycling:Gamma}
\end{equation}
where $\Gamma_i^{t_r}$ and $\tau_i^{t_r}$ are the photo-ionization rate and
the optical depth at the beginning of the
sub-cycling, given by Eqs.~\ref{Eq:Photoionizationrate} and \ref{Eq:AposterioriOpticalDepth}, and 
$\tau_i^{t_i}  = \tau_i^{t_r} \eta_i^{t_i} / \eta_i^{t_r}$.
\par
The number of ionizations $\Delta \mathcal{N}_{sub, i}$ occurring over $\Delta t_r$ is then
$\Delta \mathcal{N}_{sub, i} = (m_i^{t_r} X_i^{t_r} / {m_H}) \sum \Gamma_i^{t_i}
\eta_i^{t_i}  \delta t_i$, where the sum is over all sub-steps $\delta t_i$ in
$(t_r, t_r + \Delta t_r)$.
We set $\delta t_i \equiv \min(f \tau_{eq,i}^{t_i}, t_r + \Delta t_r - t_i)$, 
where $f< 1$ is a dimensionless factor. It can be demonstrated that 
for this choice of $\delta t_i$  
Eq.~\ref{Photoionization:RateEquation:Discretization} is a sufficiently good 
approximation to Eq.~\ref{Photoionization:RateEquation} that the
neutral fraction never violates the physical bound $0 \leq \eta_i \leq 1$. Because 
the neutral fraction changes during the sub-cycling,
the number of ionizations $\Delta \mathcal{N}_{sub, i}$ can be less than the 
number of photons $\Delta \mathcal{N}_{abs,i}$ that have been removed due to
absorptions  during the radiation transport over the time step $\Delta t_r$
based on the assumption of a constant neutral fraction.
We then explicitly conserve photons by
adding $\Delta \mathcal{N}_{abs,i} - \Delta \mathcal{N}_{sub, i}$ photons 
to the photon transport in the next radiative transfer step.
\par
When either the photo-ionization rate or recombination rate is high,   
$\tau_{eq}$ and hence $\delta t$ will be very small (dropping the
particle index $i$ for simplicity). For $\delta t  \ll \Delta t_r$ the sub-cycling becomes
computationally very expensive. We find, however, that photo-ionization equilibrium is 
typically reached after only a few sub-cycles. Once photo-ionization equilibrium is reached, 
an explicit integration of the rate equation is 
no longer necessary. For a photon-conserving transport we still require 
knowledge about the number of photo-ionizations and recombinations occuring during the equilibrium phase.
Both can, however, be obtained in a stroke, based on the number of photo-ionizations and recombinations
occuring during the last sub-cycle step over which the rate equation was explicitly integrated.
\par
The importance of properly following the evolution of the photo-ionization
rate in the presence of an evolving
neutral fraction in the optically thick\footnote{
  For $\tau \ll 1$ Eq.~\ref{Subcycling:Gamma} implies that the photo-ionization rate is constant,  
  $\Gamma_i^{t_i} = \Gamma_i^{t_r}$.} regime has been pointed out 
by \cite{Mellema:2006}. There, a time-averaged photo-ionization rate
obtained from an iterative procedure was employed. 
The sub-cycling procedure (Eq.~\ref{Subcycling:Gamma}) presented here has the advantage that it is 
well-motivated also in the presence of recombinations.
\par
We note that whenever  one is only interested in
obtaining the equilibrium neutral fraction, the detailed handling of the
photo-ionization rate is rather unimportant. This is because ionization equilibrium implies that the number of
photo-ionizations $d\mathcal{N}_{in} / dt (1-e^{-\tau}) \Delta t_r$ 
over the time interval $\Delta t_r$ exactly cancels the number of recombinations 
$(1-\eta)^2 \mathcal{N}_H n_H \alpha \Delta t_r$ over that same time interval. This balance,
however, has a unique (and stable; see 
Eq.~\ref{Photoionization:RateEquation:Reformulation:Solution}) 
solution for the neutral fraction. The equilibrium neutral fraction then also
implies the correct photo-ionization rate, via Eq.~\ref{Eq:Photoionizationrate}.
When one is interested in following the details
of the evolution of the neutral fraction towards photo-ionization equilibrium,
on the other hand, the dependence of the photo-ionization rate on the neutral
fraction needs to be taken into account, as presented above.
\par

\subsubsection{The time step $\Delta t_r$}
\label{Sec:Timestep}
Our main consideration when choosing the size of the radiative transfer time step 
for the simulations we are presenting in this work, 
is that we wish to accurately reproduce the analytical and numerical
reference results we are comparing with. These results include the
time-dependence of the size of ionized regions around
ionizing sources. At early times, just after the sources start to emit
ionizing photons, 
the ionized regions expand quickly into the neutral hydrogen field surrounding the sources. 
To accurately reproduce this early phase of rapid growth, we necessarily have
to employ time steps $\Delta t_r$ that are relatively small. The phase of rapid growth
is, however, only of relatively short duration. The subsequent evolutionary stages of modest
resp. slow growth, which account for most of the (simulation)
time, are often more interesting. We show in
Section~\ref{Section:Test1} that whenever we are not interested in the
very early phase of rapid growth we can, thanks to the photon-conserving nature of
\traphic, choose substantially larger time steps without affecting the outcome of our simulations.
\par
For all but one of the simulations we present in this work, we will be concerned with solving the 
time-independent radiative transfer equation. In these simulations,  
we choose to propagate photon packets downstream from their location of emission over only a single
inter-particle distance per radiative transfer time step, unless stated otherwise. This approach is equivalent to 
solving the time-independent radiative transfer equation in the limit of small radiative transfer time steps, $\Delta t_r \to 0$. 
We have explicitly checked for all our simulations that the time step was chosen sufficiently 
small to be in agreement with this limit.
\par
Our treatment of the time-independent radiation transport reduces
the computational effort for the simulation of problems for which the time-step 
has been fixed to a small value, e.g. by considerations like those presented in the beginning of this section. 
In the limit that radiation completely fills the simulation box, 
the computational effort required to solve the time-independent radiative transfer equation over the time
interval $T$ by propagating photon packets only over a single inter-particle
distance per radiative transfer time step $\Delta t_r$ is proportional to  
(cp. Section~\ref{Sec:TRAPHIC:Merging})  
$N_{SPH} \times \tilde{N}_{ngb} \times N_c \times T / \Delta t_r$. This has to be compared to the computational 
effort required to follow all photon packets over each time step until they leave the simulation box, which is proportional to
$N_{SPH} \times \tilde{N}_{ngb} \times N_c  \times N_{SPH}^{1/3} \times T / \Delta t_r$. This is larger by a factor of
$N_{SPH}^{1/3}$, which for typical simulations reaches values of the order of $100$.
\par
In Sec.~\ref{Sec:Test3} we will present one simulation in which we solve the
time-dependent radiative transfer equation. In this simulation we will employ the photon clock (Section~\ref{Sec:Evolution}) to 
accurately control the distance over which photon packets are propagated 
over each time step $\Delta t_r$ to match the light crossing distance $c
\Delta t_r$. In this context it is interesting to note 
that in the case of small time steps, i.e. $c \Delta t_r < L_{box}$, it is 
less expensive to solve the time-dependent  
than the time-independent radiative transfer equation. This is because 
the computational effort to solve the time-dependent equation 
(again in the limit that radiation completely fills the simulation box and assuming that its boundaries are photon-absorbing) 
scales as  $N_{SPH} \times \tilde{N}_{ngb} \times N_c \times c \Delta t_r / L_{box} \times N_{SPH}^{1/3} \times T / \Delta t_r$, 
which is smaller than the computational effort for obtaining the time-independent solution (assuming that over each 
radiative transfer time step photon packets are transported until they leave the box) 
by the factor\footnote{Observe that for larger time steps, i.e. $\Delta t_r \ge  L_{box} / c$, and assuming 
  photon-absorbing boundaries, the computational 
  effort for solving the time-dependent radiative transfer equation equals the
  computational effort for solving the time-independent radiative transfer
  equation. It exceeds the computational effort for solving the time-independent radiative transfer equation
  with photon packets propagating only a single inter-particle distance per radiative transfer time-step 
  if $c \Delta t_r >  L_{box} /  N_{SPH}^{1/3}$. } 
$c \Delta t_r / L_{box}$.
\par
\subsubsection{Effective multi-frequency description}
In our current implementation we use only a single frequency. Thus, we either
assume that the ionizing radiation is mono-chromatic, or we assume
ionizing radiation 
with a frequency spectrum $J_\nu$ and provide an effective multi-frequency 
description using only a single frequency bin (for a textbook introduction 
to the numerical treatment of multi-frequency radiation, see 
e.g. \citealp{Mihalas:2000}).
\par
For the latter case, we define a frequency-independent (grey) photo-ionization cross-section
$\bar{\sigma}$ such that the total photo-ionization rate (Eq.~\ref{Photoionzation:Rate}) is conserved, 
\begin{equation}
  \Gamma =   \int_{0}^\infty d\nu\   \frac{4 \pi J_\nu(\nu)}{h_p\nu}
  \sigma_\nu  \equiv  \bar{\sigma} \int_{\nu_0}^\infty d\nu\   \frac{4 \pi
    J_\nu(\nu)}{h_p\nu},
\end{equation}
or,
\begin{equation}
  \bar{\sigma} = \int_{0}^\infty d\nu\ \frac{4 \pi J_\nu(\nu)} 
    {h_p\nu} \sigma_\nu \times \left [\int_{\nu_0}^\infty d\nu\ \frac{4 \pi J_\nu(\nu)}{h_p\nu} \right]^{-1}.
\label{Eq:Photoionization:Crosssection:Grey}
\end{equation}

\subsection[]{Tests}
\label{Sec:Application:Tests}
In this section we report on the performance of our implementation in simple
and well-defined test problems. The tests comprise the evolution
of the ionized region around a single star in a homogeneous medium
(Section~\ref{Section:Test1}), the casting of a shadow behind an opaque
obstacle (Section~\ref{Section:Test2}) and the propagation of ionization
fronts driven by the ionizing radiation of multiple stars in an inhomogeneous
density field obtained from a cosmological simulation
(Section~\ref{Sec:Test3}). We compare the results obtained with \traphic\
with analytic solutions, where available. For most physical
settings, however, no analytic solution to the radiative transfer problem is known,
mainly due to the complex geometries involved. We have therefore designed the
test problems in Sections~\ref{Section:Test1} and \ref{Sec:Test3} to closely follow the description given in the 
Cosmological Radiative Transfer Codes Comparison Project (\citealp{Iliev:2006b}),  
which provides a set of very useful numerical reference solutions to compare with. 
\par
Throughout we will assume that the density field is static. To facilitate a direct
comparison with \cite{Iliev:2006b}, we present results after mapping
physical quantities defined on the SPH particles to a regular grid, unless
stated otherwise.  This is done using a mass-conserving
SPH interpolation similar to the one described in \cite{Alvarez:2006}. We opted
for the SPH interpolation since it is most consistent with the SPH
simulation we are employing. For comparison, we repeated our analysis using TSC 
mass-conserving interpolation (\citealp{Hockney:1988}) but found no significant 
differences. 
\par 
For all tests reported in this section we employ the on-the-spot approximation
(e.g.~\citealp{Osterbrock:1989}), in order to allow a direct comparison with \cite{Iliev:2006b}. 
In the on-the-spot approximation diffuse photons emitted during recombinations
to the hydrogen ground energy level are assumed to be immediately
re-absorbed by neutral hydrogen atoms close to the location of
emission. The effect of diffuse recombination radiation can then be simply taken into account
by setting the recombination coefficient $\alpha$ to the so-called case B
value $\alpha_B$, which for the temperature range of interest can be well
approximated by 
\begin{equation}
  \alpha_B(T) = 2.59 \times 10^{-13} 
  \left(\frac{T}{10^4\K}\right)^{-0.7} \cmc \si.
\end{equation}
We will report on the study of diffuse radiation in which we will be explicitly following
the scattering of diffuse photons instead of employing the on-the-spot approximation in future work.
\par
In their simulations, \cite{Iliev:2006b} assumed that the speed of light is infinite, 
i.e. they solved the time-independent radiative transfer equation. For the comparison with these simulations 
we will therefore make the same assumption (recall Sec.~\ref{Sec:Timestep} for the discussion of how we solve the
time-independent radiative transfer equation). From now on, when referring to the radiative transfer equation,
we therefore assume it to be of the time-independent form, unless stated otherwise. 
We will repeat one of the simulations presented in Sec.~\ref{Sec:Test3} to solve the time-dependent radiative transfer equation by
employing the photon packet clocks as described in Sec.~\ref{Sec:Evolution}. 

\subsubsection{Test 1: Spherically-symmetric HII region expansion}
\label{Section:Test1}

  \begin{figure} 
  \includegraphics[width=84mm, clip=]{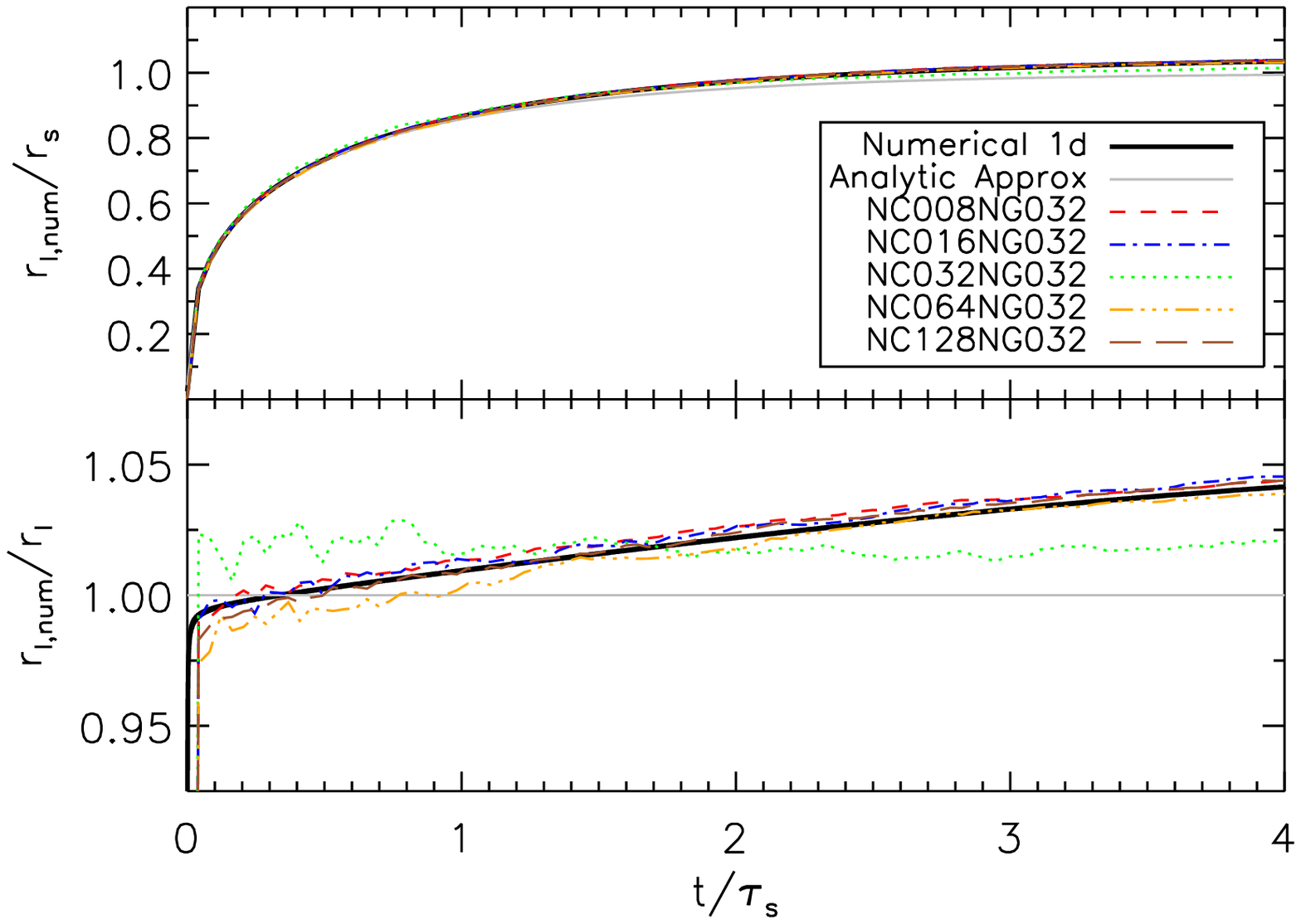} 
  \caption{Test 1. The evolution of the ionization front   
    for the angular resolutions $N_c = 8, 16,32,64$ and $128$, as indicated in the legend. The
    spatial resolution is fixed ($N_{SPH} = 64^3$, $\tilde{N}_{ngb} = 32$). 
    The top panel shows the  position of the ionization front $r_{I, num}$
    normalized by the Str\"omgren radius $r_s$ as a function of
    time. The thick black solid curve shows a numerical reference solution obtained
    with a one-dimensional, grid-based radiative transfer code (see text). The
    grey curve shows the analytic reference solution, Eq.~\ref{Test1:IF}, which has been
    obtained by assuming $\chi \equiv 1$ throughout the ionized region. The results from the numerical
    simulations employing \traphic\ closely match the numerical 
    reference solution. The bottom panel shows the position of the ionization
    fronts of the top panel divided by the analytic reference solution. Note that the
    analytic reference solution slightly differs from the numerical reference
    solution, due to the simplifying assumptions inherent to the analytic approach
    (see also the discussion of Eq.~\ref{Eq:Test1:EqNeutralFraction1}).
  } 
  \label{Fig:Test1:IF} 
  \end{figure} 

\begin{figure*}

  \includegraphics[trim = 40mm 0mm 40mm 0mm, width=0.19\textwidth, clip=true]{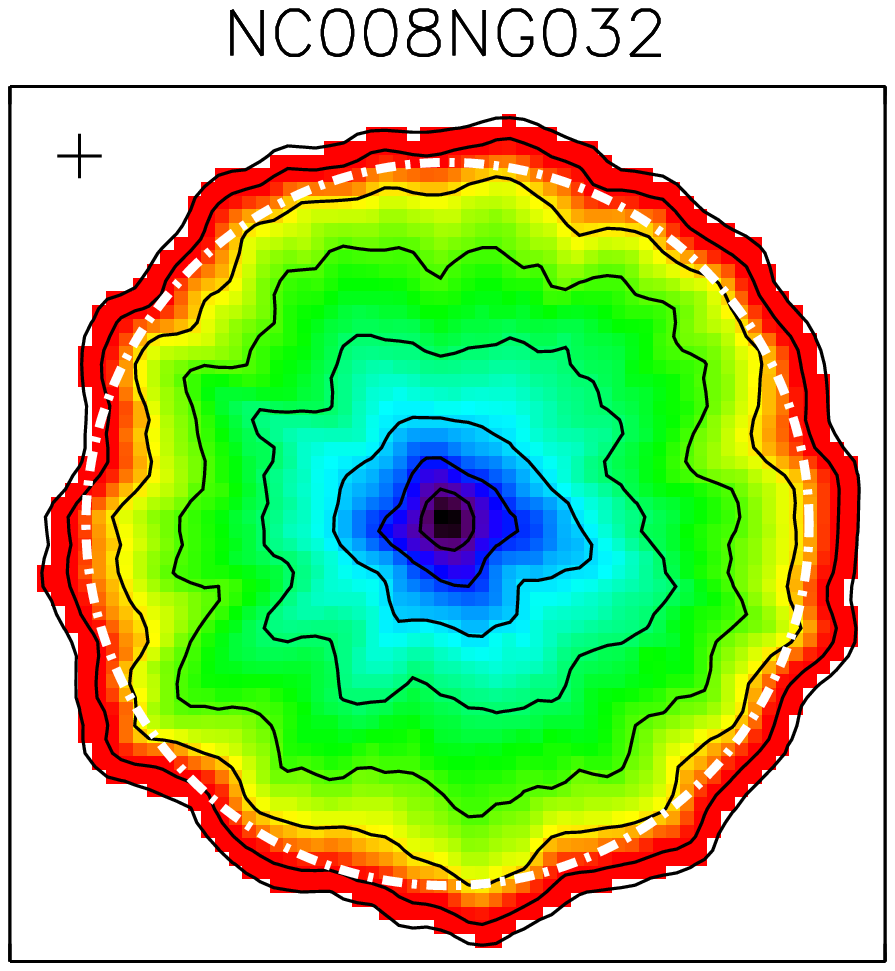}      
  \includegraphics[trim = 40mm 0mm 40mm 0mm, width=0.19\textwidth, clip=true]{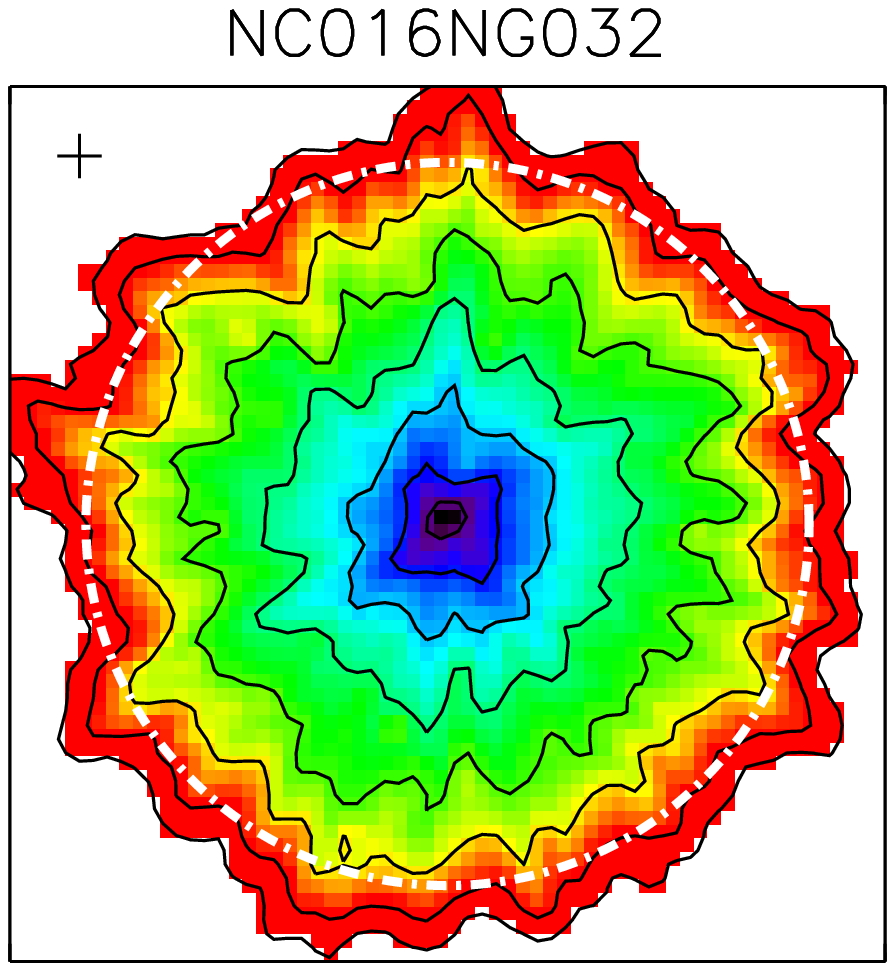}      
  \includegraphics[trim = 40mm 0mm 40mm 0mm, width=0.19\textwidth, clip=true]{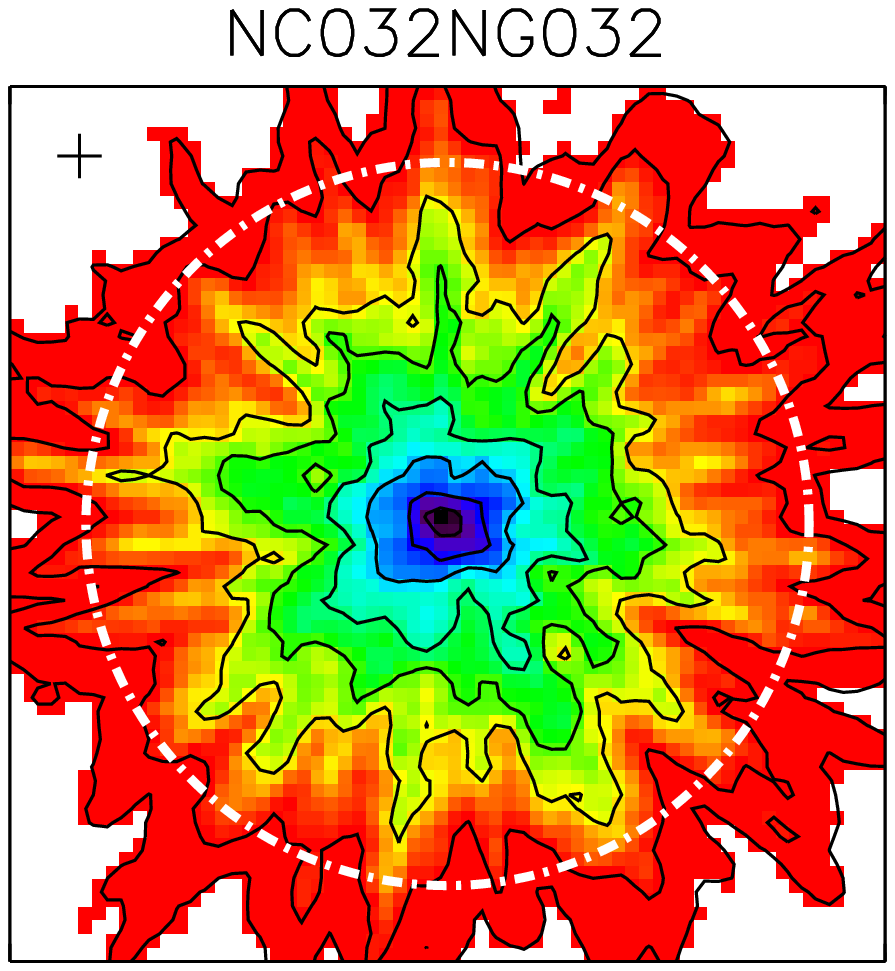}      
  \includegraphics[trim = 40mm 0mm 40mm 0mm, width=0.19\textwidth, clip=true]{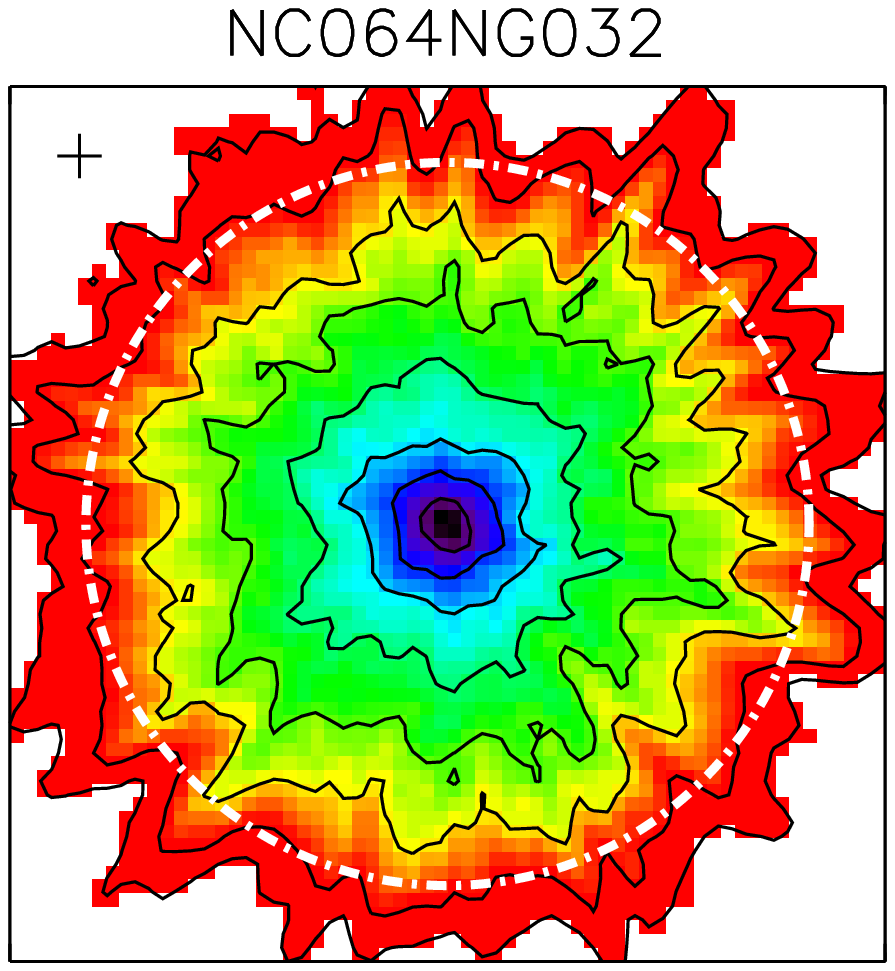}      
  \includegraphics[trim = 40mm 0mm 40mm 0mm, width=0.19\textwidth, clip=true]{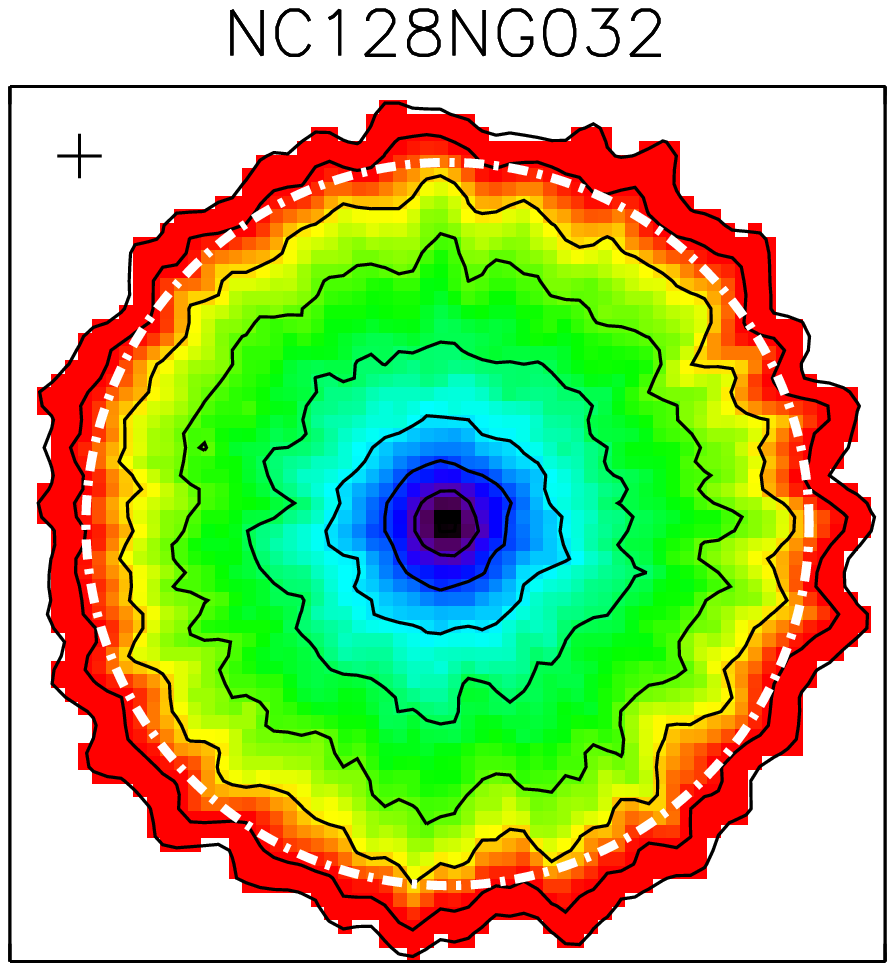}      \\
  \includegraphics[trim = 40mm 0mm 40mm 0mm, width=0.19\textwidth, clip=true]{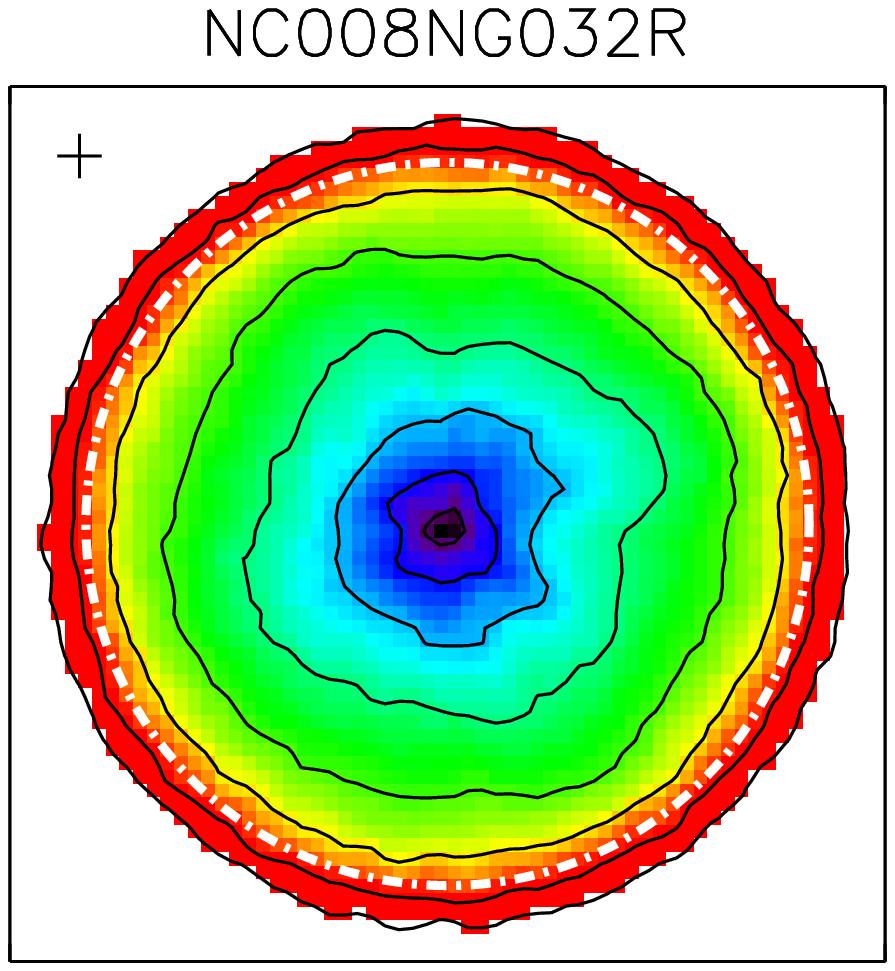}      
  \includegraphics[trim = 40mm 0mm 40mm 0mm, width=0.19\textwidth, clip=true]{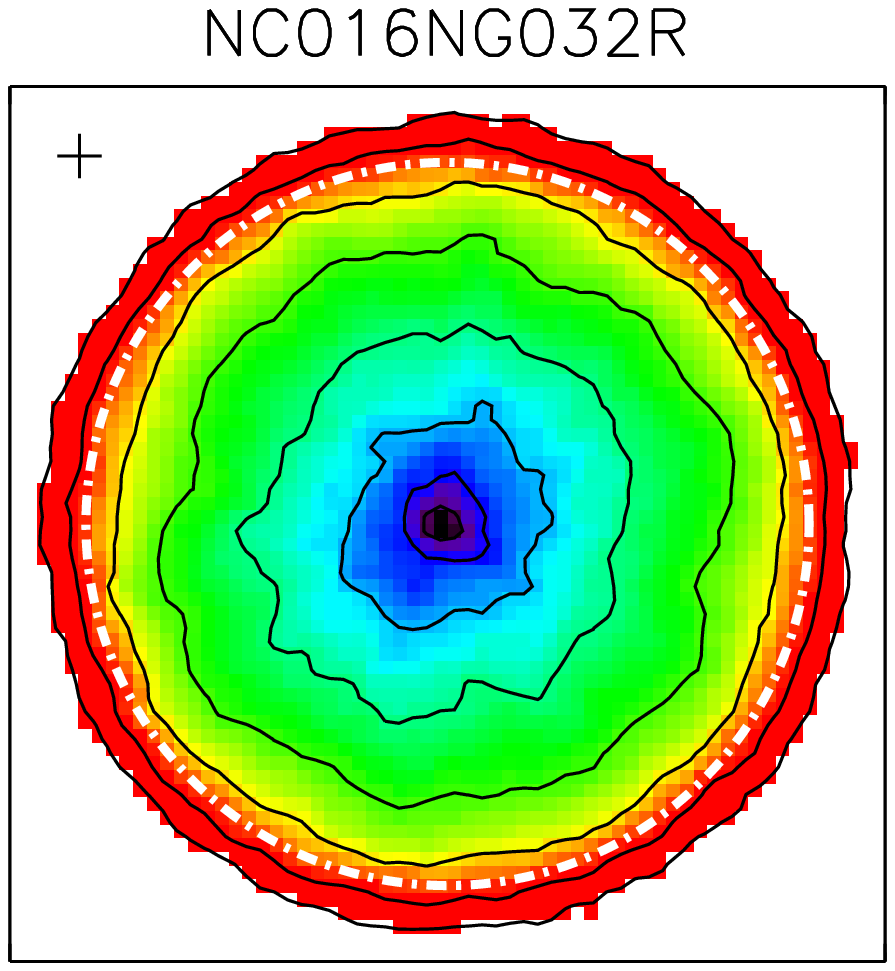}      
  \includegraphics[trim = 40mm 0mm 40mm 0mm, width=0.19\textwidth, clip=true]{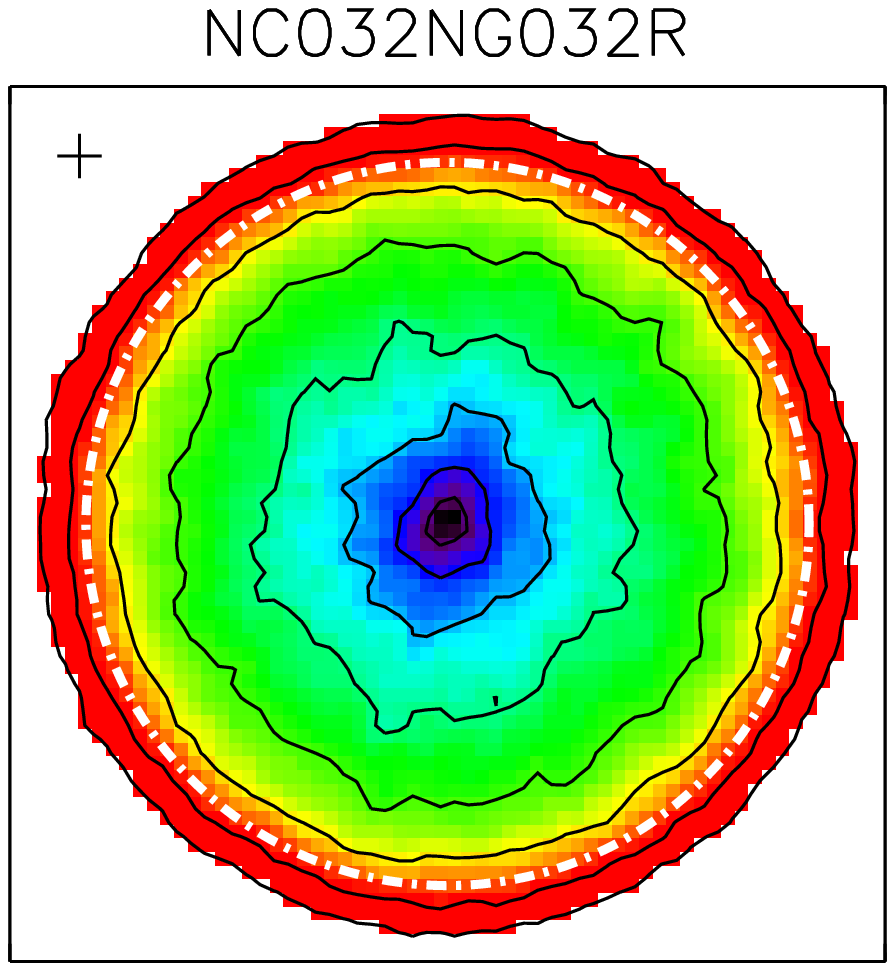}      
  \includegraphics[trim = 40mm 0mm 40mm 0mm, width=0.19\textwidth, clip=true]{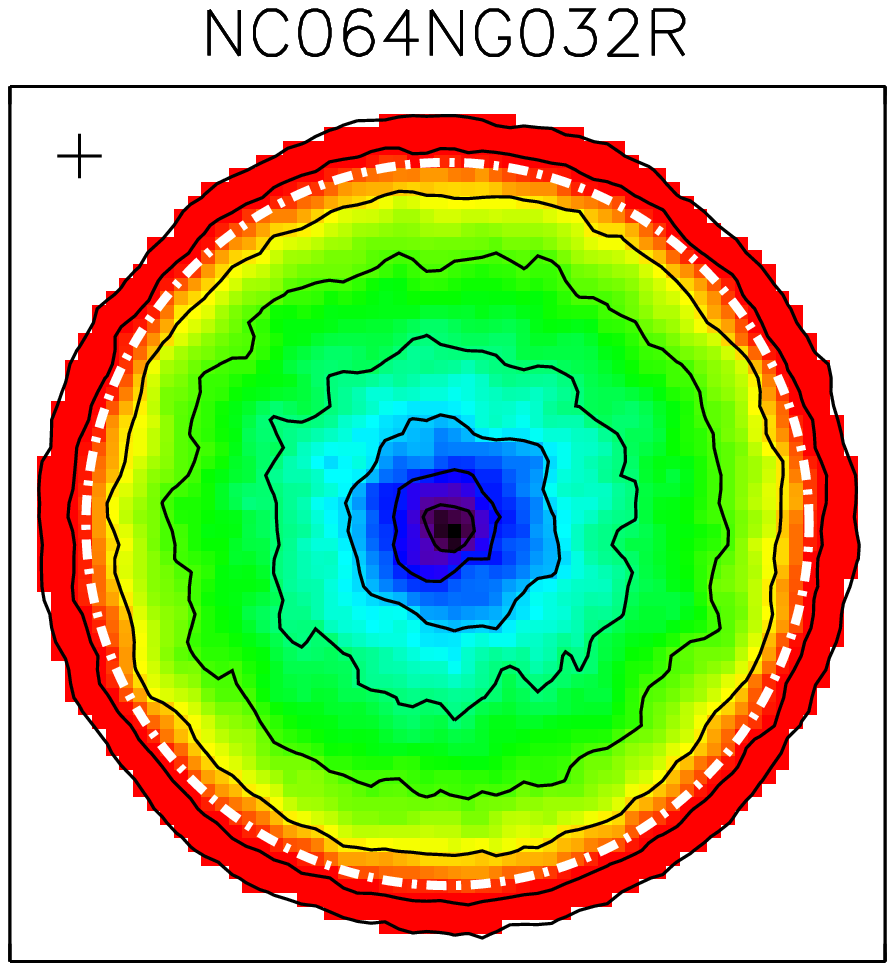}      
  \includegraphics[trim = 40mm 0mm 40mm 0mm, width=0.19\textwidth, clip=true]{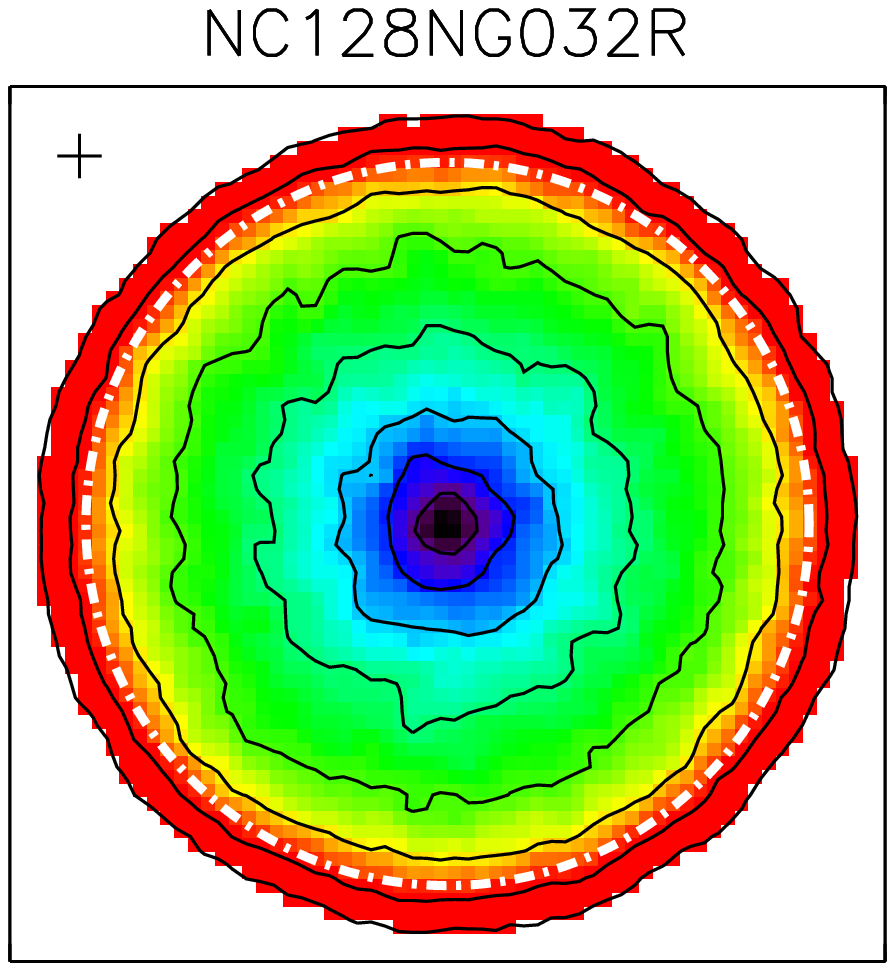}     

  \caption{Test 1. Slice through the simulation box at $z = L_{box} / 2$ showing the neutral
    fraction at the end of the simulation ($t_r = 500 \Myr$), i.e. in
    photo-ionization equilibrium, for simulations with (bottom row) and without
    (top row) resampling of the density field. The angular resolution increases from $N_c =
    8$ in the left-most to $N_c = 128$ in the right-most column, as indicated
    in the panel titles. The spatial resolution is fixed to $N_{SPH} = 64^3$,
    $\tilde{N}_{ngb} = 32$ and is indicated by the cross of length
    $2\langle\tilde{h}\rangle$ in the upper left corner of each panel. Black contours show 
    neutral fractions of $\eta = 0.9, 0.5, \log \eta = -1, -1.5, -2, -2.5, -3,
    -3.5, -4$, going from the outside in. The white
    dot-dashed circle indicates the Str\"omgren sphere, which should be, and
    is, just inside to the $\eta = 0.5$ contour. The colour scale is
    logarithmic and has a lower cut-off $\log \eta  = -5$. Note that the
    resampling strongly suppresses the particle noise seen in the top-row panels.} 

  \label{Fig:Test1:NeutralFraction2d} 
\end{figure*} 

\begin{figure*}

  \includegraphics[trim = 15mm 0mm 35mm 0mm, width=0.49\textwidth, clip=true]{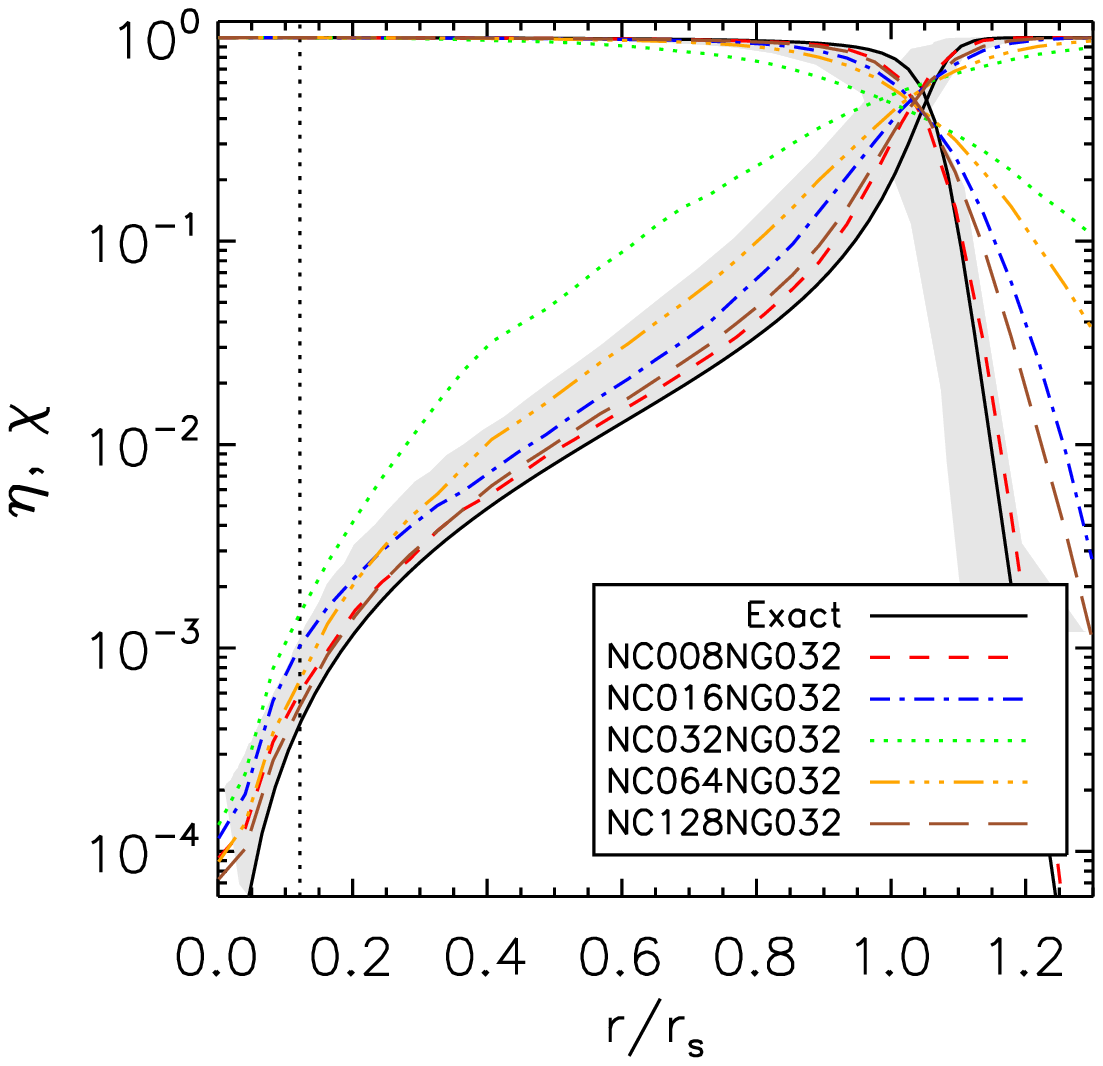}      
  \includegraphics[trim = 15mm 0mm 35mm 0mm, width=0.49\textwidth, clip=true]{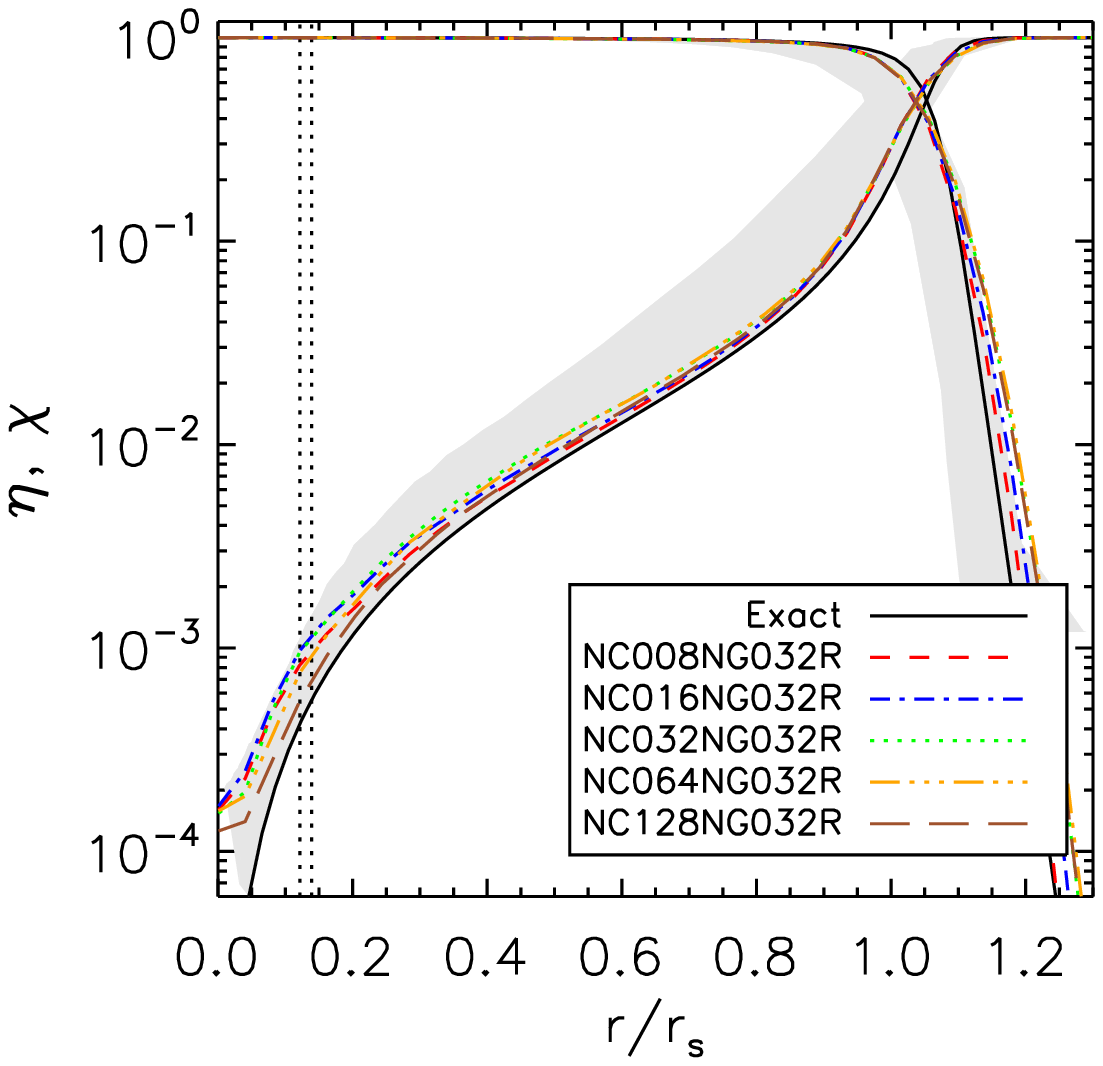}      
  
  \caption{Test 1. Spherically averaged neutral and ionized fractions within the Str\"omgren
    sphere at $t_r = 500 \Myr$ for different angular resolutions, as indicated
    in the legend. The profiles in the left-hand (right-hand) panel correspond
    to simulations without (with) resampling the density field. 
    The spatial resolution is fixed ($N_{SPH} = 64^3$, $\tilde{N}_{ngb} = 32$).
    The black solid curves corresponds to the exact profiles,
    given by the solution of Eq.~\ref{Eq:Test1:EqNeutralFraction1}. The grey bands show the range of
    neutral and ionized fractions found by other codes as reported in \protect\cite{Iliev:2006b}.
    The vertical dotted line marks the radius r = $2\langle\tilde{h}\rangle$, corresponding to the spatial
    resolution employed. In the right-hand panel, the additional
    (right-most) vertical dotted line indicates the radius corresponding to the spatial
    resolution $2\langle h \rangle$ of the SPH simulations, which is the scale
    on which particle positions are randomly displaced during the resampling. } 
  \label{Test1:Fig:NeutralFraction1d} 

\end{figure*} 

In this section we consider the problem of a steady ionizing source emitting $\dot{\mathcal{N}}_\gamma$  
mono-chromatic photons of frequency $h_p\nu=13.6\eV$ per unit time,
which is turned on in a static, initially neutral,  
uniform medium of constant hydrogen number density $n_H$. 
\par
The analytical solution to this problem is well-known (for a textbook
discussion see e.g.~\citealp{Dopita:2003}). 
In equilibrium, the number of photons  emitted by the source 
has to compensate the number of recombinations within the surrounding,
stationary, ionized region, the so-called Str\"omgren sphere.  Assuming that the Str\"omgren sphere is
fully ionized, i.e. $\chi \equiv  1$, its radius $r_s$ is therefore given by the balance equation
\begin{equation}
\dot{\mathcal{N}}_\gamma = \frac{4}{3}\pi r_s^3 \alpha_B(T)  n_H^2.
\label{Eq:StroemgrenRadius}
\end{equation}
\par
The evolution of the spherical, fully ionized region towards the  
equilibrium Str\"omgren sphere is described by the following equation for its
radius $r_I$, the ionization front,  
\begin{equation} 
4\pi r_I^2n_H \frac{dr_I}{dt}=\dot{\mathcal{N}}_\gamma-\frac{4}{3}\pi r_I^3 \alpha_B(T) n_H^2. 
\label{Test1:IonizationFront:Position}
\end{equation} 
Introducing the new variables $\xi \equiv r_I/r_s$ and $\tau \equiv t/\tau_s$, where  
the Str\"omgren time-scale $\tau_s=1/(\alpha_B n_H)$ is the recombination
 time in a fully ionized gas, we arrive at the differential equation 
\begin{equation} 
\frac{d\xi}{d\tau}=\frac{1-\xi^3}{3\xi^2}, 
\end{equation} 
the solution of which reads 
\begin{equation} 
r_I(t) = r_s(1-e^{-t/\tau_s})^{1/3}.
\label{Test1:IF}
\end{equation} 
Hence, the ionization front reaches the Str\"omgren sphere after a few Str\"omgren 
times $\tau_s$ and stays static thereafter. 
\par
In reality the neutral fraction inside the ionized region is, however, not zero,
but varies smoothly with the distance to the ionizing source. We therefore invoke the
commonly employed definition of the ionization front 
as the radius at which the neutral fraction equals $50$ per cent, i.e. $\eta = 0.5$. 
The equilibrium neutral fraction $\eta_{eq} (r) = \lim_{t\to\infty} \eta (r)$
can be obtained by solving the equation (e.g. \citealp{Osterbrock:1989}) 
\begin{equation}
  \frac{\eta_{eq}(r) n_{H}}{4\pi r^2}\int d\nu\ \dot{\mathcal{N}}_\gamma(\nu)
  e^{-\tau_{\nu}}\sigma_{\nu} = \chi^2_{eq}(r)n_{H}^2  \alpha_B
  \label{Eq:Test1:EqNeutralFraction1},
 \end{equation}
where the optical depth $\tau_{\nu}(r)$ is given by 
\begin{equation}
   \tau_{\nu}(r) =  n_{H}\sigma_{\nu} \int_0^r  dr^{\prime}\ \eta_{eq}(r^{\prime}) \label{Eq:Test1:EqNeutralFraction2}.
\end{equation}
We refer to the neutral fraction $\eta_{eq} (r)$  obtained by direct numerical
integration of Eq.~\ref{Eq:Test1:EqNeutralFraction1} as the exact neutral 
fraction profile and to $\chi_{eq}(r) = 1 - \eta_{eq} (r)$ as the exact
ionized fraction profile. These profiles are shown as black solid curves in
Fig.~\ref{Test1:Fig:NeutralFraction1d}. Note that for our choice of
parameters, $r_{I, eq} = 1.05\ r_s$. The ionization front obtained from the solution to
Eq.~\ref{Eq:Test1:EqNeutralFraction1} is thus at a slightly larger radius than 
the equilibrium ionization front obtained assuming the Str\"omgren sphere is fully ionized (Eq.~\ref{Test1:IF}).
\par
The last observation indicates that the analytic solution given by
Eq.~\ref{Test1:IF} generally fails to provide an accurate reference
solution that can be used to judge the validity of the numerical results
obtained with a new radiative transfer scheme like \traphic, 
due to its simplification of the problem. We therefore employ a one-dimensional, explicitly photon-conserving, 
grid-based radiative transfer scheme that we have implemented ourselves to obtain an accurate
numerical reference solution. We will make use of this numerical reference
solution in our comparisons below.
\par
In the following we present a suite of radiative transfer simulations
demonstrating that \traphic\ is able to accurately follow the evolution of the
ionization front around a single ionizing point source. For the setup of the numerical
simulations we closely follow the description of Test 1 presented in \cite{Iliev:2006b}, the
only difference being the spatial resolution we employ. This allows us to
directly compare our results to the results presented in the code comparison project. In particular,  
we choose a number density $n_H=10^{-3}\cmci$
and an ionizing luminosity of $\dot{\mathcal{N}}_\gamma  = 5\times
10^{48}~\mbox{photons}~\si$. The gas is assumed to have a constant temperature $T=10^4\K$. 
With these values, $r_s = 5.4\kpc$ and $\tau_s = 122.4\Myr$. 
\par
We set up the initial conditions in a simulation box of length $L_{box}=13.2 \kpc$ 
containing $N_{SPH} = 64^3$ SPH particles\footnote{We note that \cite{Iliev:2006b} employed
  $N_{cell} = 128^3$ cells, with the ionizing source located in one of the
  corners of the box.}. The ionizing source is located in the
centre. The box boundary is photon-transmissive. We assign each SPH particle a mass 
$m = n_H m_H L_{Box}^3/ N_{SPH}$. The positions of the SPH particles are chosen
to be glass-like\footnote{The glass-like distribution of particles is
achieved by first placing them randomly in the simulation box and thereafter
letting them freely evolve under the influence of a reversed-sign
(i.e. repelling) gravitational force until they settle down into an equilibrium
configuration, see \cite{White:1996}.}. Glass-like initial conditions yield a more regular
distribution of the particles within the box as compared to Monte Carlo
sampling of the density field. The SPH smoothing kernel is computed and the 
SPH densities are found using the SPH formalism implemented in \gadget, with
$N_{ngb} = 48$. 
\par
We perform  5 simulations, increasing
the angular resolution in factors of two from $N_c = 8$ to $N_c = 128$. The number of neighbours
employed for the transport of radiation is fixed to
$\tilde{N}_{ngb}=32$. Hence all 5 simulations employ the same spatial
resolution. The time step is set to $\Delta t_r = 10^4 \yr.$
In Fig.~\ref{Fig:Test1:IF} we show the evolution of the ionization front
radius, which we determined by taking the average over
the positions of all particles that have a neutral fraction $0.4 < \eta <
0.6$. For all 5 simulations, the position of the ionization front never deviates more
than $5$ per cent from the analytic solution, Eq.~\ref{Test1:IF}, 
comparable to what has been found with other codes
as reported in the Cosmological Radiative Transfer Code Comparison Project
(\citealp{Iliev:2006b}). Note that the deviations from the analytic solution
can mainly be attributed to the fact that the analytic approach 
provides only an approximate expression for the radius of the ionization
front, because it erroneously assumes $\chi\equiv 1$.  
In fact, all simulations (except for the $N_c =
\tilde{N}_{ngb}$ run, which shows a small deviation of less than two percent,
the reason for which will become clear in our discussion below), 
nearly perfectly follow the numerical reference solution and approach the proper
asymptotic limit $r_{I, eq} = 1.05\ r_s$.
\par

\begin{figure*}

  \includegraphics[trim = 15mm 0mm 35mm 0mm, width=0.49\textwidth, clip=true]{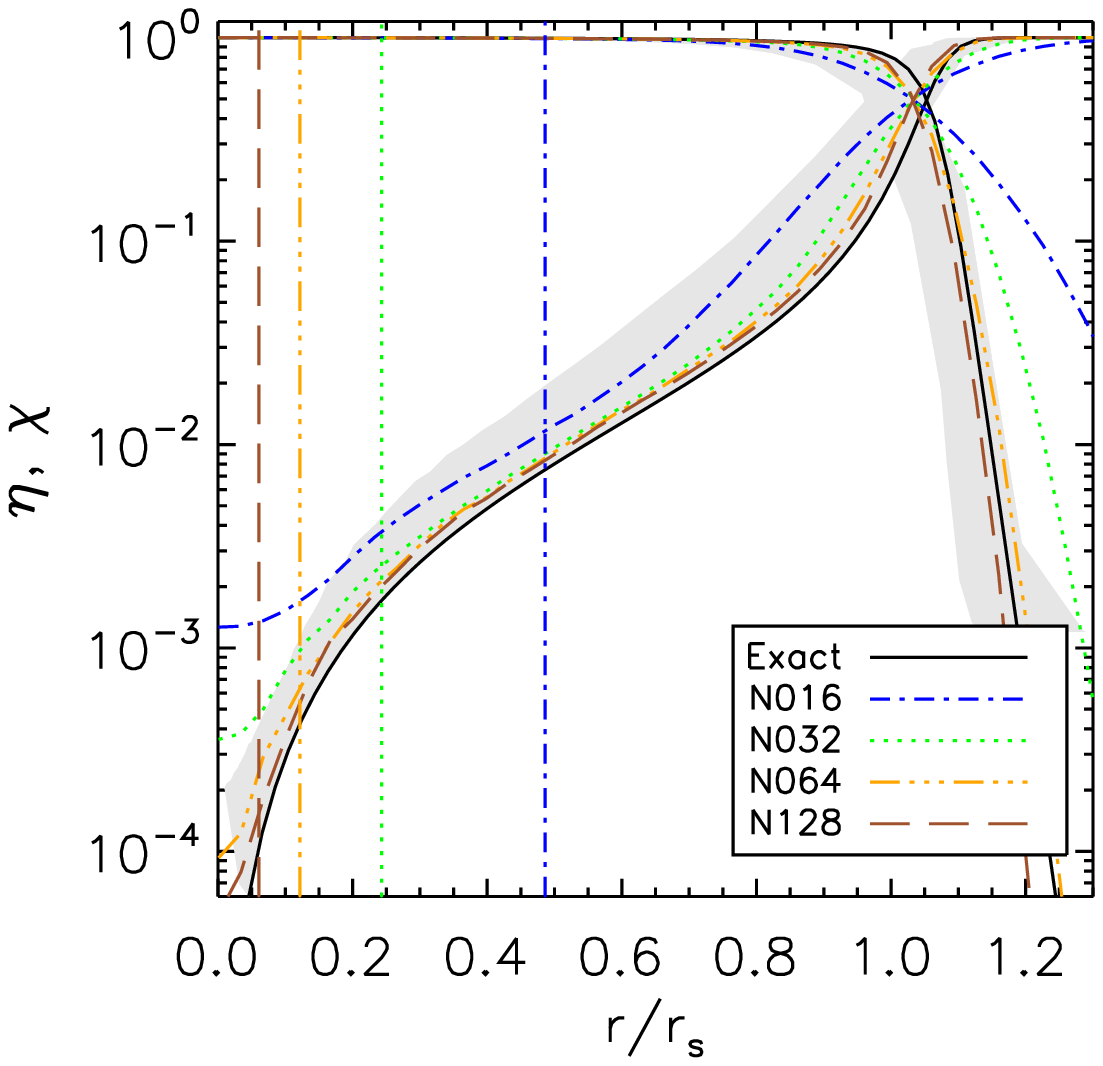}      
  \includegraphics[ trim = 15mm 0mm 35mm 0mm, width=0.49\textwidth, clip=true]{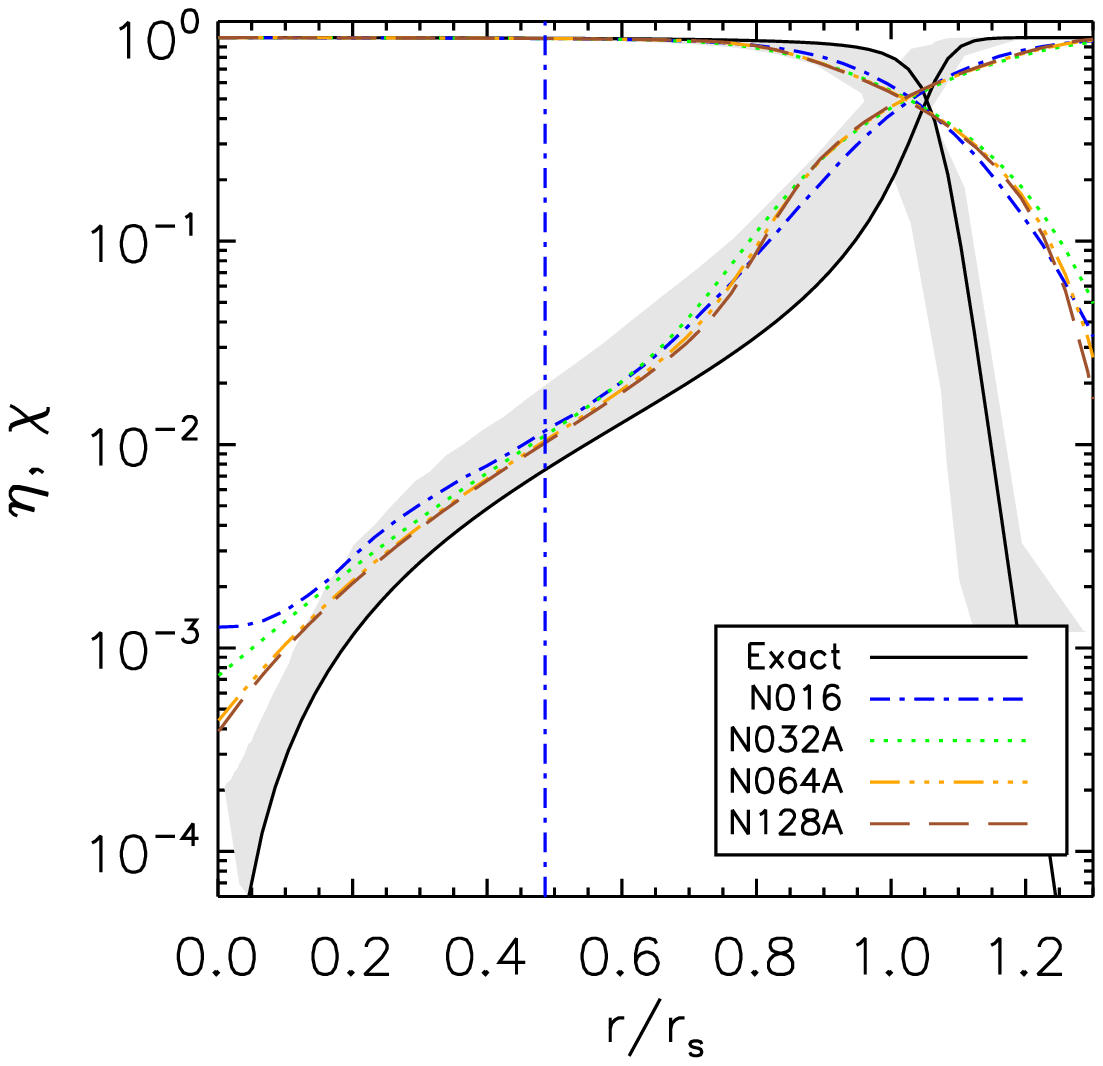}

  \caption{Test 1. Spherically averaged neutral and ionized fractions within the Str\"omgren
    sphere at $t_r = 500 \Myr$. The simulations all have the same angular
    resolution $(N_c = 8)$ and employ the same number of neighbours $(\tilde{N}_{ngb}
    = 32)$, but use a different number of SPH particles, increasing from
    $N_{SPH} =16^3$ to $N_{SPH} =128^3$ in factors of $2^3$. 
    The black solid curves corresponds to the exact profiles, obtained from
    Eq. \ref{Eq:Test1:EqNeutralFraction1}. The grey bands show the range of
    neutral  and ionized fractions found by other codes as reported in
    \protect\cite{Iliev:2006b}.
    Left-hand panel: For each simulation the spatial resolution is marked by a vertical line
    (in colour and line-style identical to the corresponding profile) at
    radius $2\langle \tilde{h} \rangle$. The higher the spatial resolution,
    the more closely the exact profile is approached. Right-hand panel: The profile
    corresponding to the lowest spatial resolution simulation ($N_{SPH}
    =16^3$, blue dot-dashed line) is repeated from the left-hand panel. The
    profiles of all other simulations have been averaged
    over the spatial resolution element $2\langle \tilde{h} \rangle$
    of the lowest spatial resolution simulation, corresponding to the radius
    marked by the vertical line. The profiles become nearly identical after
    smoothing them to the same resolution. 
  }
   
  \label{Test1:Fig:NeutralFraction1dRes} 
\end{figure*}
 
The top row of Fig.~\ref{Fig:Test1:NeutralFraction2d} shows the neutral fraction in a slice
through the centre of the simulation box at $t_r = 500 \Myr$, which marks the end of
the simulation, for each of the 5 simulations. As we already noted, the ionization front is at the expected 
position. As is true for all other surfaces of constant neutral fraction
shown, the ionization front clearly exhibits the expected spherically symmetric
shape, although it is noisy in some of the simulations. The amount of noise depends on the
ratio of the angular and spatial resolution employed. For $N_c = 8$ (left-most
panel in the top row), the average number of neighbours
per emission or transmission cone is high, $\tilde{N}_{ngb}/N_c = 4$ and, as a result, numerical
noise arising from the  representation of the continuous density field with
discrete SPH particles are
suppressed. With increasing angular resolution the average number of neighbours
per cone decreases, and the contours become more noisy. The noise level
reaches a maximum at $N_c = \tilde{N}_{ngb}$ (middle panel in the top row). For higher angular resolutions, the probability of finding
no neighbours inside an emission or transmission cone becomes high and a
large number of ViPs are created. The ratio of the number of ViPs to
the number of SPH particles enclosed by the ionization front for the
simulation with angular resolution $N_c = 8, 16,32,64$ and $128$ is $\approx 0,
0.003, 0.06, 0.5$ and $0.9$, resp. The ViPs
placed in empty cones regularize the neutral fraction of the ionized
density field by distributing the photons they absorb amongst their
$\tilde{N}_{ngb}$ SPH neighbours using (photon-conserving) SPH interpolation.
\par
In the left-hand panel of Fig.~\ref{Test1:Fig:NeutralFraction1d} we plot the
neutral and ionized fraction averaged in spherical shells as a function of distance to
the star, for all 5 simulations. The grey bands indicate the neutral and
ionized fraction profiles obtained with other codes
as reported in the cosmological radiative transfer code comparison project
(\citealp{Iliev:2006b}). Except for the $N_c
= 32$ run, for which the neutral contours were most noisy (see
Fig.~\ref{Fig:Test1:NeutralFraction2d}), 
all simulations agree very well with the results
published in the comparison project. The deviations from the exact
equilibrium neutral fraction profile, i.e. the result of the numerical
integration of Eq.~\ref{Eq:Test1:EqNeutralFraction1},
can be explained by the particle noise seen in
Fig.~\ref{Fig:Test1:NeutralFraction2d}. Due to the fuzzy structure exhibited by the neutral
fraction contours, a range of neutral fractions can
simultaneously be found within each spherical shell. The profiles obtained
from the numerical simulation with \traphic\ should therefore not be directly
compared to the exact profile, i.e. the solution of 
Eq.~\ref{Eq:Test1:EqNeutralFraction1}, but to the 
profile that results after locally averaging the exact profile
along the radial direction. 
\par
To investigate the effect of particle noise on the neutral fraction
profile we apply the resampling technique introduced in
Section~\ref{Sec:Method:Regularization}. We perform a series of 5 simulations that are identical to the simulations presented above,
except that every 10th radiative transfer time
step  the particle positions  are randomly perturbed within their SPH spheres of influence. The
numerical implementation of the resampling is described in
Appendix~\ref{Sec:Resampling:Implementation}. The densities are not
recalculated after the positions have been changed due to the resampling,
because this would generate fluctuations in
the neutral hydrogen density which would increase the recombination rate and
lead to a smaller Str\"omgren sphere. The resulting neutral fraction profiles are
shown in the right-hand panel of Fig.~\ref{Test1:Fig:NeutralFraction1d}. All profiles are now in
close agreement with each other and with the exact result. The panels in the
bottom row of Fig.~\ref{Fig:Test1:NeutralFraction2d} show the neutral fraction in a slice
through the centre of the simulation box from the simulations with resampling. Clearly, 
resampling strongly suppresses the particle noise visible in the panels in the
top row of Fig.~\ref{Fig:Test1:NeutralFraction2d}, yielding nearly spherical
neutral fraction contours.
\par
We now investigate the dependence of the equilibrium neutral and ionized fraction
profile on the spatial resolution by varying $N_{SPH}$, the number of particles 
employed in the simulation. Because \traphic\ is explicitly
photon-conserving, we expect that the radiative transfer in a homogeneous medium is essentially
independent of the spatial resolution (see e.g. the discussion in
\citealp{Mellema:2006}), except for a trivial averaging. 
For each of the simulations with angular resolution $N_c =
8, 32$ and $128$ and $\tilde{N}_{ngb} = 32$ presented above, we performed three
additional simulations, at lower $(N_{SPH} = 16^3, 32^3)$ and higher 
($N_{SPH}= 128^3$) spatial resolutions, but otherwise identical to the fiducial
$(N_{SPH} = 64^3)$ case. We will focus on the $N_c = 8$ runs, but note that
the $N_c = 32$ and $N_c = 128$ series shows the same behaviour.
\par
The equilibrium neutral and ionized fraction profiles  are shown 
in the left-hand panel of Fig.~\ref{Test1:Fig:NeutralFraction1dRes}. For all
spatial resolutions the ionization front is at nearly the same radius. It can
furthermore be seen that when  the spatial resolution is increased, the
equilibrium neutral fraction  follows the exact result more
closely. The simulation employing the fiducial spatial resolution
($N_{SPH}=64^3$) is almost converged. The differences in the neutral
fraction profiles between the simulations using different numbers of particles 
are fully consistent with the corresponding spatial resolutions, as is
demonstrated in
the right-hand panel Fig.~\ref{Test1:Fig:NeutralFraction1dRes}. There, we average the
neutral fraction profiles obtained in the simulations employing $N_{SPH} =
32^3,64^3$ and $128^3$ particles  over the spatial resolution element of the lowest resolution
simulation $(N_{SPH} = 16^3)$, the size of which is indicated by the vertical line. The averaged profiles match the neutral fraction
profile obtained in the low spatial resolution simulation almost exactly. This shows that 
decreasing the spatial resolution does not introduce any artefacts. The
solution obtained by \traphic\ is the converged solution averaged over the adopted
spatial resolution.
\par
\begin{figure} 
  \includegraphics[width=84mm, clip=]{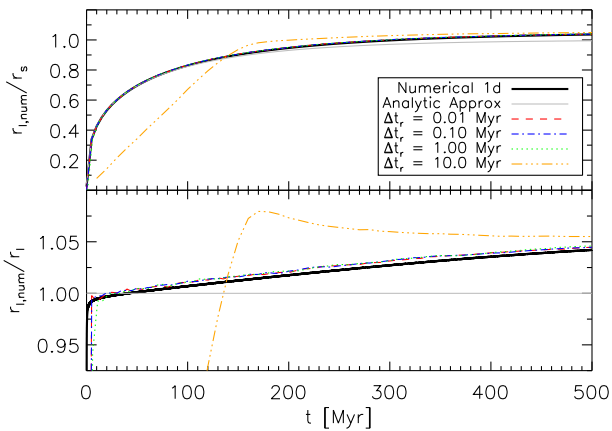} 
  \caption{Test 1. Effect of the size of the time step. We show the evolution
    of the ionization fronts for four simulations with $N_c = 8$,
    $N_{SPH} = 64^3$, $\tilde{N}_{ngb} = 32$ and time steps $\Delta t_r = 0.01, 0.1, 1$ and
    $10 \Myr$, resp, as indicated in the legend. After an initial phase,
    the evolution of the ionization fronts becomes independent of the size of the 
    radiative transfer time step.} 
  \label{Test1:Fig:Timestep} 
\end{figure} 
Finally, we show how the size of the time step $\Delta t_r$ affects the outcome
of our simulations. We again concentrate on the simulation with
angular resolution $N_c = 8$ (and $N_{SPH} = 64^3$, $\tilde{N}_{ngb} =
32$), noting that the simulations of higher angular resolution exhibit a
similar behaviour. In Fig.~\ref{Test1:Fig:Timestep} we show the evolution of
the ionization front for four different choices for the size of
the radiative transfer time step, $\Delta t_r = 0.01, 0.1, 1 $ and $10 \Myr$. Note that the
simulation with $\Delta t_r = 0.01$ is identical to the simulation discussed
in Fig.~\ref{Fig:Test1:IF}. In order to keep the angular sampling the same,
at each radiative transfer step we split the emission of
photons over 10, 100 and 1000 random orientations of the emission cone tessellation of
the ionizing source for the simulations employing  $\Delta t_r = 0.1, 1$ and
$10\Myr$, resp. Photon packets that are emitted by the source in a certain orientation
are transmitted further downstream and can propagate over multiple inter-particle distances. 
We follow each photon packet until it has either been absorbed or left the simulation box, 
to properly solve the time-independent radiative transfer equation 
for the large time steps under consideration.
\par
From Fig.~\ref{Test1:Fig:Timestep} we see that the evolution of the 
ionization front for the simulations with time step $\Delta t_r = 0.1, 1$ and
$10\Myr$ is delayed with respect to the evolution of the ionization front for
the simulation with time step $\Delta t_r = 0.01\Myr$. This delay increases
with the size of the time step, being barely visible for the simulation using 
$\Delta t_r = 0.1\Myr$. The delay arises because
the neutral fraction is only updated at the end of each
radiative transfer time step. Photons that have been absorbed during the transport over a single time step 
but that have not been used to advance the neutral fraction during the subsequent sub-cycling of the rate equation 
are therefore only re-inserted in the transport process at the beginning of
the next time step\footnote{We note that we have also
  tried to re-insert these photons immediately after they have been
  absorbed, by integrating the rate equation already at the end of
  each transport cycle (without updating the neutral fraction) to obtain the number of photons that have
  been erroneously counted for being absorbed (because of the assumption of a constant neutral fraction). 
  The re-insertion of these photons in the transport process within the same radiative transfer time step 
  over which they have been absorbed indeed reduced the delay of the ionization front observed in Fig.~\ref{Test1:Fig:Timestep}.
}, and are thus delayed.
From Fig.~\ref{Test1:Fig:Timestep} it can, however, be
seen that after a few time steps (here $\approx 15$), the ionization front
catches up to agree with the ionization front obtained in
the simulation using $\Delta t_r = 0.01 \Myr$. We have convinced
ourselves that from then on the neutral fraction profile around the ionizing
source is also nearly identical to the profile obtained in the simulation with
$\Delta t_r = 0.01 \Myr$ (Fig.~\ref{Test1:Fig:NeutralFraction1d}). 
\par
In summary, in this section we showed that \traphic\ is able to reproduce the
expected equilibrium neutral fraction around an ionizing source embedded in a homogeneous
medium, as well as the dynamical evolution towards it. Because the radiative
transfer is  explicitly photon-conserving, the spatial resolution only
determines the scale over which the converged solution is averaged. We have
furthermore seen that the performance of \traphic\ is stable with respect to
variations in the size of the time step. Particle noise due to the discrete nature of SPH simulations is small
except for the choice of parameters $N_c = \tilde{N}_{ngb}$. The noise can
be successfully  suppressed by applying the resampling technique of
Section~\ref{Sec:Method:Regularization}, i.e. by periodically perturbing the positions of the SPH particles
within their spheres of influence. 
\par
In the next section we employ a test with broken
symmetry to demonstrate the ability of \traphic\ to correctly produce
shadows behind opaque obstacles and we study further how the choice of
the parameters $N_c$ and $\tilde{N}_{ngb}$ affects the outcome of the numerical simulations. 

\subsubsection{Test 2: Breaking the spherical symmetry}
\label{Section:Test2}
\begin{figure*}

  \includegraphics[trim = 40mm 0mm 40mm 0mm, width=0.19\textwidth, clip=true]{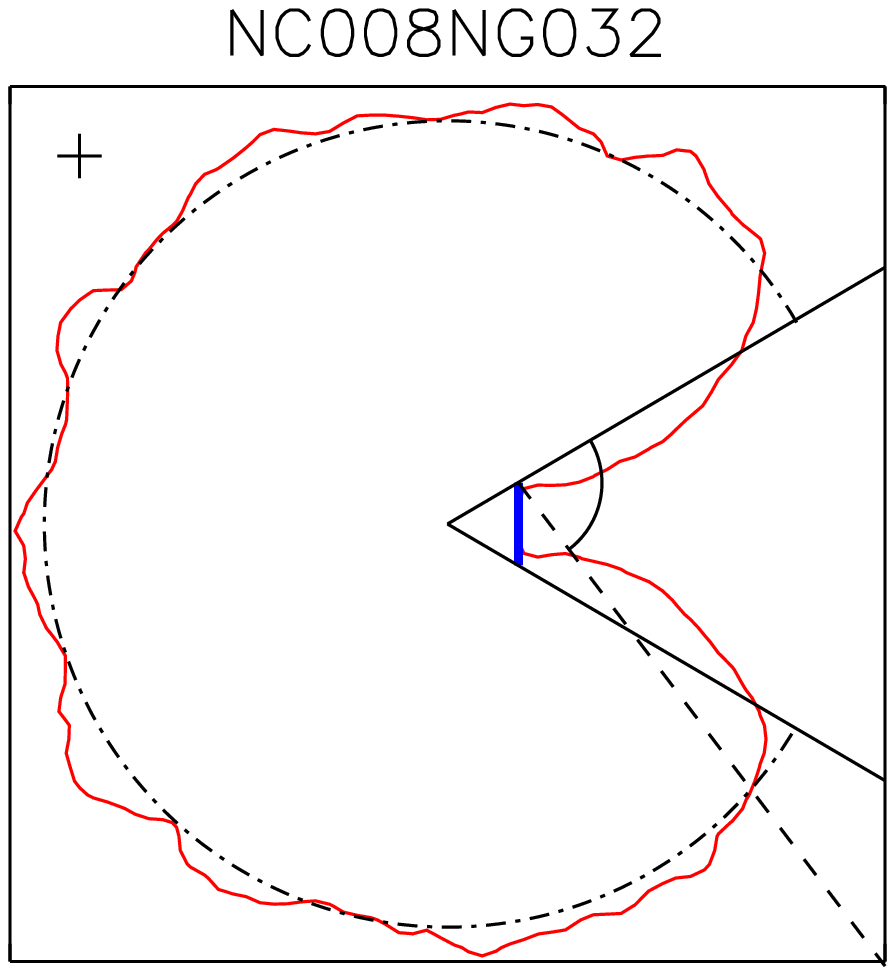}      
  \includegraphics[trim = 40mm 0mm 40mm 0mm, width=0.19\textwidth, clip=true]{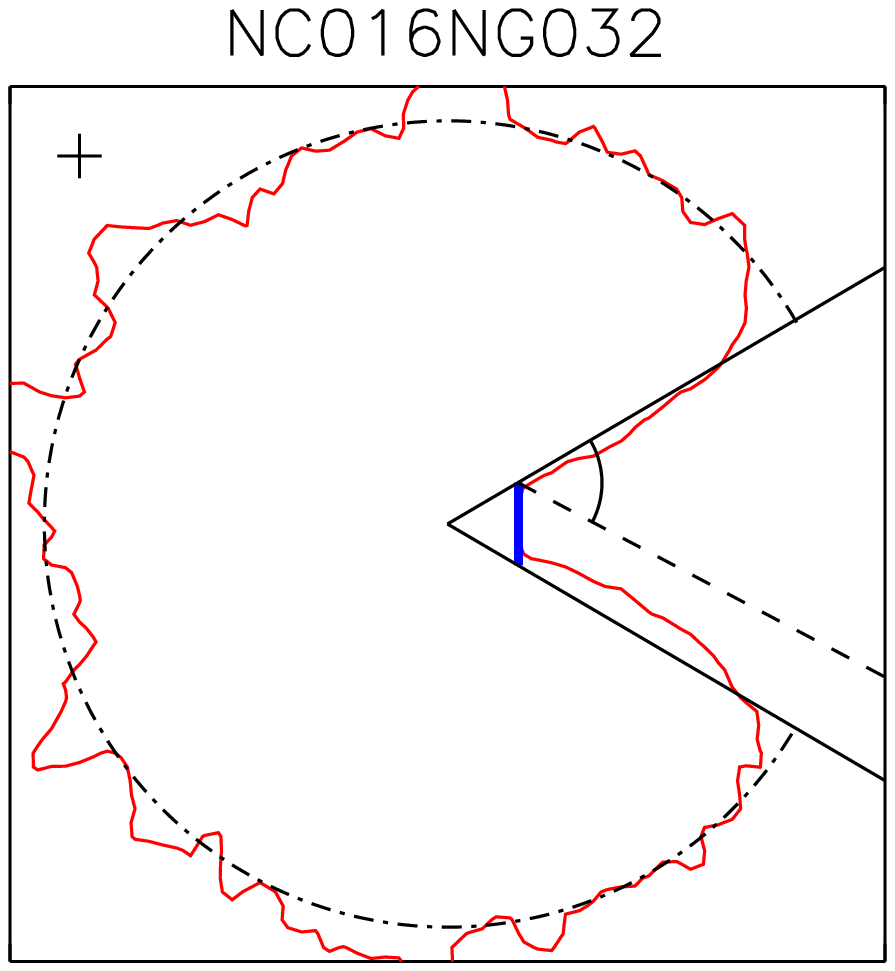} 
  \includegraphics[trim = 40mm 0mm 40mm 0mm, width=0.19\textwidth, clip=true]{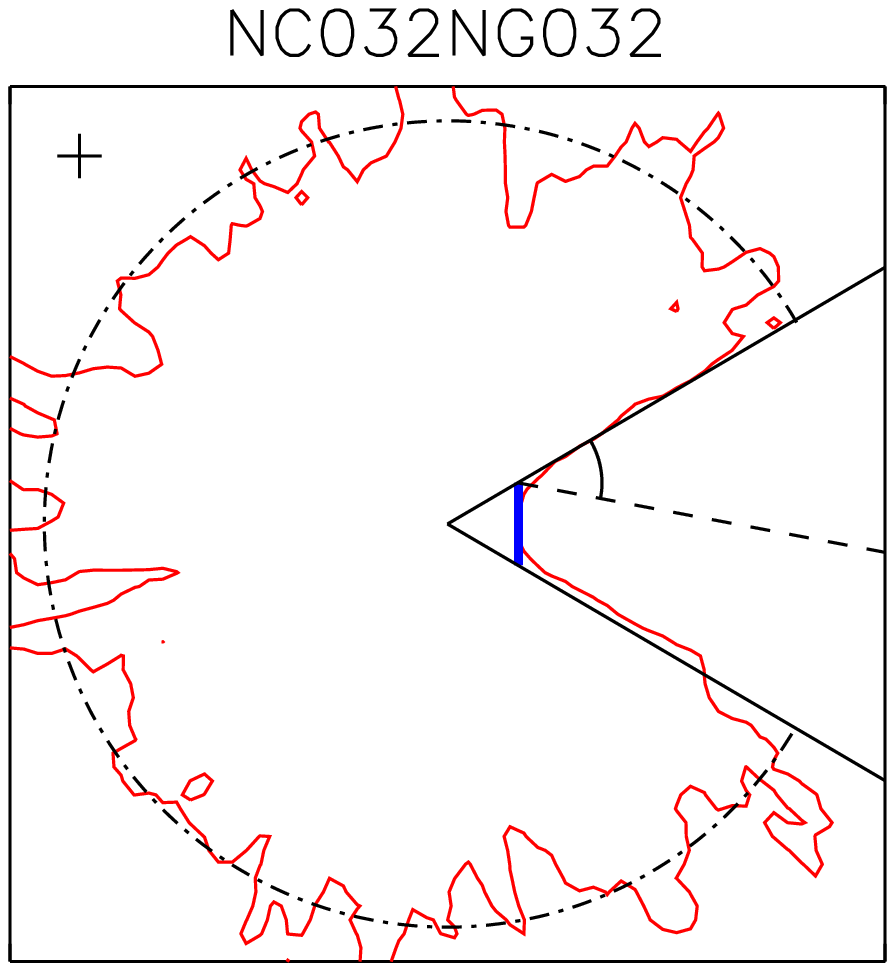} 
  \includegraphics[trim = 40mm 0mm 40mm 0mm, width=0.19\textwidth, clip=true]{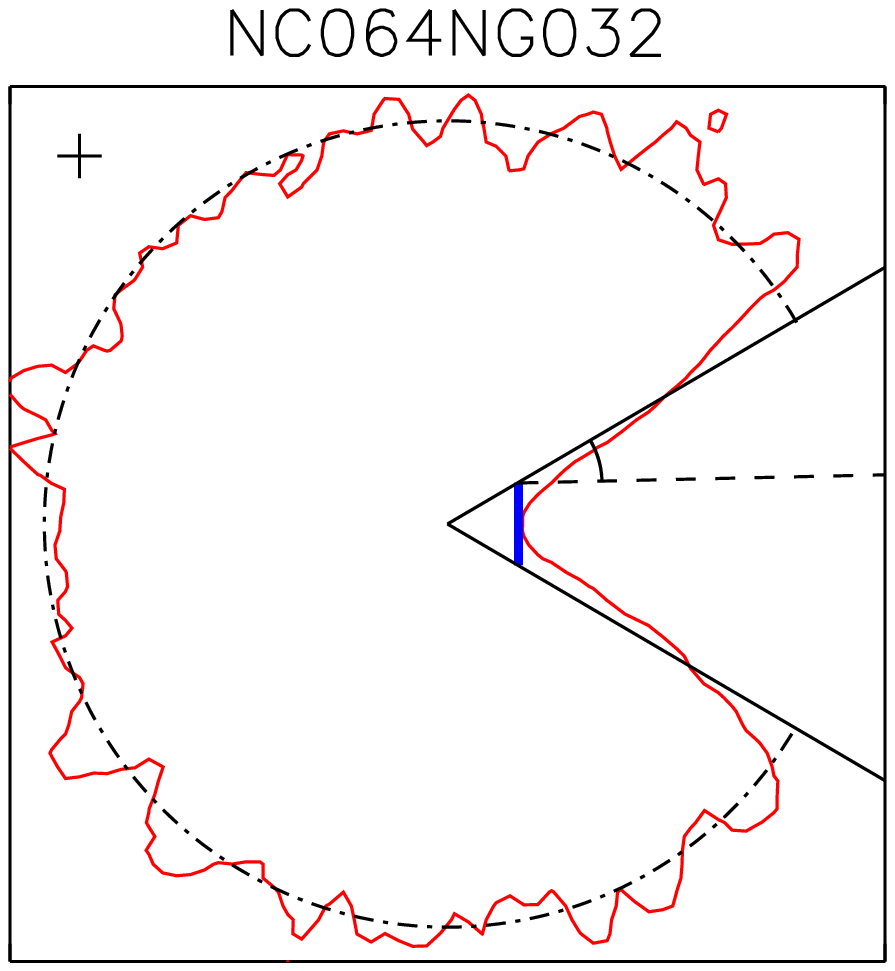} 
  \includegraphics[trim = 40mm 0mm 40mm 0mm, width=0.19\textwidth, clip=true]{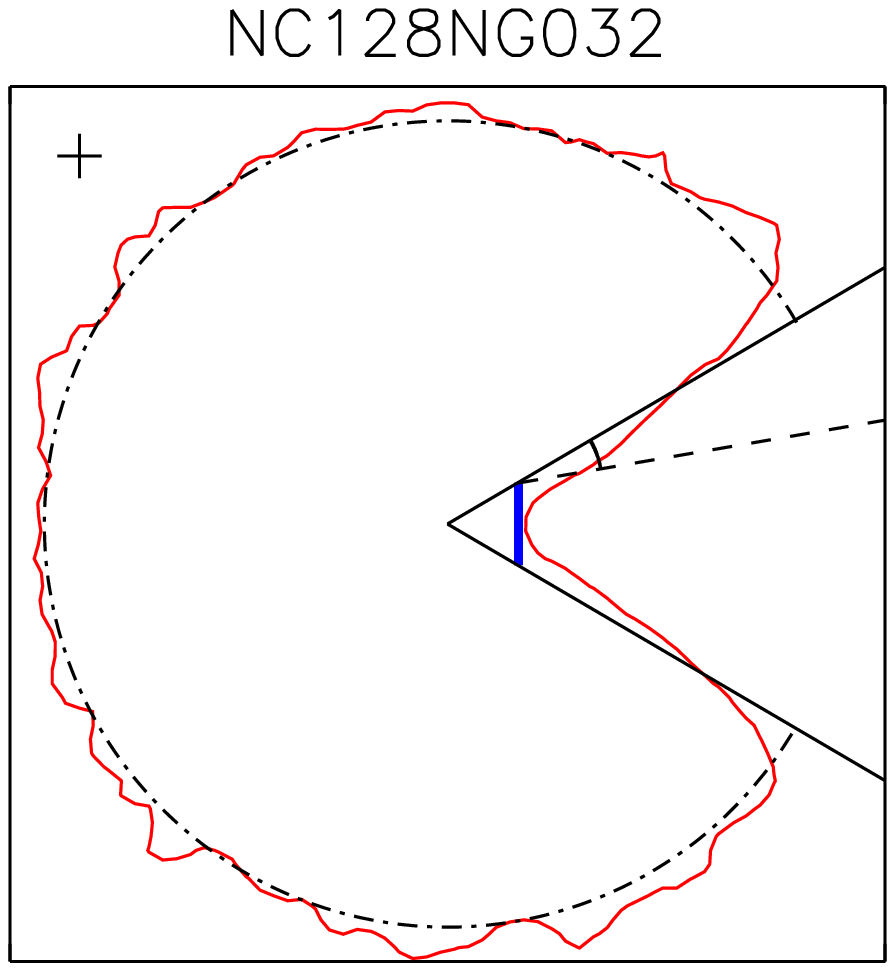}

  \caption{Test 2. Slice through the simulation box at $z=L_{box}/2$ showing the ionization front
  (red solid line) at time $t_r = 80\Myr$ around an ionizing source sitting
  in the centre of the simulation box. The black dot-dashed line shows the expected
  ionization front position. The thick blue vertical line indicates an
  obstacle opaque to ionizing photons and the black solid lines trace out the
  boundaries of the shadow this obstacle is expected to throw. The cross and the black dashed
  line indicate the spatial and angular resolution, respectively, as described
  in the text. The spatial resolution is fixed to $N_{SPH} = 64^3, \tilde{N}_{ngb} = 32$. The
  angular resolution increases from $N_c = 8$ in the left-most panel to $N_c
  =128$ in the right-most panel, in factors of 2.} 
  \label{Test2:Fig:1} 
  
\end{figure*} 

\begin{figure*}

  \includegraphics[trim = 40mm 0mm 40mm 0mm, width=0.19\textwidth, clip=true]{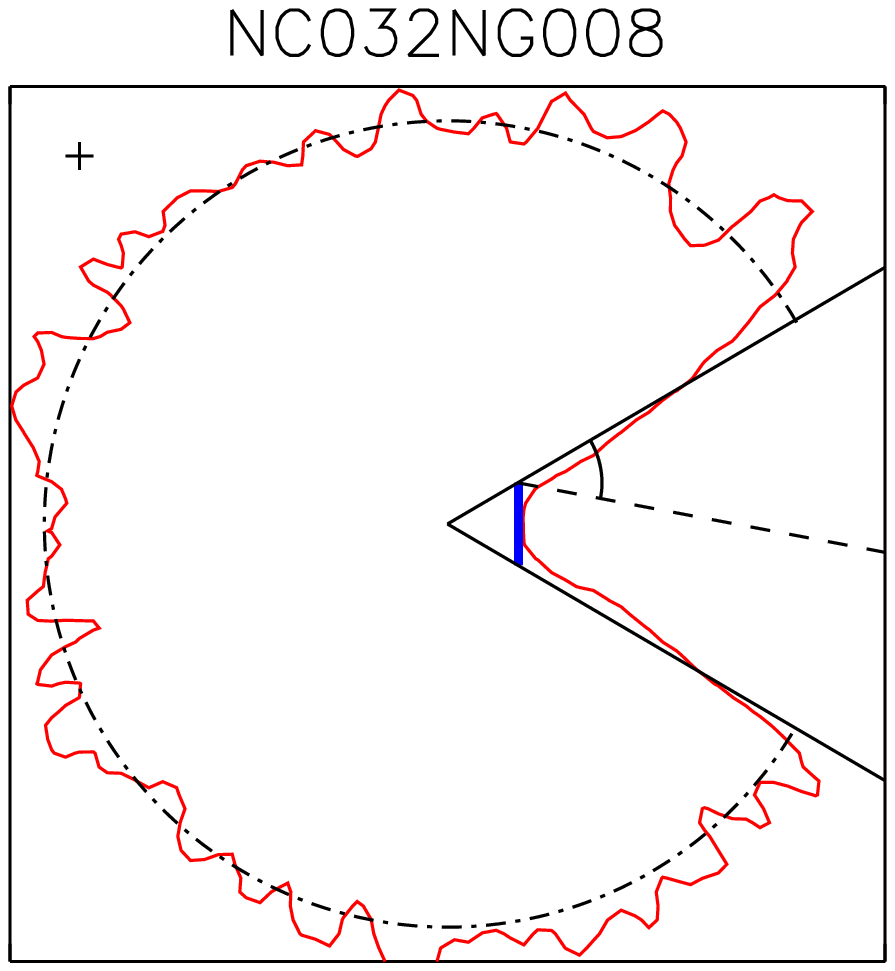}
  \includegraphics[trim = 40mm 0mm 40mm 0mm, width=0.19\textwidth, clip=true]{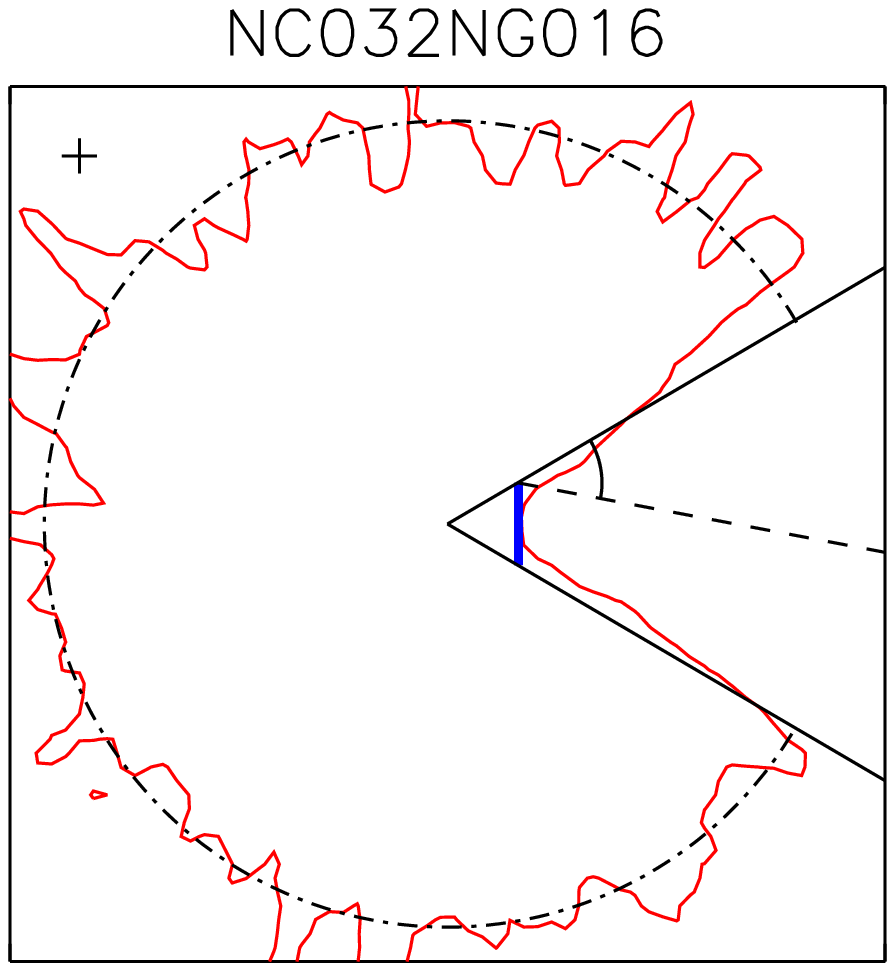} 
  \includegraphics[trim = 40mm 0mm 40mm 0mm, width=0.19\textwidth, clip=true]{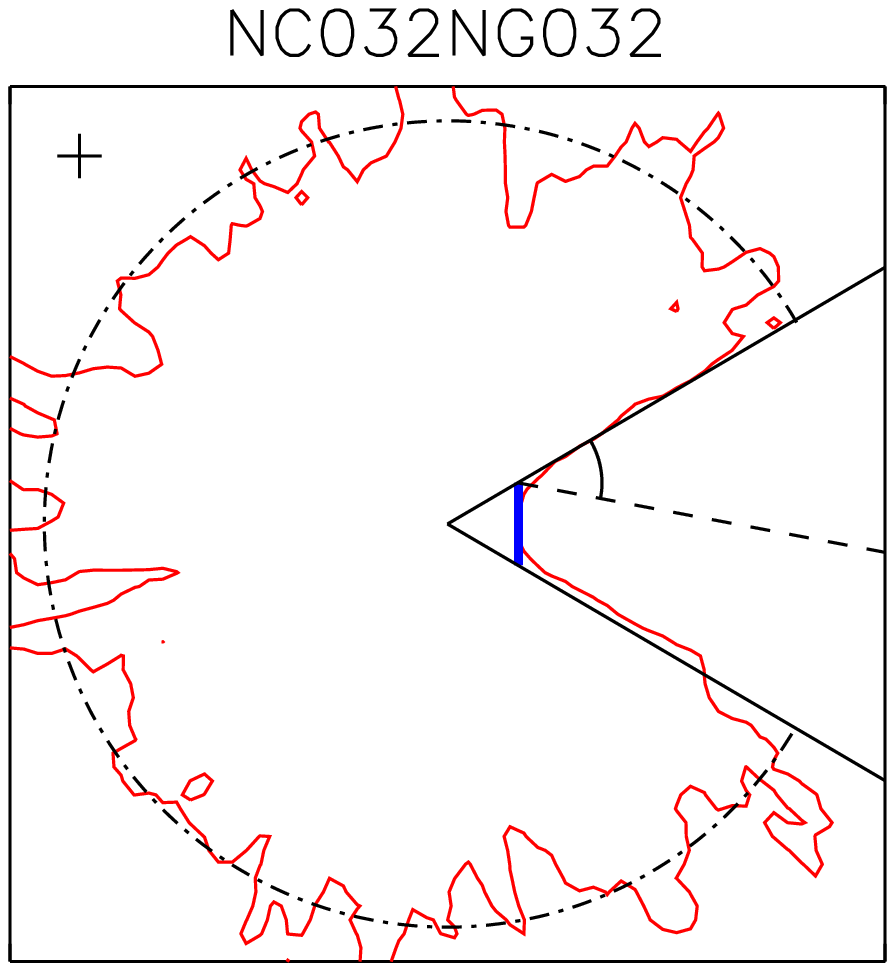} 
  \includegraphics[trim = 40mm 0mm 40mm 0mm, width=0.19\textwidth, clip=true]{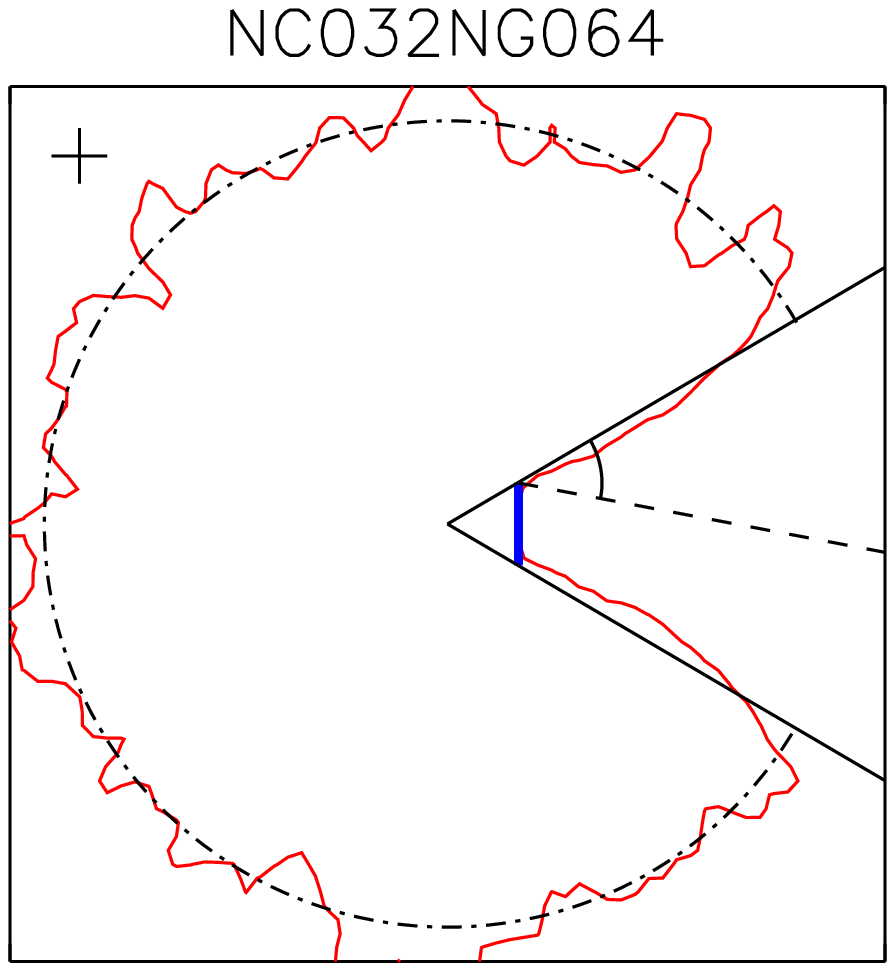}
  \includegraphics[trim = 40mm 0mm 40mm 0mm, width=0.19\textwidth, clip=true]{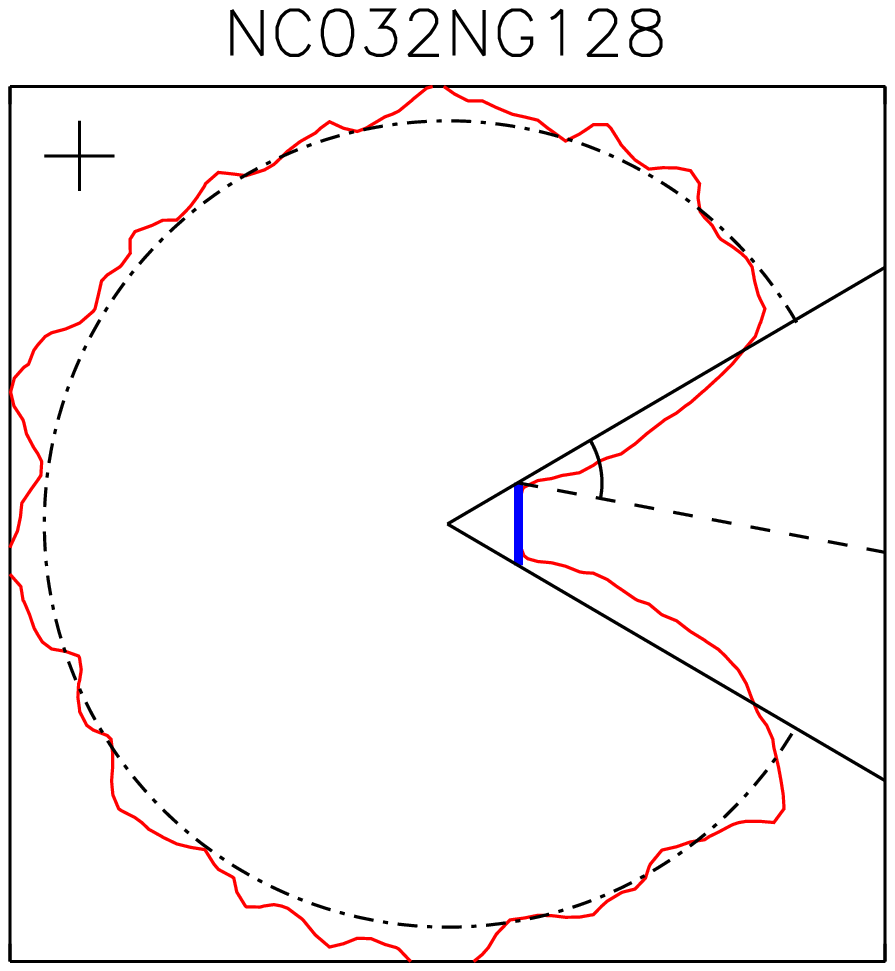}

  \caption{Test 2. Same as Fig.~\ref{Test2:Fig:1}, but with the angular
  resolution fixed to $N_c = 32$ and the spatial resolution decreasing, from
  $N_{SPH} = 64^3, \tilde{N}_{ngb} = 8$ in the left-most panel to $N_{SPH} =
  64^3, \tilde{N}_{ngb} = 128$ in the right-most panel, increasing the number
  of neighbours $\tilde{N}_{ngb}$ in factors of 2.} 
  \label{Test2:Fig:3} 
  
\end{figure*} 

In the absence of scattering interactions, photons propagate along
straight lines into the direction set at the time of their emission. Consequently, opaque
obstacles throw sharply defined shadows. In this
section we are mainly interested in studying the properties of the shadow thrown by an opaque obstacle exposed to
ionizing radiation from a single point source, as obtained with \traphic. At the
same time, we will extend the study of particle noise presented in
Section~\ref{Section:Test1} to include other choices for the parameter $\tilde{N}_{ngb}$.
\par
\begin{figure*}

  \includegraphics[trim = 40mm 0mm 40mm 0mm, width=0.24\textwidth, clip=true]{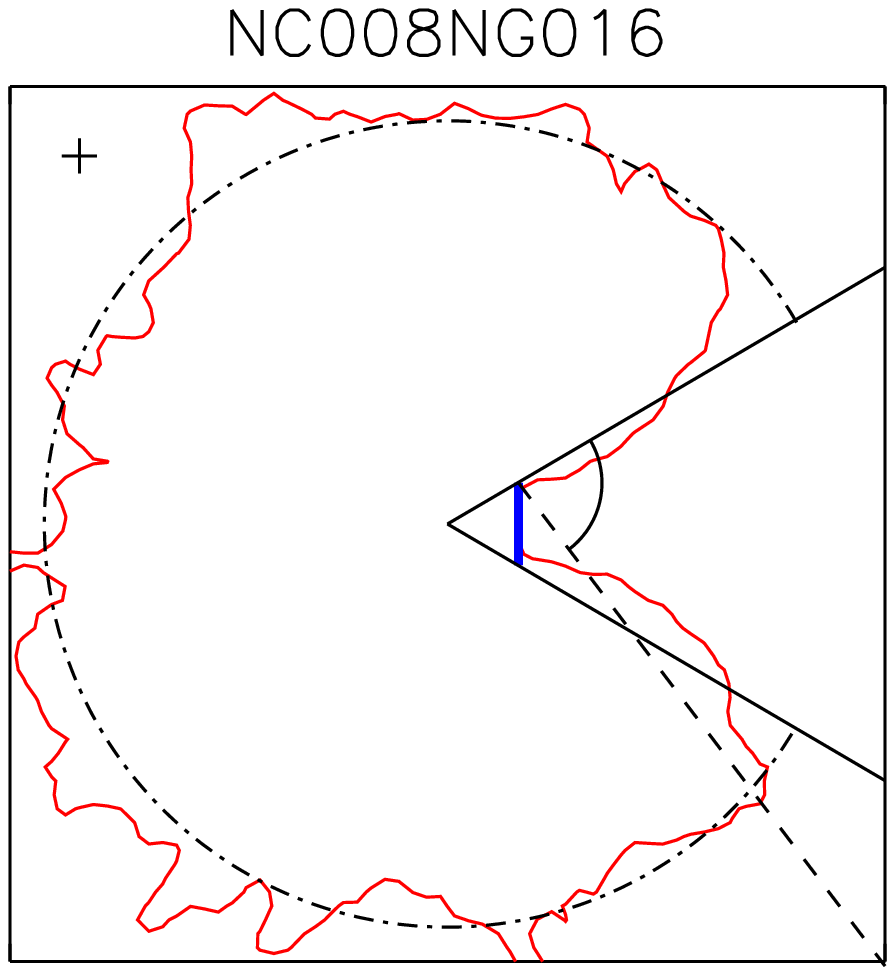}
  \includegraphics[trim = 40mm 0mm 40mm 0mm, width=0.24\textwidth, clip=true]{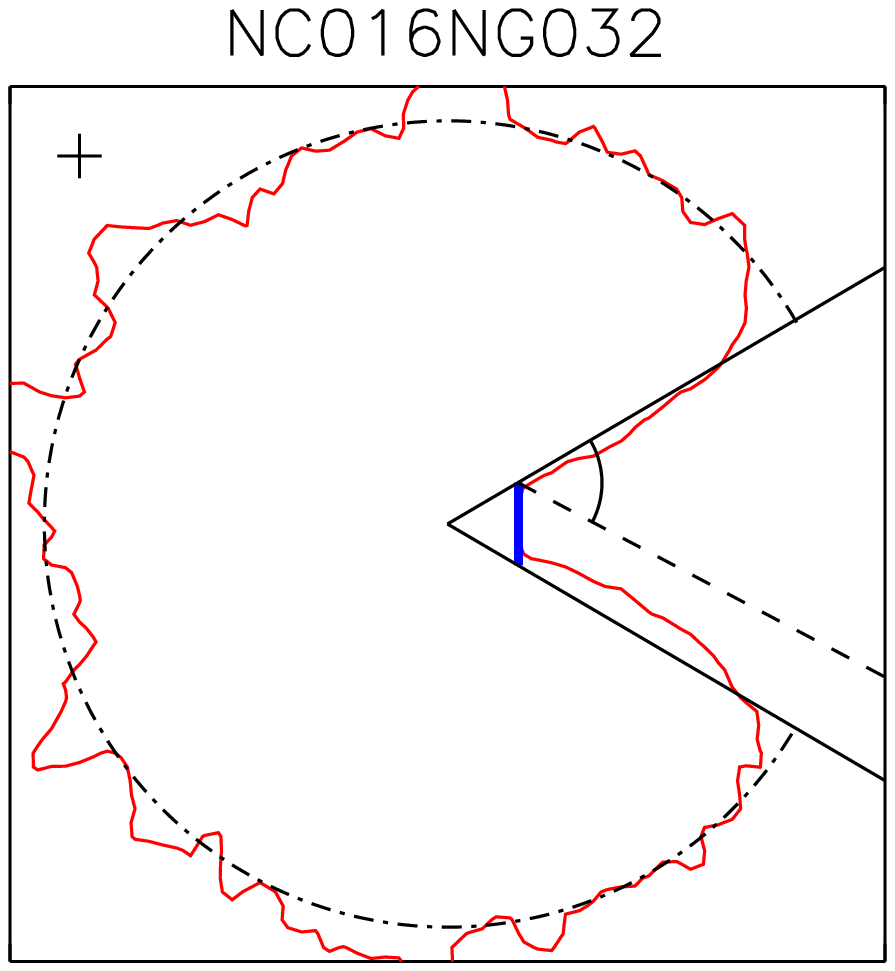} 
  \includegraphics[trim = 40mm 0mm 40mm 0mm, width=0.24\textwidth, clip=true]{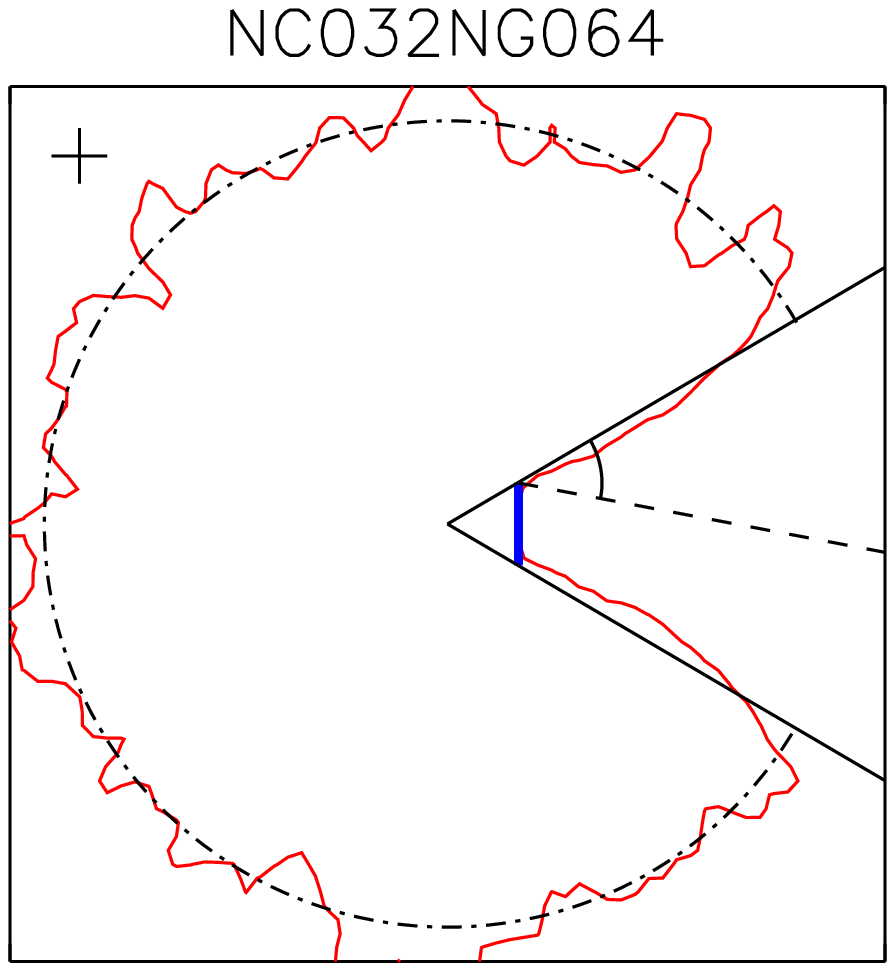} 
  \includegraphics[trim = 40mm 0mm 40mm 0mm, width=0.24\textwidth, clip=true]{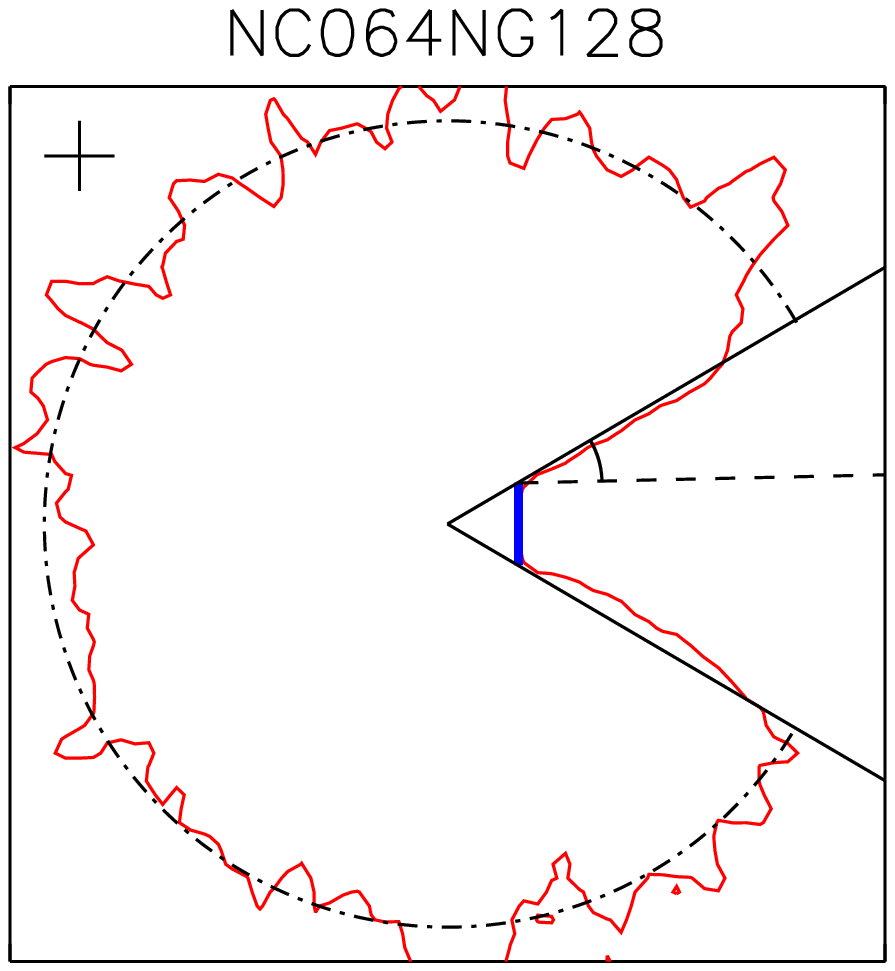} 

  \caption{Test 2. Same as Fig.~\ref{Test2:Fig:1}, but fixing the ratio
  between spatial and angular resolution to $\tilde{N}_{ngb} / N_c = 2$ ($N_{SPH} = 64^3$).} 
  \label{Test2:Fig:5} 
  
\end{figure*}
\begin{figure*}

  \includegraphics[trim = 40mm 0mm 40mm 0mm, width=0.24\textwidth, clip=true]{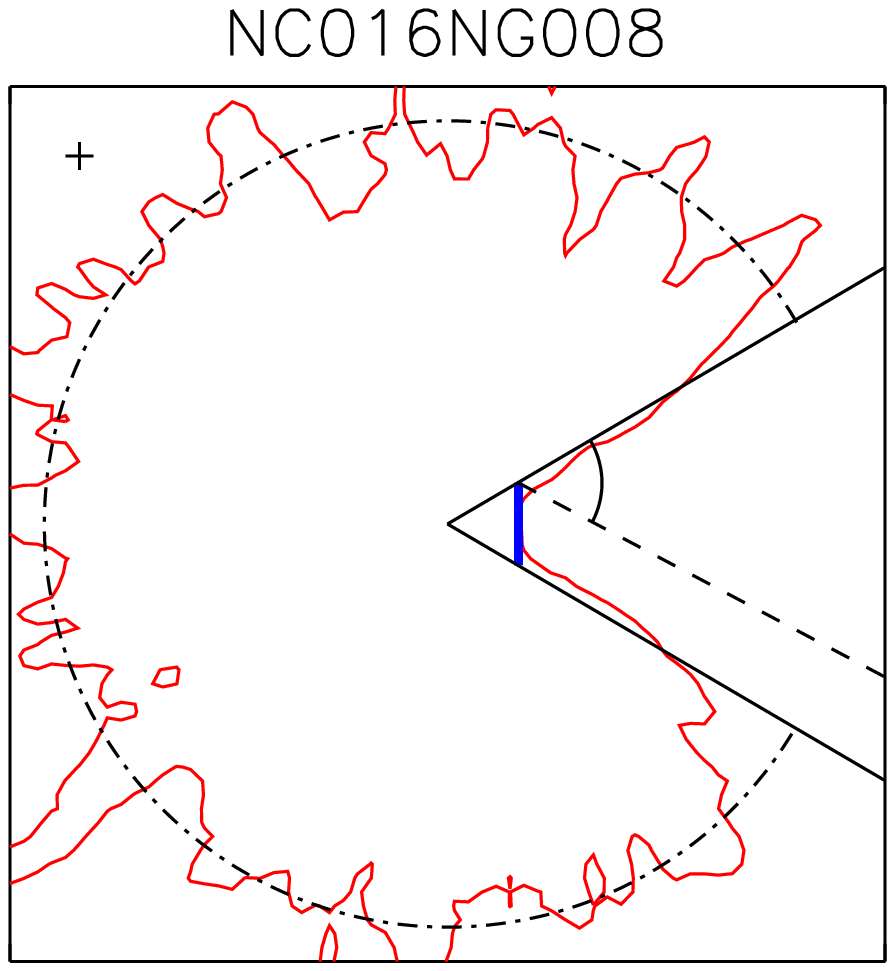}
  \includegraphics[trim = 40mm 0mm 40mm 0mm, width=0.24\textwidth, clip=true]{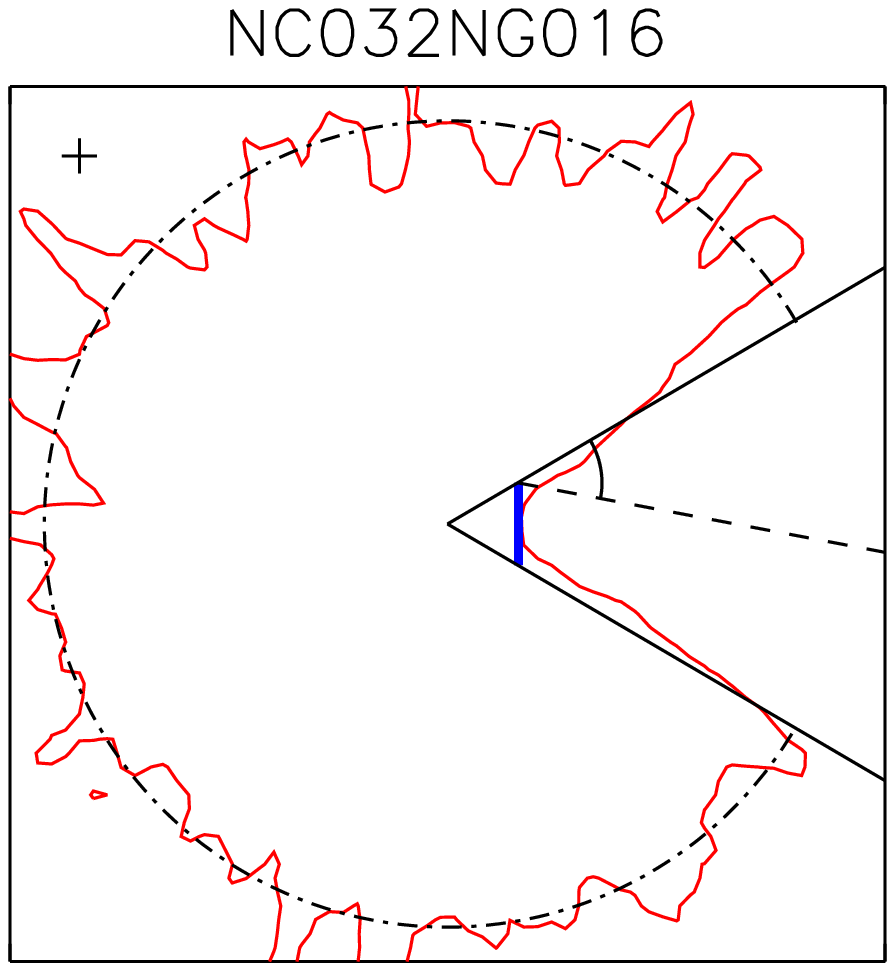} 
  \includegraphics[trim = 40mm 0mm 40mm 0mm, width=0.24\textwidth, clip=true]{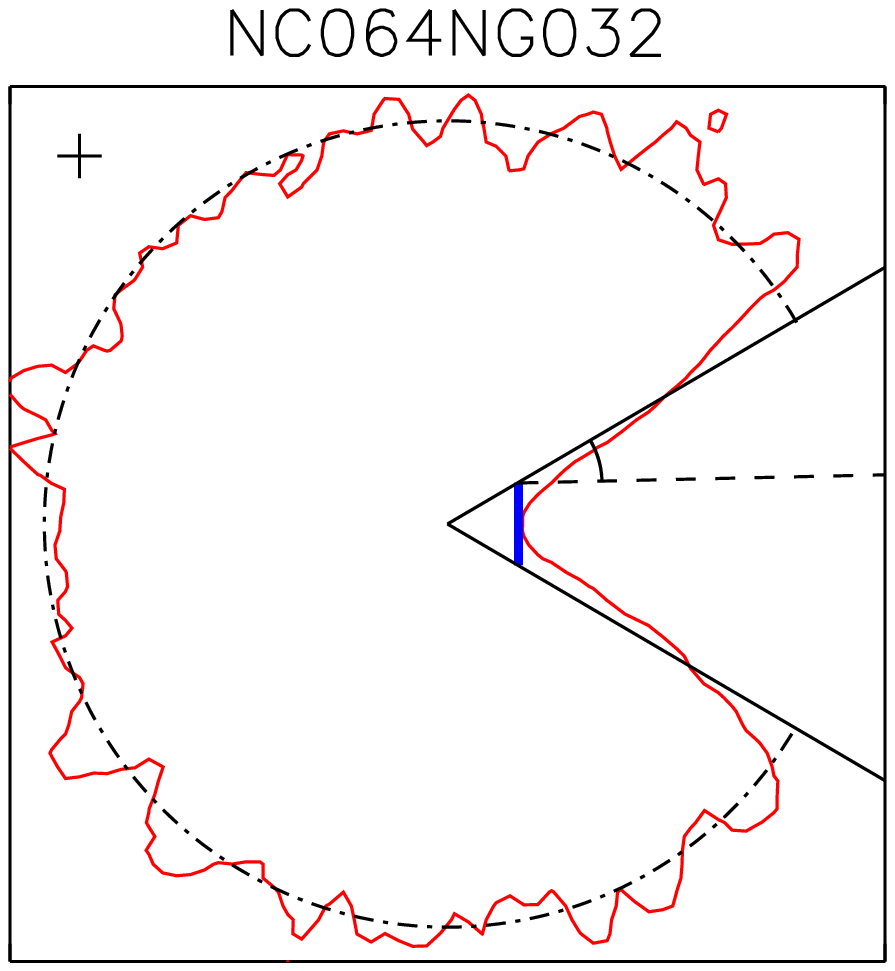} 
  \includegraphics[trim = 40mm 0mm 40mm 0mm, width=0.24\textwidth, clip=true]{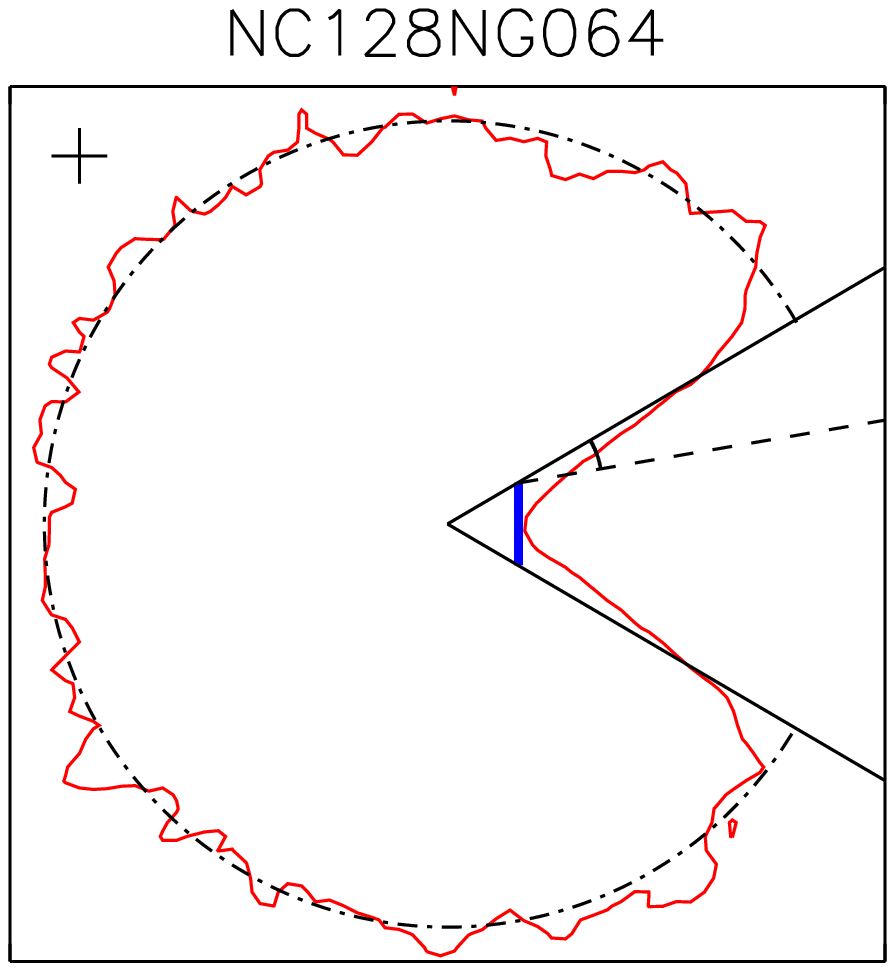} 
  
  \caption{Test 2. Same as Fig.~\ref{Test2:Fig:1}, but fixing the ratio
  between spatial and angular resolution to $\tilde{N}_{ngb} / N_c = 1 / 2$ ($N_{SPH} = 64^3$). } 
  \label{Test2:Fig:6} 
  
\end{figure*}
\begin{figure}

  \includegraphics[trim = 40mm 0mm 40mm 0mm, width=0.23\textwidth, clip=true]{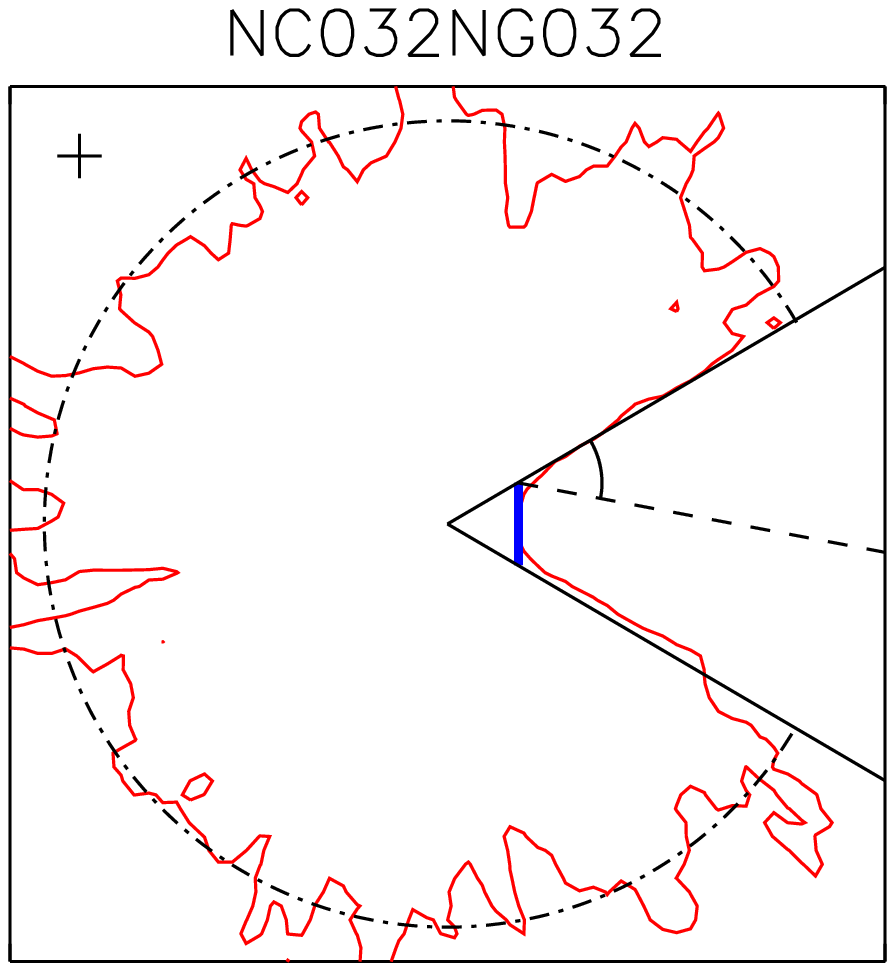}
  \includegraphics[trim = 40mm 0mm 40mm 0mm, width=0.23\textwidth, clip=true]{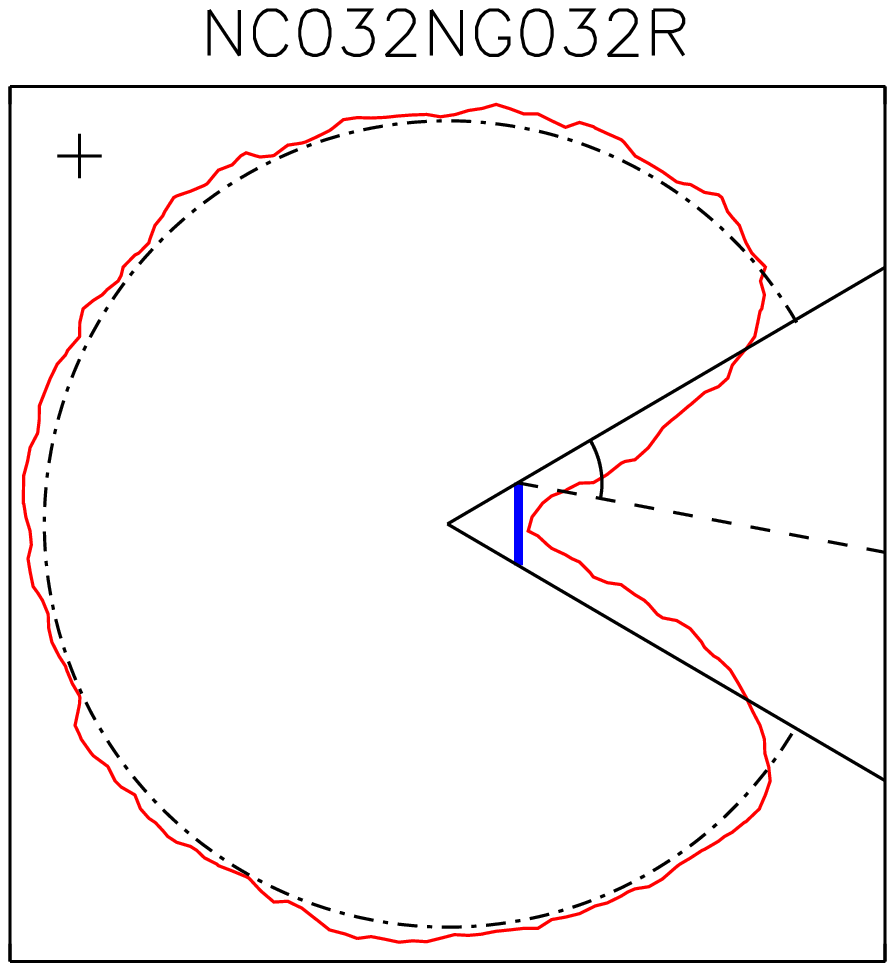}

\caption{Test 2. Left-hand panel: Identical to the middle panel of
  Fig.~\ref{Test2:Fig:1} and \ref{Test2:Fig:3}.
  Right-hand panel: Same as left-hand panel, except for the fact that in the simulation we
  periodically resampled the density field, resulting in a strong suppression of
  the particle noise seen in the left-hand panel. Note the (small amount of) diffusion of photons across
  the shadow boundary due to the motion of the transmission cone apexes.} 
  \label{Test2:Fig:2} 
  
\end{figure}

The geometry of our test problem closely follows the description of the shadow
test in \cite{Mellema:2006}. We consider a source emitting
$\dot{\mathcal{N}}_\gamma = 10^{54}~\mbox{photons}~\si$, each of hydrogen-ionizing
energy $h_p\nu=13.6\eV$, residing in an initially neutral, static hydrogen-only field
of constant number density $n_H=1.87\times10^{-4}\cmci$ and temperature $T = 10^4\K$. The
Str\"omgren radius (Eq.~\ref{Eq:StroemgrenRadius}) corresponding to this set
of parameters is $r_s = 0.965\Mpc$ and the Str\"omgren time is $\tau_s =654.3\Myr$.
For the numerical simulation a star particle is placed in the centre of a box of size
$L_{box}=1\Mpc$. The boundaries of the box are photon-transmissive. 
The density field is sampled using $N_{SPH}= 64^3$ gas
particles with mass $m = n_H m_H L_{Box}^3/ N_{SPH}$ at glass-like positions. The particle densities are found
using the SPH interpolation implemented in \gadget, with $N_{ngb} = 48$.
We furthermore place an infinitely thin opaque disc perpendicular to the $x$-axis
at a distance of $0.08\Mpc$ along the $x$-axis from the star (thick blue
vertical lines in Figs.~\ref{Test2:Fig:1}-\ref{Test2:Fig:2}). The $y$ and $z$
coordinates of the disc centre are identical to those coordinates of
the star. Photons that cross the disc are removed. 
\par
We performed a series of radiative transfer simulations (with time step $\Delta t_r = 10^4
\yr$), using different choices for the parameters $\tilde{N}_{ngb}$, which sets the
spatial resolution if the total number of SPH particles is fixed, and $N_c$,
which sets the angular resolution. The results are shown in
Fig.~\ref{Test2:Fig:1} - \ref{Test2:Fig:2}, displaying a slice 
through the simulation box at $z=L_{box} / 2$. In each panel, the black
dash-dotted line shows the expected position of the ionization front
(Eq.~\ref{Test1:IF}) at time $t_r = 80\Myr$, which marks the end of the simulation. The black solid lines emerging
from the star at the centre of the slice show the
boundaries of the shadow expected to be thrown by the opaque disc (thick blue vertical line).
In the top-left corner of each panel we indicate the spatial resolution by a cross of length
$2 \langle\tilde{h}\rangle$ corresponding to the average diameter of the sphere containing
$\tilde{N}_{ngb}$ neighbours. The angular resolution is
indicated by the angle $\omega$ enclosed by the black dashed 
line and the upper shadow boundary, where
$\omega$ is determined using Eq.~\ref{Eq:Apex}. It indicates the {\it maximum}
angle photons can theoretically diverge from the shadow boundary into the shadow region,
given the chosen angular resolution $N_c$.
\par
The position of the ionization front (the iso-surface for which the 
neutral fraction $\eta = 0.5$)  at time $t_r = 80\Myr$ as obtained with
\traphic\ is shown by the red solid line. In
all panels, that is for all spatial and angular resolutions, the
ionization front is found at the proper position, in agreement with our
findings of Section~\ref{Section:Test1}. The shadow thrown by the
opaque disc is always sharp. We now
discuss the dependence of the results on the chosen spatial and angular resolutions.
\par
In Fig.~\ref{Test2:Fig:1} we show the ionization front obtained in simulations employing a
fixed spatial resolution, $\tilde{N}_{ngb} = 32$, but varying angular
resolution, ranging from $N_c = 8$ in the left-most to $N_c = 128$ in the
right-most panel. The most prominent difference between
the results of the different simulations is the noisiness of the contour tracing out
the ionization front. The angular resolution study presented here is very similar to the
one in the last section. For the lowest angular resolution, $N_c=8$, the ionization front is very
smooth due to the large number of neighbours within each emission and
transmission cone. The noise increases with the angular resolution until
it reaches a maximum for 
$N_c = \tilde{N}_{ngb}$. For higher angular resolutions, particle noise 
is efficiently suppressed due to the large number of ViPs
that are placed in empty cones and that distribute the photons they absorb amongst their
$\tilde{N}_{ngb}$ SPH neighbours using (photon-conserving) SPH interpolation.
\par
From Fig.~\ref{Test2:Fig:1} it can furthermore be seen how the 
sharpness of the shadow thrown by the opaque disc depends on the angular resolution. For the lowest
angular resolution, the shadow is somewhat blurred, though not nearly as much
as the angular resolution would imply. Increasing the angular resolution yields
slightly sharper shadows. However, if the angular resolution is increased beyond
$N_c=\tilde{N}_{ngb}$, the shadows become slightly less sharp. This is
because the photons absorbed by ViPs are distributed amongst the
neighbouring gas particles using SPH interpolation and the
interpolation procedure does not know about the shadow region. The slight diffusion
of photons across the shadow boundary is in this case consistent with the spatial resolution.
\par

In Fig.~\ref{Test2:Fig:3} we show the results of the simulations where we fixed the
angular resolution to $N_c = 32$, but varied the spatial resolution by changing $\tilde{N}_{ngb}$. The trends visible in
Fig.~\ref{Test2:Fig:1} can also be observed here. The ionization front is
most noisy and the shadow is sharpest for $N_c = \tilde{N}_{ngb}$.   
For $\tilde{N}_{ngb} > N_c$, 
noise due to the discreteness of the particles employed for the transport
of photons is suppressed by the large number of neighbours per cone, but the
shadow is slightly blurred. The shadow becomes sharper for a smaller number of
neighbours, since generally not all of the solid angle will be covered by
the neighbours, an effect that becomes more important for smaller numbers
of neighbours.  For $\tilde{N}_{ngb} < N_c$  
the ionization front becomes smoother due to the
regularizing effect of the SPH interpolation from ViPs, which
also leads to a small diffusion of photons across the shadow boundary,
consistent with the spatial resolution.
\par
In Figs.~\ref{Test2:Fig:5} and \ref{Test2:Fig:6} we keep the ratio 
$\tilde{N}_{ngb} / N_c$ fixed at $\tilde{N}_{ngb} / N_c = 2$ and 
$\tilde{N}_{ngb} / N_c = 1/2$, resp. In the first case
there are on average $2$ neighbours per cone, whereas in the
second case there is on average one neighbour in every second cone. 
From Fig.~\ref{Test2:Fig:5} it is clear
that the shadow does get sharper when the angular resolution is increased,
although the effect is small, since the shadow is always very sharp.
Because we keep the ratio  $\tilde{N}_{ngb} / N_c$ fixed at  $2$, the number of ViPs
employed in the simulation stays low for all angular resolutions  and the
shadows are not visibly diffused by the SPH interpolation of absorbed photons from the
ViPs. Furthermore, 
the noisiness of the ionization front remains constant
throughout the parameter range. This is because the noise is primarily set by
the ratio $\tilde{N}_{ngb} / N_c$ if $\tilde{N}_{ngb} > N_c$.
In Fig.~\ref{Test2:Fig:6}, on the other hand, there is a substantial
probability for creating a ViP per cone. Since the absolute
number of ViPs present in the simulation increases with increasing angular
resolution $N_c$, the noise decreases with $N_c$.
\par

In the last section (see bottom row of Fig.~\ref{Fig:Test1:NeutralFraction2d}) we employed a resampling technique to decrease the noise
exhibited by the neutral fraction contours. 
Recall from Section \ref{Section:Transmission} that the apexes of the transmission cones
are attached to the positions of the SPH particles. Hence, resampling results
in slight shifts in the position of each transmission cone apex, on the scale
of the spatial resolution employed in the SPH simulation. This shifting is expected to lead to a small diffusion of photons
across the expected shadow boundary. As noted earlier in
Section~\ref{Sec:Method:Regularization}, 
such a diffusion due to particle resp. apex motion will
also occur in radiation-hydrodynamical simulations because of the movement of
the SPH particles. It is therefore interesting
to study the properties of the shadow thrown by an opaque obstacle in the case of resampling.
\par
In Fig.~\ref{Test2:Fig:2} we show the results of a 
simulation which employs the same parameters as used for the simulation presented
in the middle panels of Fig.~\ref{Test2:Fig:1} and \ref{Test2:Fig:3}, but with an
additional resampling of the hydrogen density field every 10th radiative
transfer time step. As in the last section, the resampling is performed without changing
the hydrogen density, since this would lead to an enhanced recombination
rate. Numerical noise is successfully
suppressed by the random perturbations given to the positions of the SPH
particles. Consequently, the ionization front appears significantly
smoother. The degree of photon diffusion into the shadow region is small 
and does not significantly degrade the angular resolution of the radiative
transfer. This is because the diffusion scale is set by the spatial resolution employed in
the SPH simulation. Therefore, well-defined shadows will be thrown as long as
the obstacle is spatially resolved.
\par
In summary, in this section we showed that \traphic\ is able to produce a
well-defined shadow behind an opaque obstacle, with the shadow sharpness in
full agreement with the chosen spatial and angular resolutions. In fact,
the shadows are much sharper than implied by the formal angular
resolution, thanks to the angular adaptivity inherent to \traphic. 
For a fixed angular resolution, the shadows are sharpest for  $N_c= \tilde{N}_{ngb}$. They are
slightly broadened by photon diffusion for both $N_c < \tilde{N}_{ngb}$ and $N_c >
\tilde{N}_{ngb}$, due to the increased coverage of the solid angles traced out by
the transmission cones with SPH particles for an increasing number of
neighbours $\tilde{N}_{ngb}$ and the SPH interpolation of the photons
absorbed by ViPs, resp. We confirmed our finding of the last section that
unless $\tilde{N}_{ngb} = N_c$, 
noise due to the discreteness of the particles on which the transport of photons takes place is
small, since it is suppressed by either the
large number of neighbours per cone (if $N_c< \tilde{N}_{ngb}$) or the
large number of ViPs employed (if $N_c > \tilde{N}_{ngb} $). The resampling technique
presented in Section~\ref{Sec:Method:Regularization} is very effective at
suppressing particle noise. We have seen that resampling the density field does not 
severely degrade the angular resolution, even though it leads to a small shift
of the cone apexes. As long as the opaque obstacle is
spatially resolved by the SPH simulation, a well-defined shadow will still be thrown.
\par
The ability to produce sharp shadows is one of the main 
requirements a radiative transfer code has to pass. The results of this
section, together with the results of Test 1, which showed that \traphic\ is able to reproduce the expected neutral
fraction within a spherically symmetric HII region, indicate that \traphic\
can be used to perform the transport of ionizing photons in arbitrary
complex geometries. This is the subject of Test 3, which we will describe next.

\subsubsection{Test 3: Cosmological density field}
\label{Sec:Test3}

In this test we study the propagation of ionization fronts around multiple sources
in a static cosmological density field. The parameters of this test are taken from Test 4 of 
the Cosmological Radiative Transfer Code Comparison Project
(\citealp{Iliev:2006b}). For reference we
repeat the test description here. 
\par
The initial conditions are provided by
a snapshot (at redshift $z \approx 8.85$) from a cosmological N-body and
gas-dynamical simulation performed using the cosmological (uniform-mesh) PM+TVD code of
\cite{Ryu:1993}. The simulation box is $L_{box} = 0.5\cMpch$ on a side,
uniformly divided into $N_{cell} = 128^3$ cells. The initial temperature
is fixed at $T = 100\K$ everywhere. The halos in the
simulation box were found using a friends-of-friends
halo finder with a linking length of $0.25$. The ionizing sources are chosen
to correspond to the 16 most massive halos in the box. We assume that
these have a black-body spectrum $B_\nu(\nu, T)$ with temperature $T = 10^5\K$. The ionizing
photon production rate is assumed to be constant and assigned assuming that
each source lives for $t_s = 3 \Myr$ and emits $f_\gamma = 250$ photons per atom
during its lifetime. Hence, the number of ionizing photons emitted per unit
time is
\begin{equation}
  \dot{\mathcal{N}}_\gamma = f_\gamma \frac{M\Omega_b}{\Omega_0 m_H t_s},
\end{equation}
where $M$ is the total halo mass, $\Omega_0 = 0.27$, $\Omega_b = 0.043$ and $h
= 0.7$. For simplicity, all sources are assumed to switch on at the same time. The
boundary conditions are photon-transmissive. Outputs are produced at $t = 0.05, 0.1,
0.2, 0.3$ and $0.4 \Myr$. 
\par
With respect to the original test setup described above, we require three changes. First, since
our code does not yet solve for the temperature of the gas, we assume a
constant temperature of $T = 10^4\K$ for the ionized gas. Second, since our
code currently treats only a single frequency (bin), we 
assume the grey photo-ionization cross-section
Eq.~\ref{Eq:Photoionization:Crosssection:Grey}, 
for which we find $\bar{\sigma} = 1.49 \times 10^{-18}\cmsqi.$ 
The third change concerns the input density field. Since our code
works directly on the set of particles used in SPH simulations, we have to Monte Carlo sample the original input
density field in order to place particles in the box. We
replace every grid cell $i$ by $N^i_{SPH} = M_i / m$ SPH particles (randomly
distributed within the volume of the grid cell), where $M_i = \rho_i L_{box}^3
/ N_{cell}$ is the mass of the cell and $m$ is the mass of an SPH particle. 
If $N^i_{SPH}$ is not an integer, we draw a random number from a uniform
distribution on the interval (0,1) and place an additional  particle if this number is
smaller than the difference between $N^i_{SPH}$ and the nearest lower
integer. We use $N_{SPH} = N_{cell} = 128^3$. Since the Monte Carlo sampling only results in the approximate
equality $\sum_i N^i_{SPH} \approx N_{SPH}$,  we adjust the particle masses a
posteriori to conserve mass, i.e. $m \to m \times N_{SPH} / \sum_i N^i_{SPH}$.
After the particles have been placed, we calculate their densities using the SPH formalism of \gadget, with
$N_{ngb} = 48$. 
\par 
Note that Monte Carlo sampling the density field with $N_{SPH} \simeq
N_{cell}$ particles yields a smaller
effective resolution than that of the grid input field in low density regions
(many grid cells will be left empty of particles), and to a spurious higher
resolution in high density regions (cells are sampled with many
particles, even though there is no substructure on the scale of a single cell
in the input field). Note also that because the initial conditions were specified on a
uniform grid, we do not benefit from the intrinsic spatial adaptivity of
\traphic, effectively wasting computational resources.
\par

\begin{figure}

  \includegraphics[trim = 31mm 5mm 31mm 5mm, width=39mm, clip = true]{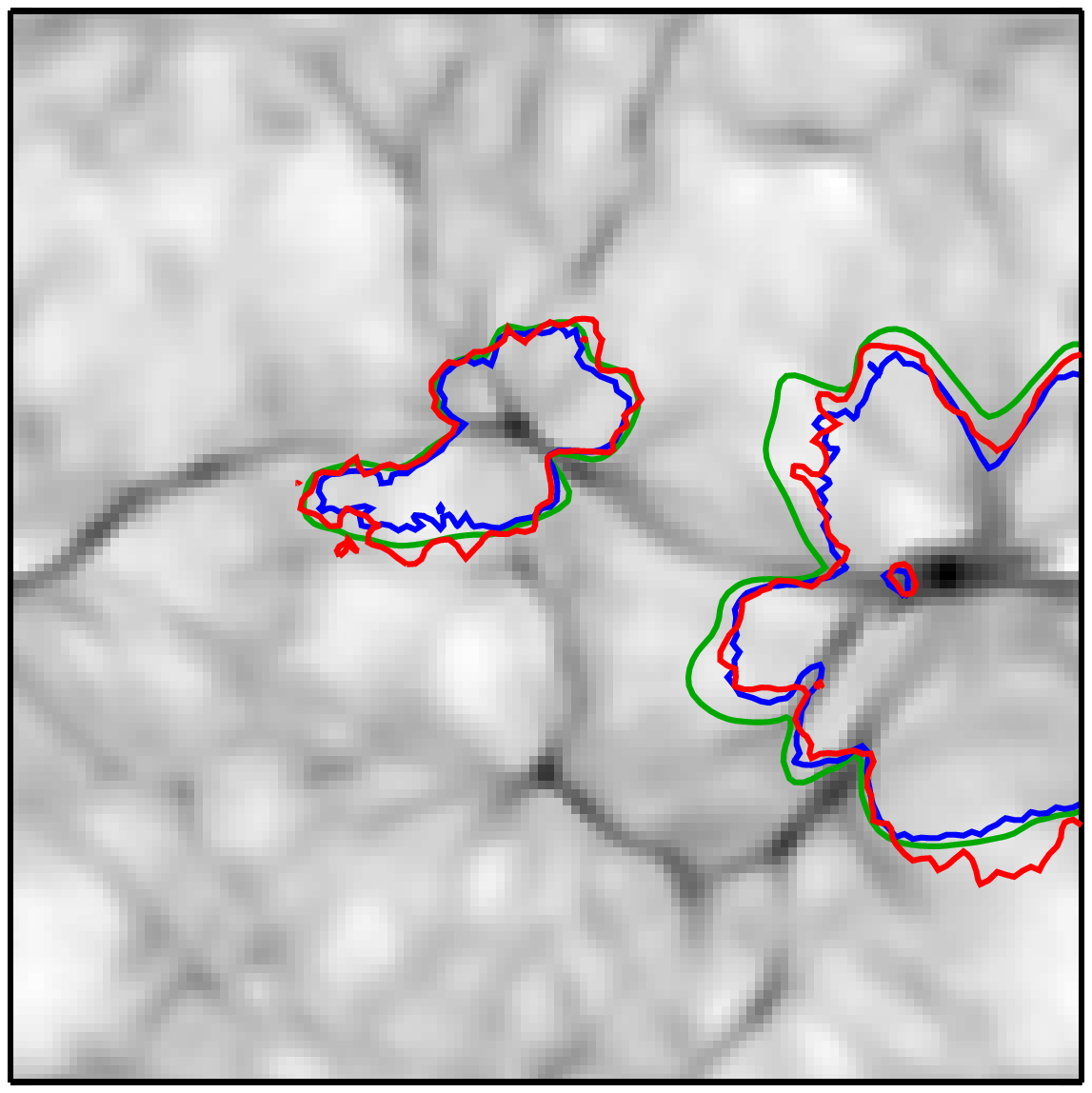}      
  \includegraphics[trim = 31mm 5mm 31mm 5mm, width=39mm, clip = true]{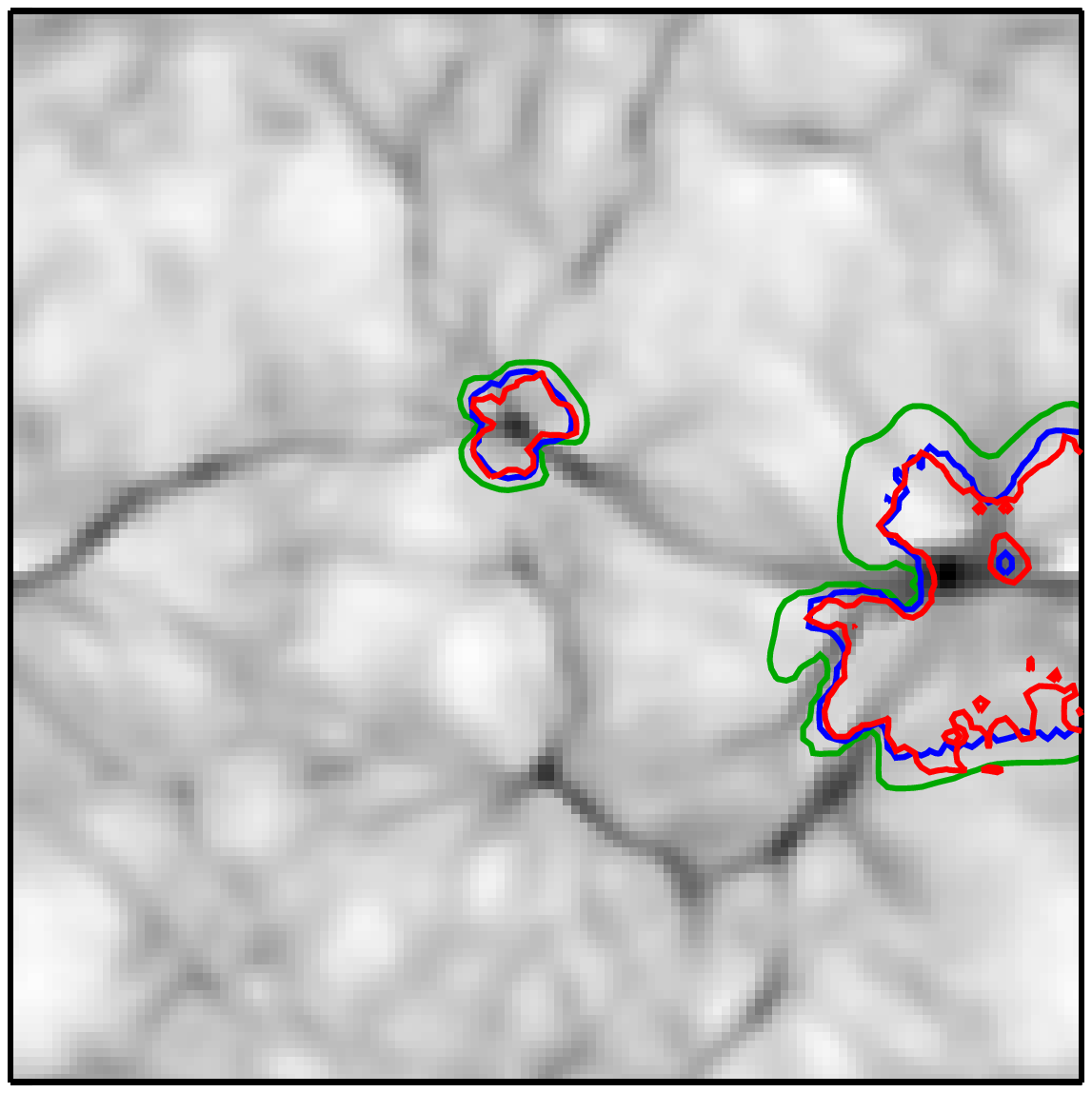} \\
  \includegraphics[trim = 31mm 5mm 31mm 5mm, width=39mm, clip = true]{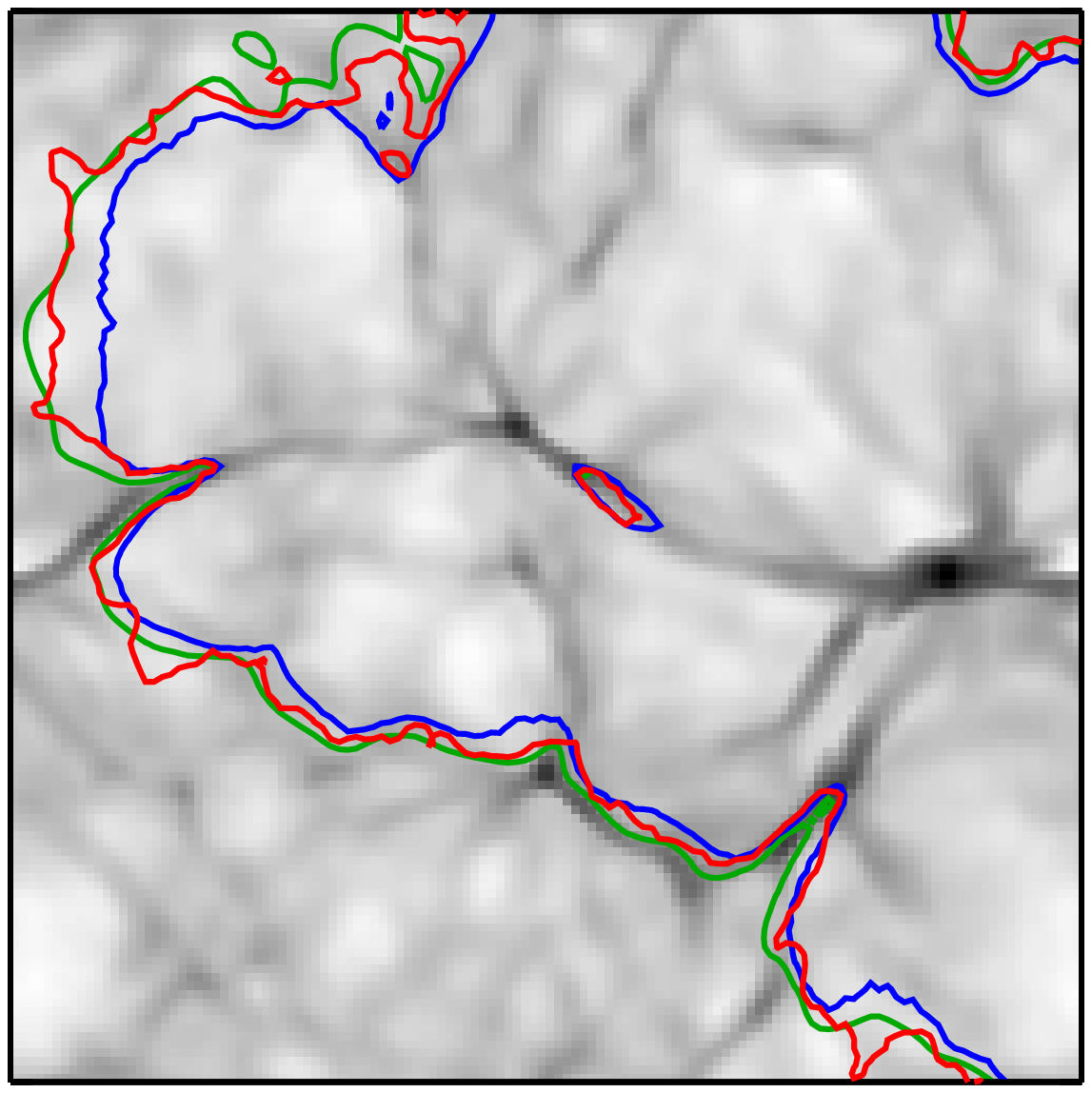} 
  \includegraphics[trim = 31mm 5mm 31mm 5mm, width=39mm, clip = true]{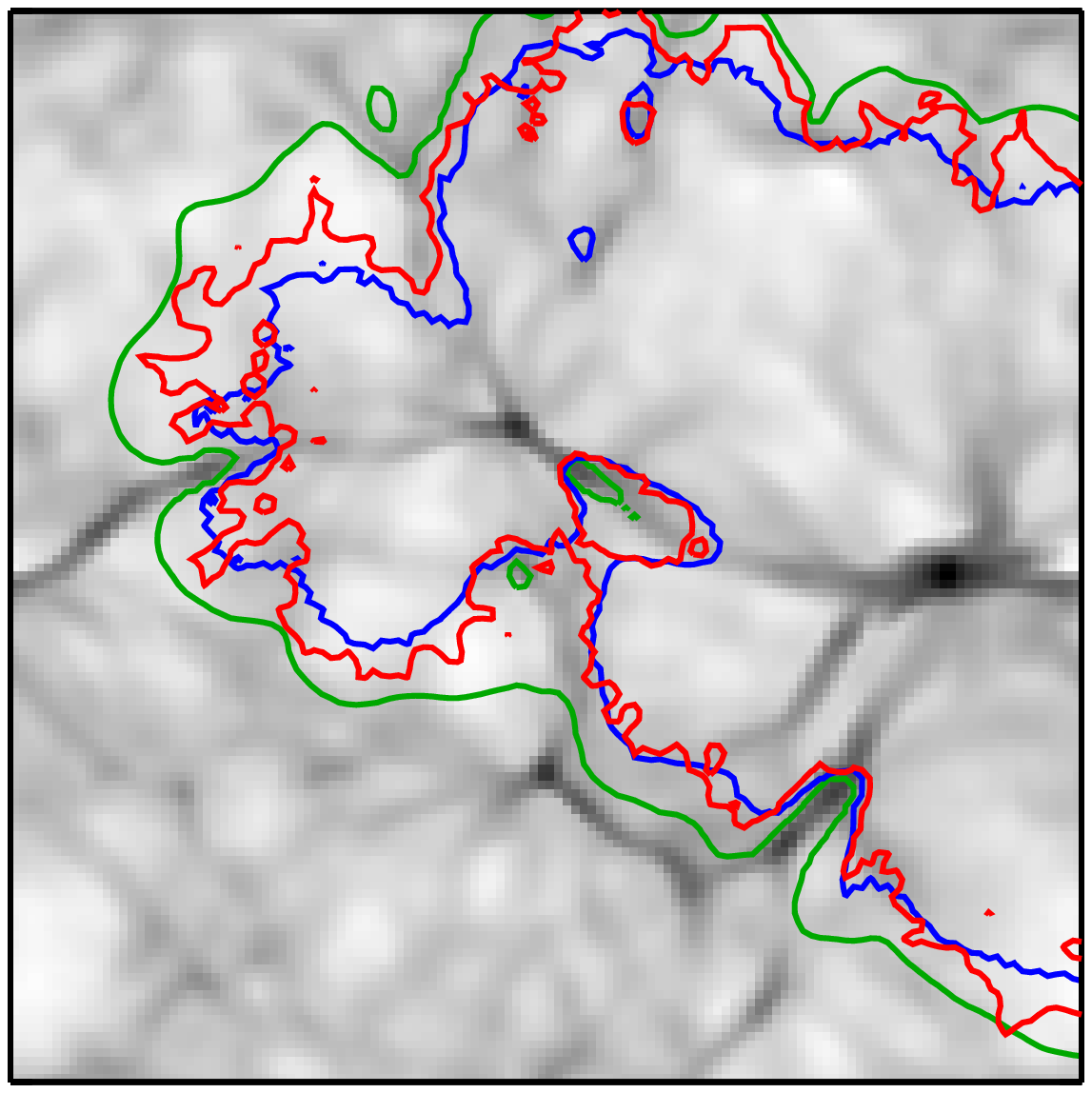} 
   
  \caption{Test 3: Slice through the density field at $z = L_{box}/2$.
    Contours show a neutral
    fraction of $\eta = 0.5$ (left-hand column) and $\eta = 0.01$ (right-hand column) at times $t =
    0.05 \Myr$ (top row) and $t = 0.2 \Myr$ (bottom row). Red contours show the results
    of our fiducial $(N_c = 32,  \tilde{N}_{ngb} = 32)$ simulation. For comparison, 
    we show the results of \ctworay\ (green) and \crash\ (blue), 
    as reported in  \protect\cite{Iliev:2006b}. The agreement is excellent.} 
    \label{Test3:Fig:Contours} 
  
\end{figure} 

\begin{figure}

  \includegraphics[trim = 31mm 5mm 31mm 5mm, width=39mm, clip = true]{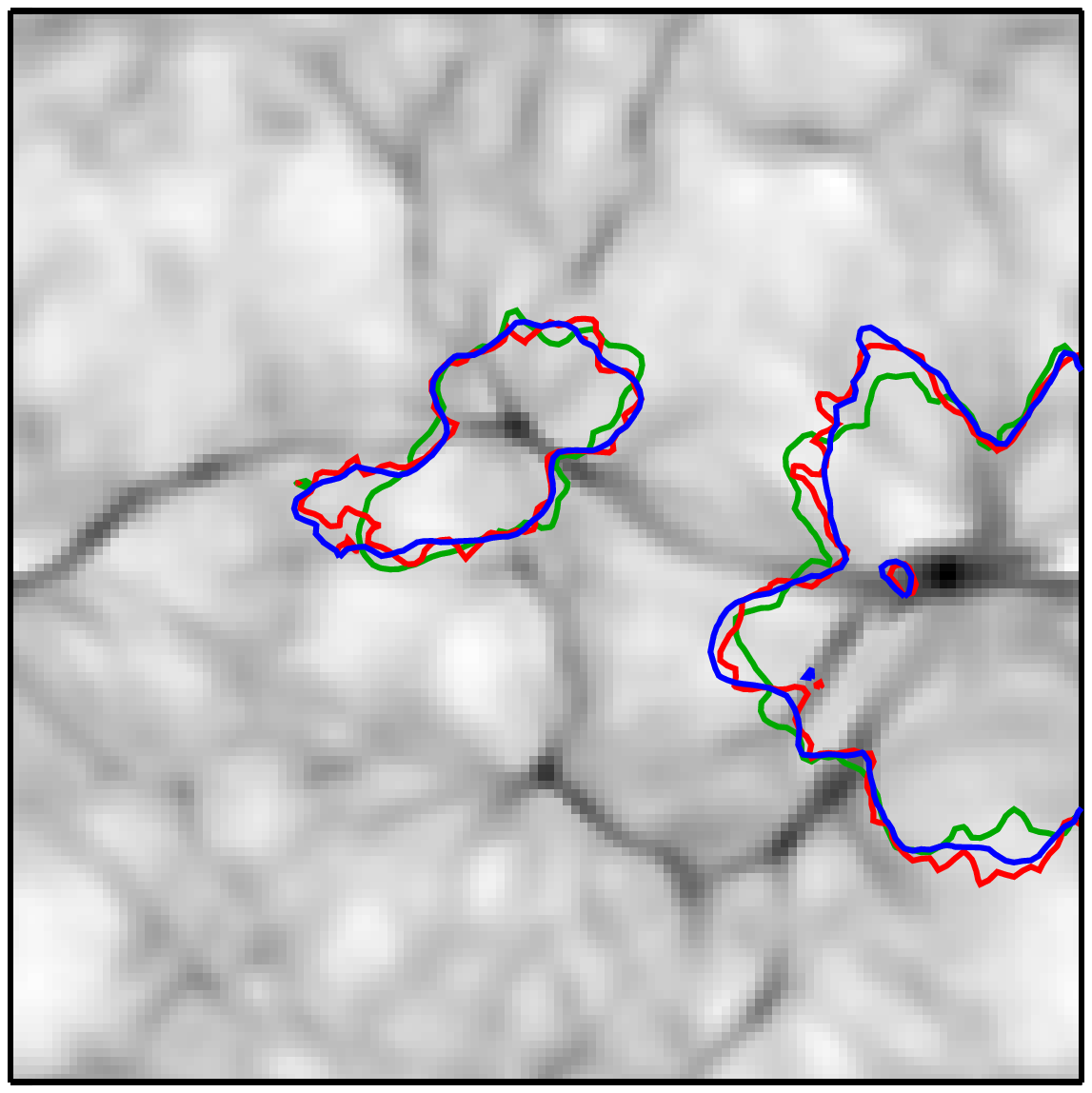}      
  \includegraphics[trim = 31mm 5mm 31mm 5mm, width=39mm, clip = true]{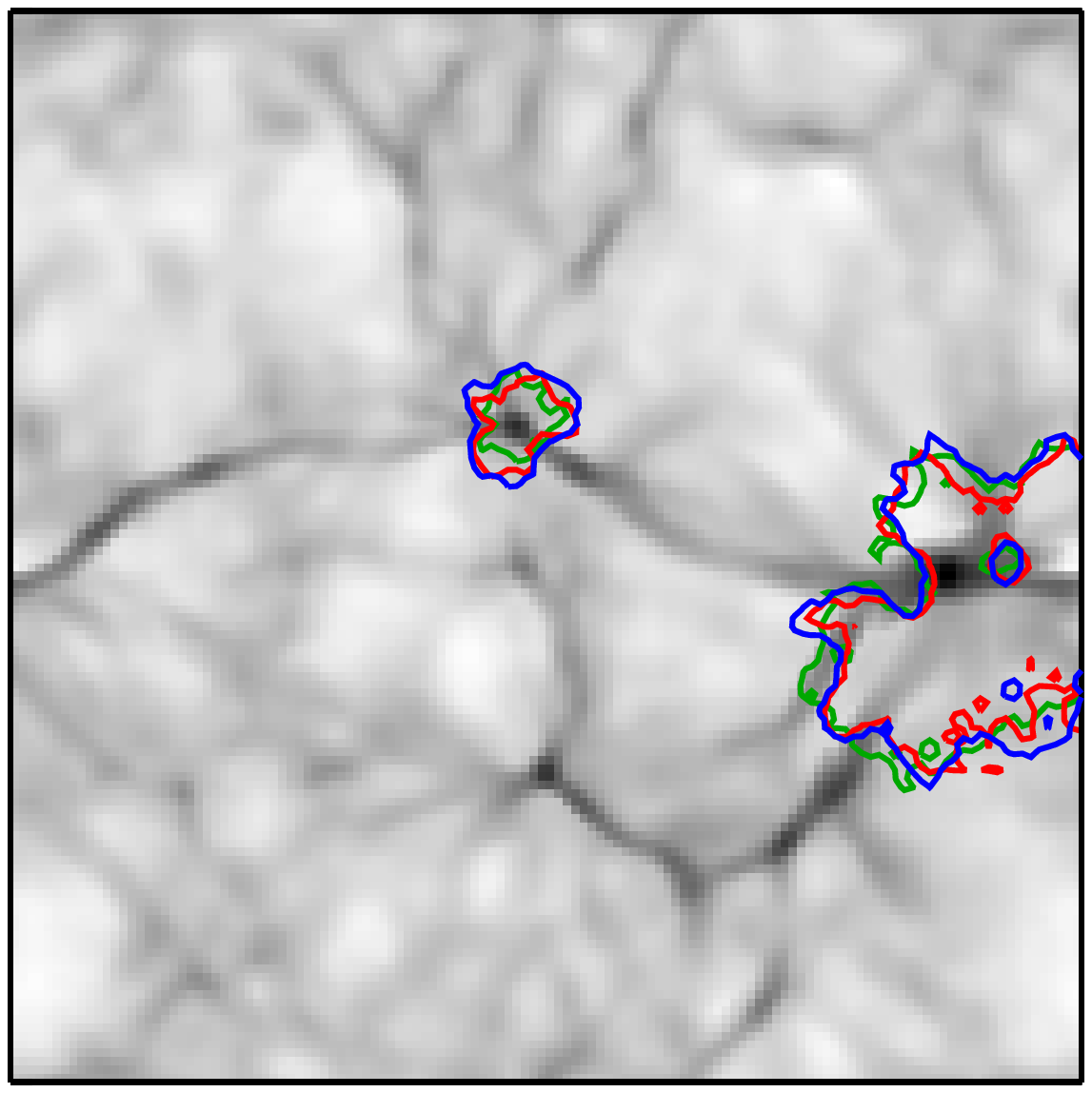} \\
  \includegraphics[trim = 31mm 5mm 31mm 5mm, width=39mm, clip = true]{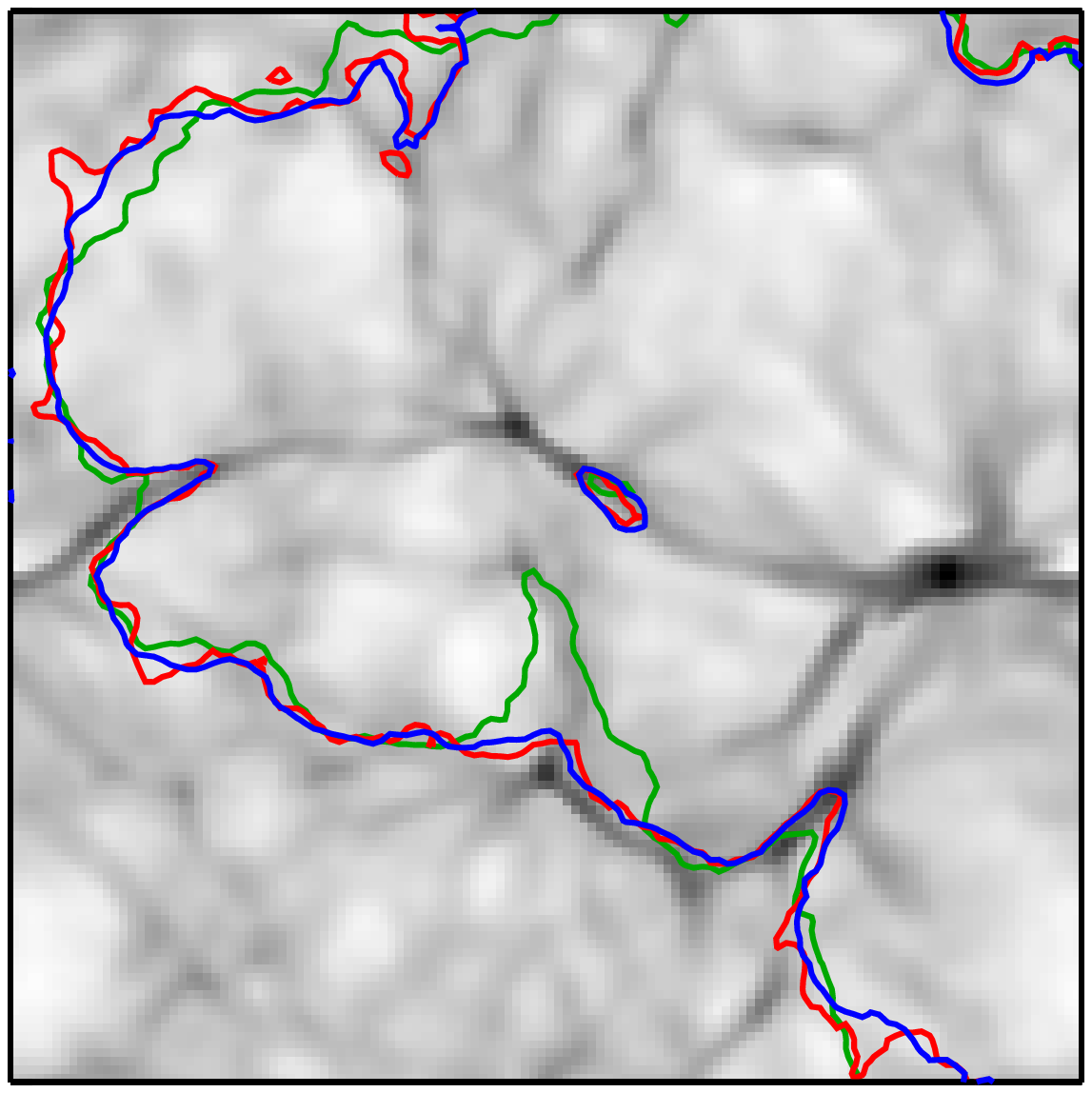} 
  \includegraphics[trim = 31mm 5mm 31mm 5mm, width=39mm, clip = true]{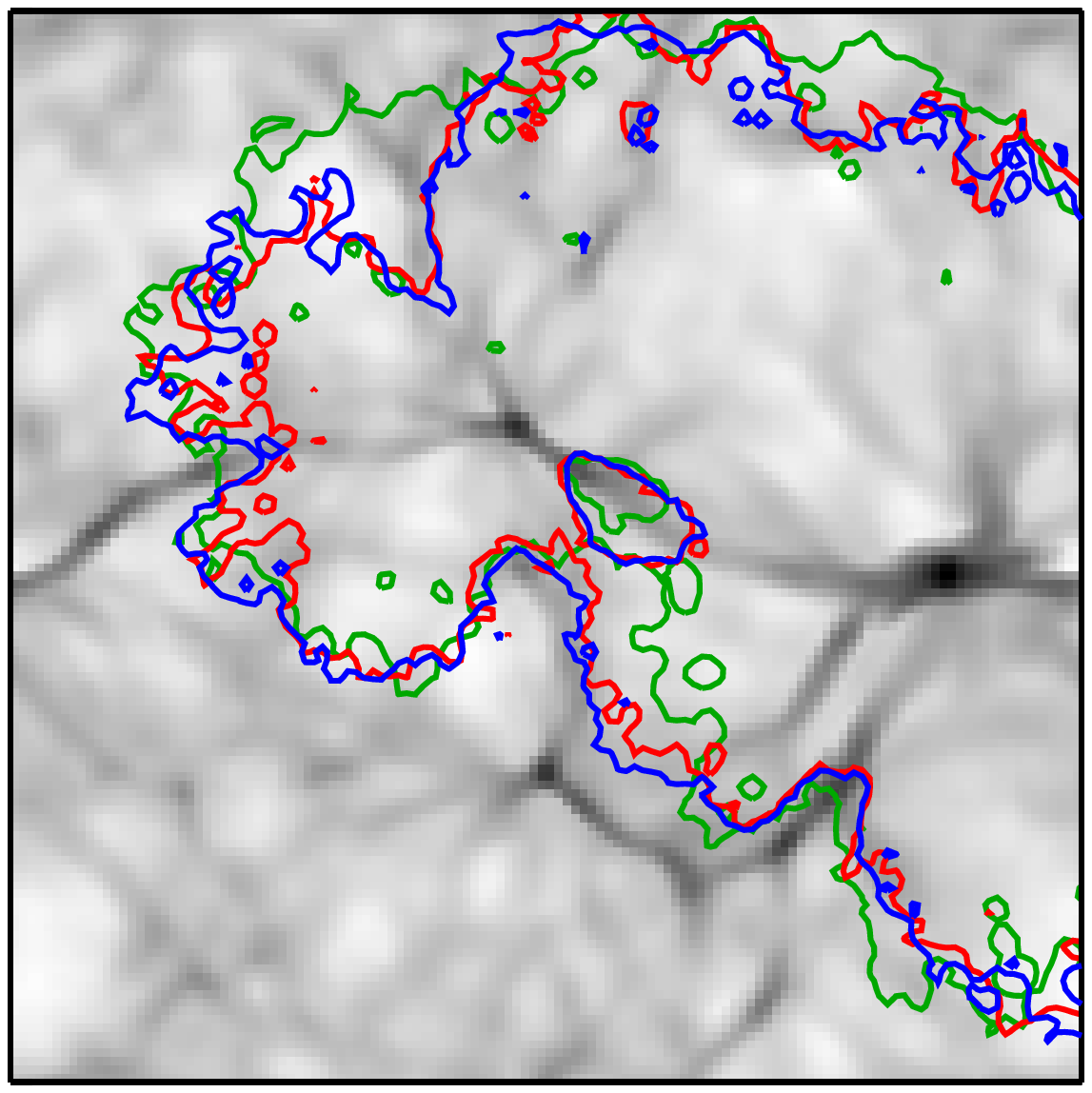}

  \caption{Test 3: Effect of angular resolution.
    The same slice as shown in Fig.~\ref{Test3:Fig:Contours}. Contours show a neutral
    fraction of $\eta = 0.5$ (left-hand column) and $\eta = 0.01$ (right-hand column) at time $t =
    0.05 \Myr$ (top row) and $t = 0.2 \Myr$ (bottom row). Green, red and blue lines correspond to the
    low ($N_c = 8$), fiducial ($N_c = 32$) and high ($N_c = 128$) angular
    resolution simulations, resp. The numbers of ViPs per SPH particle at the end of the
    simulations are $\approx 0.05, 1$ and $2.8$ for the low, fiducial and high
    angular resolution simulations, resp. Note that the fiducial simulation
    is already converged, even though its angular resolution $N_c = 32$
    corresponds to a relatively large cone opening angle of $\omega \approx
    41$ degrees.}
   \label{Test3:Fig:Contours:Resolution} 
  
\end{figure} 

\begin{figure}

  \includegraphics[trim = 31mm 5mm 31mm 5mm, width=39mm, clip = true]{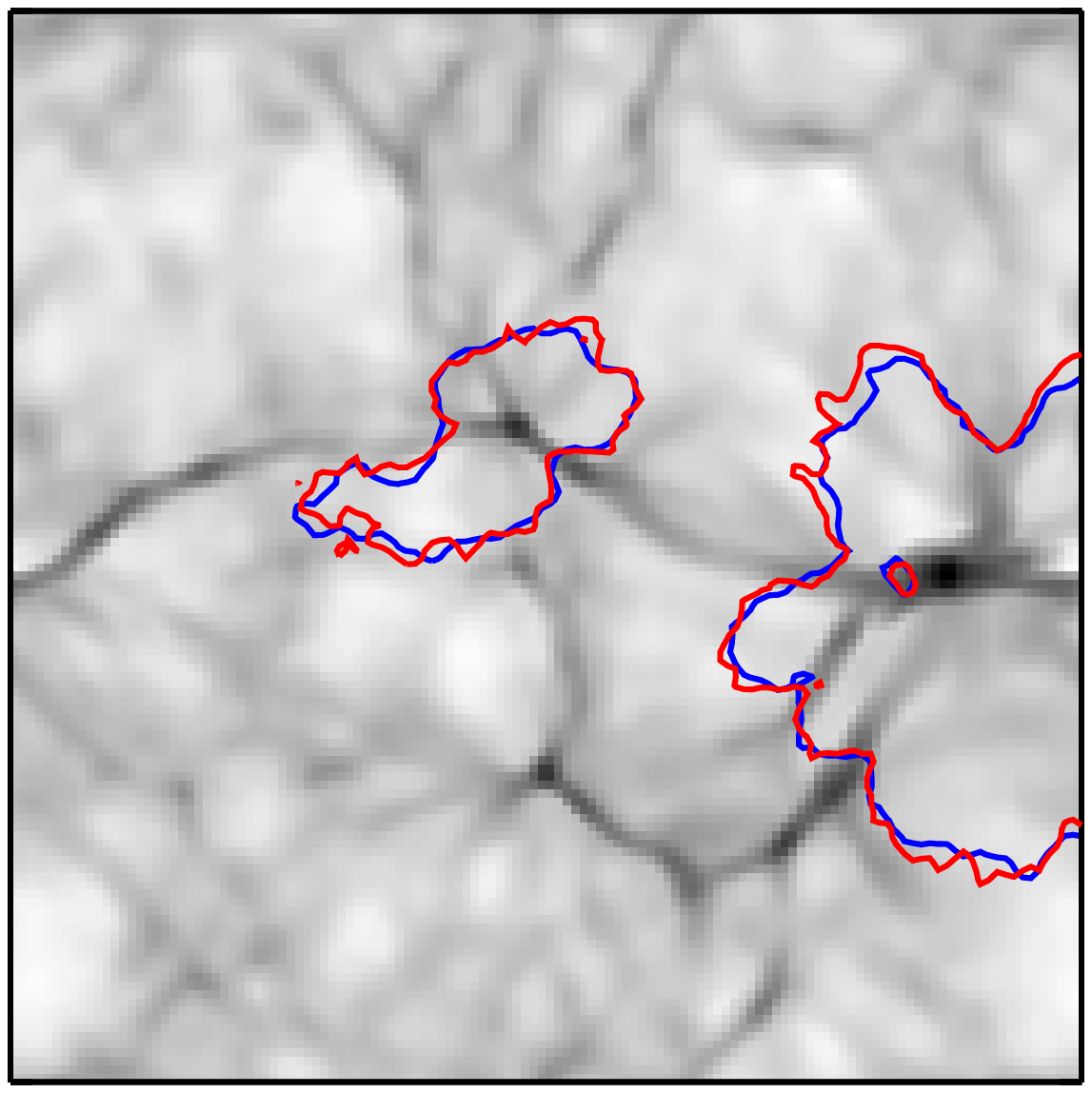}      
  \includegraphics[trim = 31mm 5mm 31mm 5mm, width=39mm, clip = true]{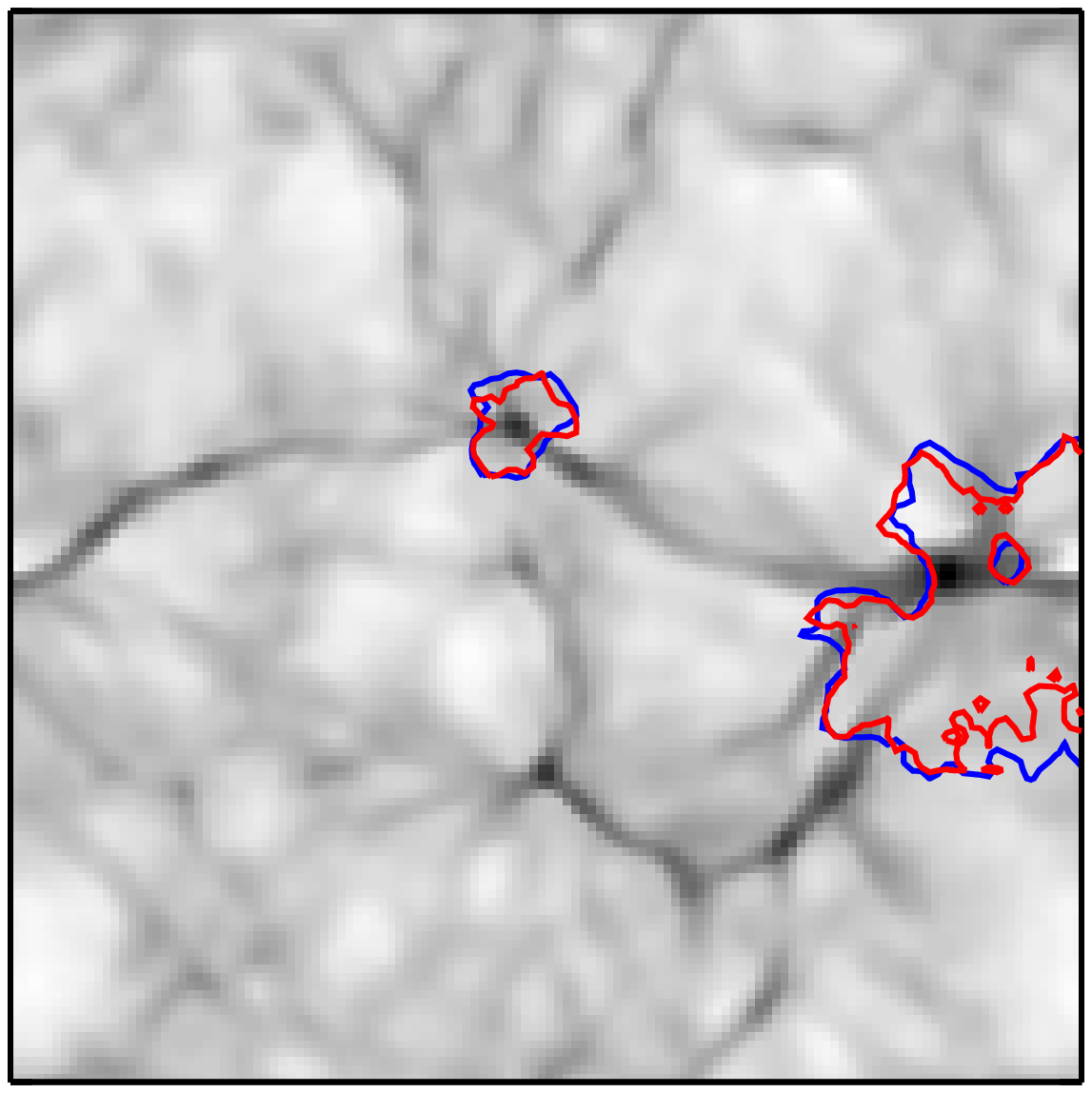} \\
  \includegraphics[trim = 31mm 5mm 31mm 5mm, width=39mm, clip = true]{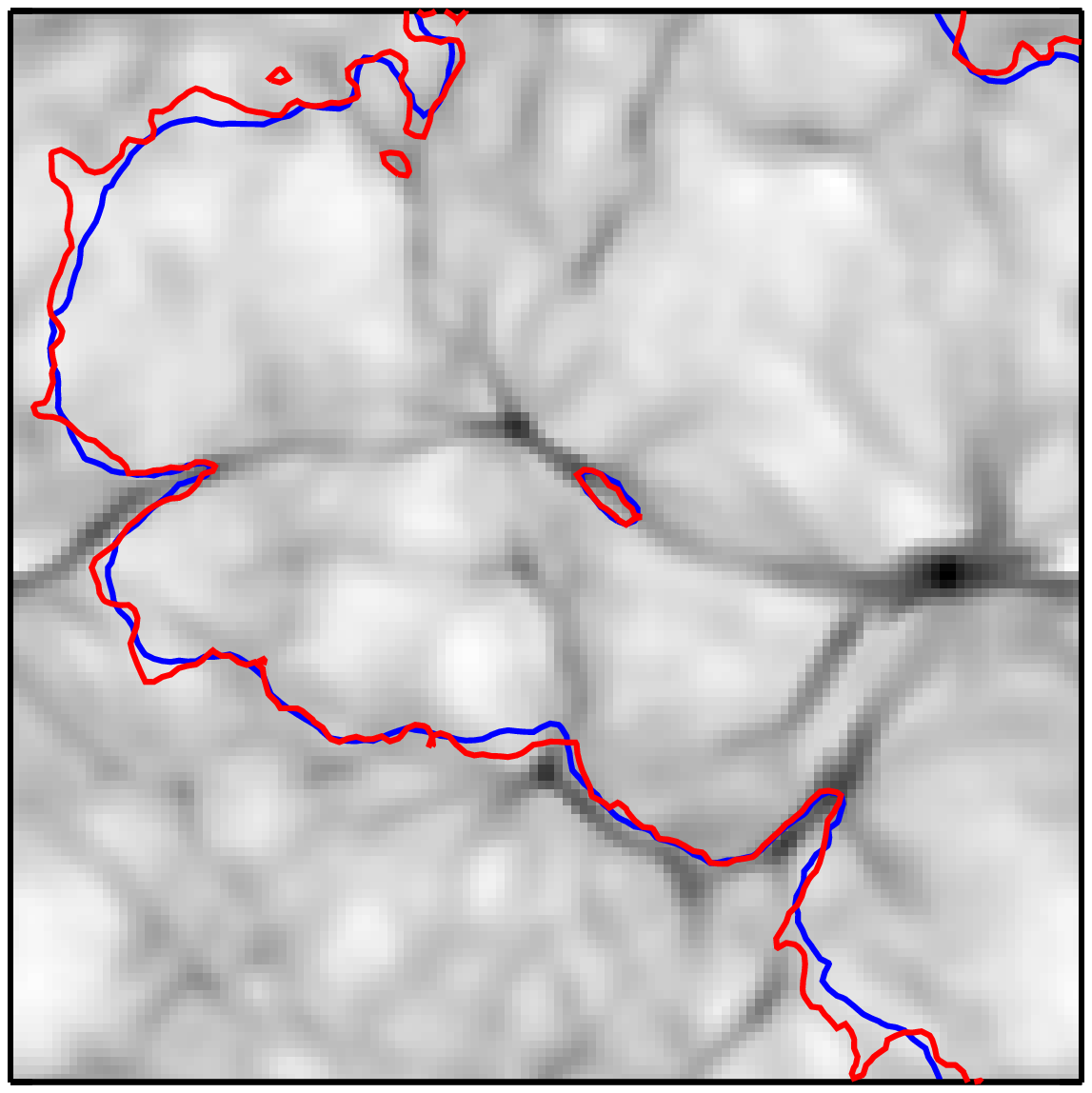} 
  \includegraphics[trim = 31mm 5mm 31mm 5mm, width=39mm, clip = true]{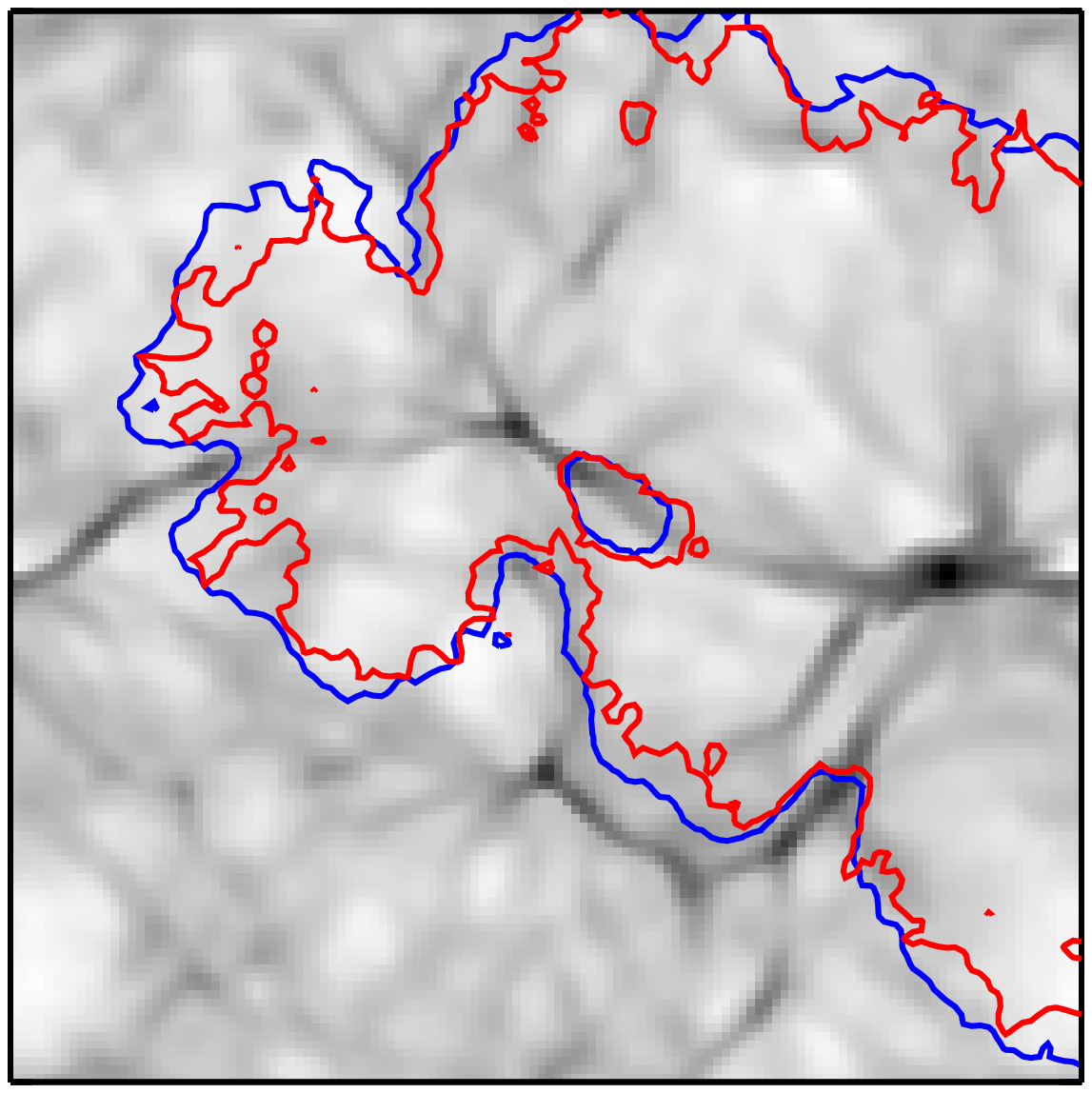}

  \caption{Test 3: Effect of resampling. The same slice as shown in Fig.~\ref{Test3:Fig:Contours}. Contours show a neutral
    fraction of $\eta = 0.5$ (left-hand column) and $\eta = 0.01$ (right-hand column) at time $t =
    0.05 \Myr$ (top row) and $t = 0.2 \Myr$ (bottom row). The red contours correspond to the
    fiducial angular resolution simulation and are identical to the red contours
    in Figs.~\ref{Test3:Fig:Contours} and
    \ref{Test3:Fig:Contours:Resolution}. The blue contours correspond to a
    simulation identical to the simulation employing the fiducial angular
    resolution, except for the fact that in this simulation we
    periodically (every 10th radiative transfer time step) resampled the
    density field to suppress the particle noise. Note that resampling does
    not visibly decrease the effective angular resolution.}
   \label{Test3:Fig:Contours:Resampling} 
  
\end{figure}

We performed a set of three radiative transfer simulations with angular
resolutions $N_c = 8, 32$ and $128$, which we refer to as the low
angular resolution, fiducial and high angular resolution simulations, resp.
Every simulation used $\tilde{N}_{ngb} = 32$ neighbours. The time step was
set to $\Delta t_r = 100 \yr.$
\par
In Fig.~\ref{Test3:Fig:Contours} we show in red the 
neutral fraction contours $\eta = 0.5$ (left-hand column) and $\eta = 0.01$ (right-hand column) 
at times $t = 0.05 \Myr$ (top row) and $t = 0.2
\Myr$ (bottom row) for the fiducial angular resolution. The ionization front (neutral fraction $\eta = 0.5$) is
delayed by the dense filaments, leading to the characteristic "butterfly" shapes of the ionized
regions. For comparison, we also show the results obtained
with two other codes, the ray-tracing scheme \ctworay\
(\citealp{Mellema:2006}; green contours) and
the Monte Carlo code \crash\ (\citealp{Maselli:2003}; \citealp{Ciardi:2001}; blue contours), as published in the cosmological radiative transfer
code comparison project (\citealp{Iliev:2006b})\footnote{
The performance of  two more codes was reported in \cite{Iliev:2006b}: \ftte\ (\citealp{Razoumov:2005}) and
\simplex\ (\citealp{Ritzerveld:2006}). For clarity and since they are very similar to the
results obtained with \ctworay\ and \crash, we do not show the results obtained with
\ftte. We do not include the results of \simplex\
in our comparison, since they differ considerably from those obtained
with all  other codes.}. Both \ctworay\ and \crash\
are mesh codes, working directly on the uniform mesh input density field
provided by the PM+TVD code of \cite{Ryu:1993}. 
\par
The agreement between the results of \traphic\ and \ctworay\ resp. \crash\ 
is very good, not only for the
ionization front, but also for smaller neutral fractions. 
The contours from \traphic\ are slightly noisier than those from \ctworay,   
which is expected since in addition to the particle noise affecting the
radiative transfer, the Monte-Carlo sampling noise imprinted on the
density field affects our simulations, particularly
in the under-sampled low density regions, as already noted earlier. The noise
level is, however, substantially lower than one would anticipate based on the
tests presented in
Sections~\ref{Section:Test1} and \ref{Section:Test2}. The most likely
explanation for this welcome surprise is that the 
presence of multiple ionizing sources leads to a regularization in the distribution of the neutral
fraction. Numerical noise arising from the representation of the
continuous density field by a discrete set of
particles is therefore reduced. Differences between our results
and those of \ctworay\ resp. \crash\ also arise through the different
treatment of the photon spectrum. Since 
the photo-ionization cross-section depends on frequency (Eq.~\ref{Photoionzation:Cross-section}),
the thickness of finite ionization fronts (e.g. defined as $0.9 < \eta <
0.1$) and hence the position of the particular contour $\eta = 0.5$ will in
part be determined by the details of the numerical
implementation of the multi-frequency transport.  
\par
In Fig.~\ref{Test3:Fig:Contours:Resolution} we show the neutral fraction
contours $\eta=0.5$ (left-hand side) and $\eta = 0.01$  (right-hand side) at times $t
= 0.05 \Myr$ (top row) and $t = 0.2 \Myr$  (bottom row) for the low (green contours) and the high
(blue contours) angular resolution simulations. For comparison, the contours
obtained in the fiducial simulation are also shown (red). 
We note that the high angular resolution simulation yields 
neutral fraction contours that are almost identical to those obtained in our fiducial
simulation, indicating numerical convergence. The
low angular resolution simulation, although still in good
agreement with the high angular resolution simulation, fails to
properly reproduce the expected neutral fraction contours when
scrutinised in detail. In the low
angular resolution simulation, neutral fraction contours are often slightly advanced
instead of delayed by the dense filaments. Although the effect is small,
it becomes apparent when the contours are compared to the corresponding contours of the
high angular resolution simulation. 
\par
The last observation agrees
with the discussion of anisotropies in particle-to-neighbour transport
schemes sketched in Section~\ref{Section:EmissionBySourceParticles} and presented in detail 
in Appendix~\ref{Appendix:A}. In Appendix~\ref{Appendix:A} we demonstrate that when photons are
transported from a given particle to its neighbours, the net transport
direction is generally strongly correlated with the direction towards the
centre of mass of the neighbouring particles. As a result, the transport
is partly governed by the spatial distribution of the SPH particles. For cosmological
simulations this implies that photons are preferentially transported along
dense filaments. \traphic\ propagates photons in cones to overcome this
problem. The cones confine the photons to the
solid angles they were emitted into, ensuring a correct transport of radiation 
on the scale of the chosen angular resolution. If the angular resolution is
chosen too low to properly resolve the structures in the SPH density
field, the transport is no longer independent of the geometry of the SPH
simulation and artefacts may occur, as we see in
Fig.~\ref{Test3:Fig:Contours:Resolution} for the low angular resolution
simulation. Even with an angular resolution as low as $N_c = 8$, however, the artefacts
are small. Nevertheless, it is clear that in order to properly solve the radiative transfer equation,
the angular resolution must be chosen high enough to establish 
numerical convergence. Fig.~\ref{Test3:Fig:Contours:Resolution} shows that an
angular resolution of $N_c =32$, which corresponds to a relatively large
opening angle of $\omega \approx 41$ degrees (Eq.~\ref{Eq:Apex}),
is already converged. The reason why a relatively poor formal angular resolution 
suffices, is, as already noted in the discussion of the sharpness of shadows
thrown by opaque obstacles in Section~\ref{Section:Test2}, that the photon transport with \traphic\ is intrinsically adaptive in angle.
\par
In Fig.~\ref{Test3:Fig:Contours:Resampling} we present results for a
simulation that used the resampling technique presented in Section~\ref{Sec:Resampling:Implementation}, but
which was otherwise identical to the fiducial simulation
presented above. The neutral hydrogen densities of the SPH particles
were not re-calculated according to the perturbed positions resulting from the
resampling, but kept constant to avoid additional scatter in the density. 
The resampling leads to a reduction in the particle noise, which is most
apparent for the $\eta = 0.01$ contours, since in case of the $\eta = 0.5$ the
noise is already low, as noted above. Note that the resampling does not noticeable
degrade the angular resolution.
\par
\begin{figure} 
\includegraphics[width=84mm, clip=]{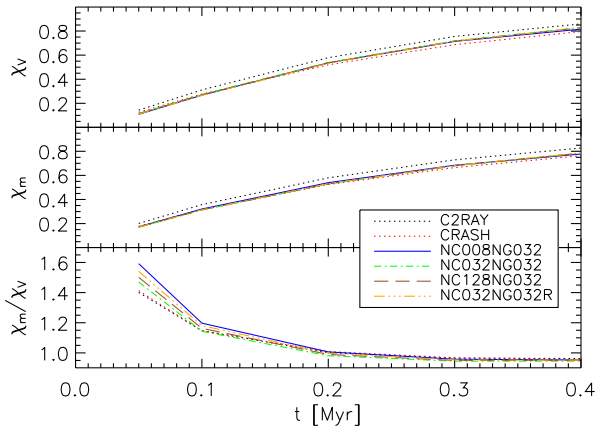} 
\caption{Test 3: The volume- and mass-weighted mean ionized fractions, $\chi_V$
  and $\chi_m$, resp., averaged over the whole simulation box as a function of time, 
  for the low, fiducial (without and with resampling of the density) and high angular resolution
  simulation, as indicated in the legend. All results fall nearly on top of each
  other. Differences in $\chi_m / \chi_V$ are only noticeable when $\chi_V$ is
  small. For comparison, we also show the results obtained with \ctworay\
  and \crash\, as reported in  \protect\cite{Iliev:2006b}. } 
  \label{Test3:Fig:Average} 
\end{figure} 
In Fig.~\ref{Test3:Fig:Average} we show the evolution of the mean (over
all particles $i$) ionized fraction, both volume-weighted, i.e. $\chi_v = \sum_i
h_i^3 \chi_i / \sum_i h_i^3$, and mass-weighted, i.e. 
$\chi_m = \sum_i m_i \chi_i / \sum_i m_i$. Again, the results obtained with \traphic\
are in excellent agreement with the results obtained with \ctworay\ and
\crash. For the latter two we obtained the mean ionized fraction by
averaging the ionized fraction reported in the cosmological radiative transfer
code comparison project (\citealp{Iliev:2006b}) over all grid cells $i$, 
i.e. $\chi_v = \sum_i  \chi_i / \sum_i 1$ and  $\chi_m = \sum_i \rho_i \chi_i
/ \sum_i \rho_i$. 
\par
The ratio of mass-weighted and volume-weighted mean
ionized fractions is at early times slightly larger for the low angular
resolution simulation than for the high angular
resolution simulation, as can be seen in the bottom panel of
Fig.~\ref{Test3:Fig:Average}. 
This is another manifestation of the fact that 
particle-to-neighbour transport is generally not independent of the geometry of
the SPH simulation, resulting in photons being preferentially transported into
high (particle) density regions. It once more underlines the importance of the
concept of emission and transmission cones (with sufficiently small solid angles) 
which \traphic\ uses to accomplish the transport of radiation 
independently of the spatial distribution of the SPH particles.
\par
\begin{figure*}
  \includegraphics[trim = 31mm 5mm 31mm 5mm, width=43mm, clip = true]{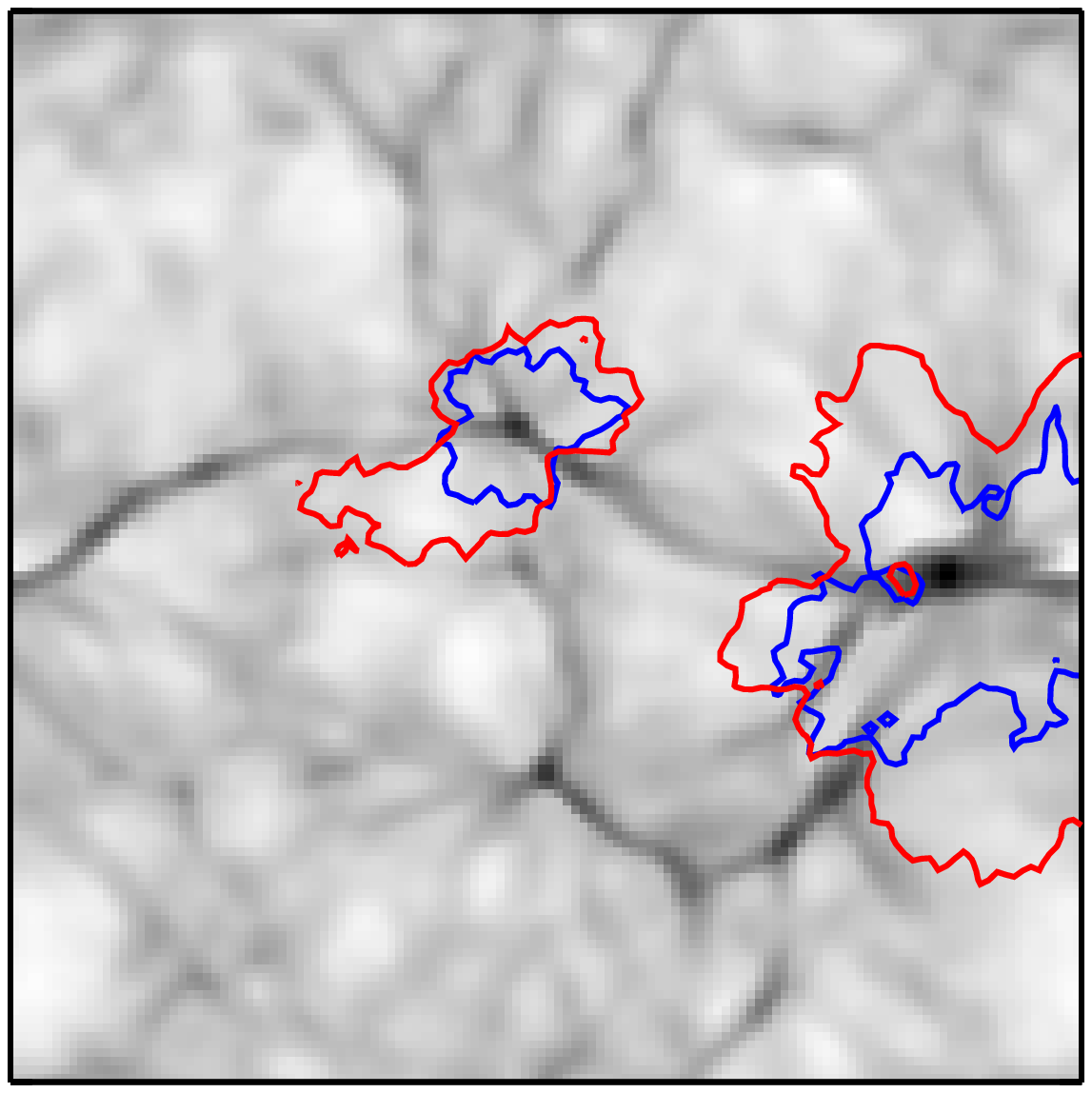}      
  \includegraphics[trim = 31mm 5mm 31mm 5mm, width=43mm, clip = true]{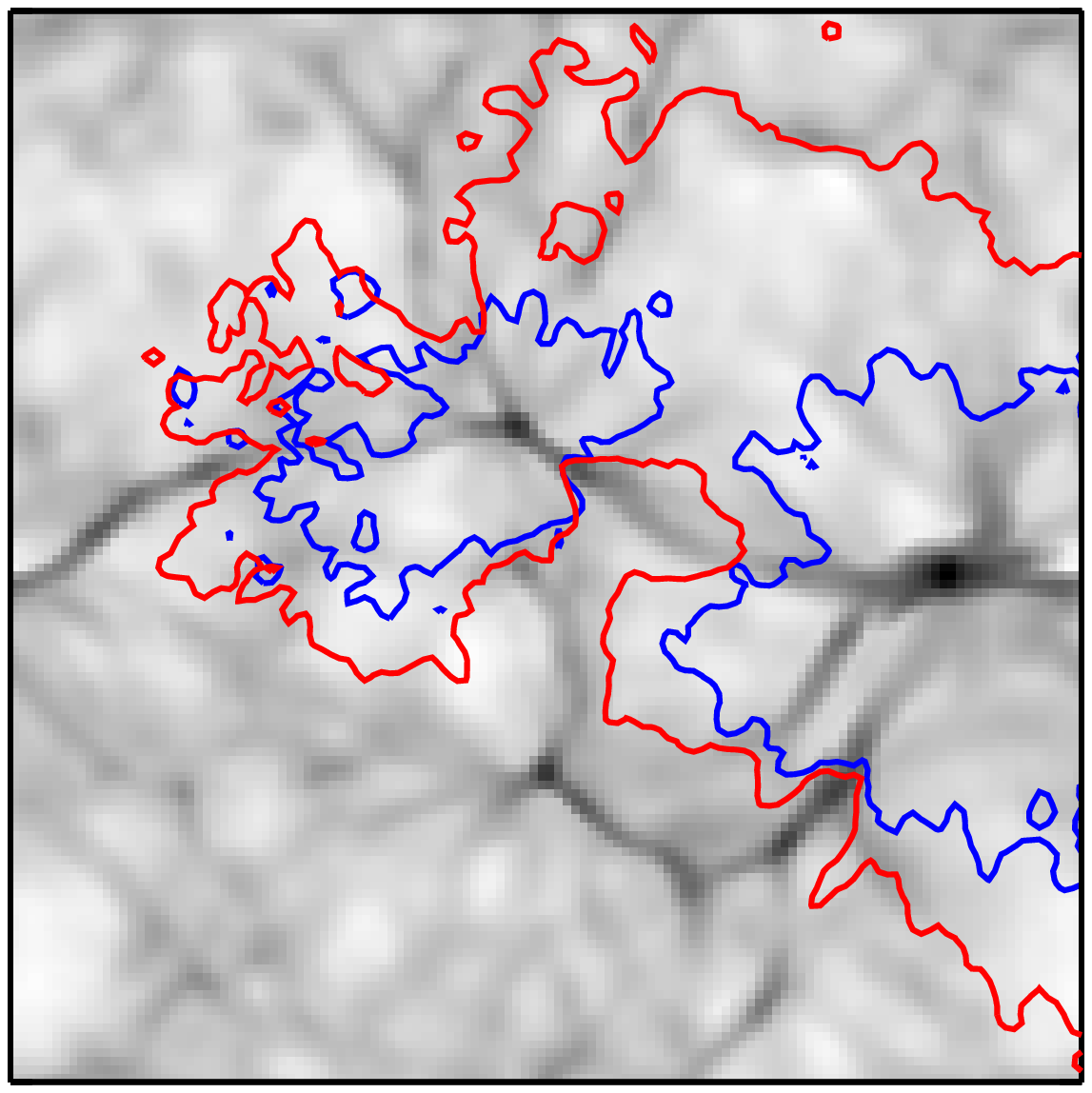} 
  \includegraphics[trim = 31mm 5mm 31mm 5mm, width=43mm, clip = true]{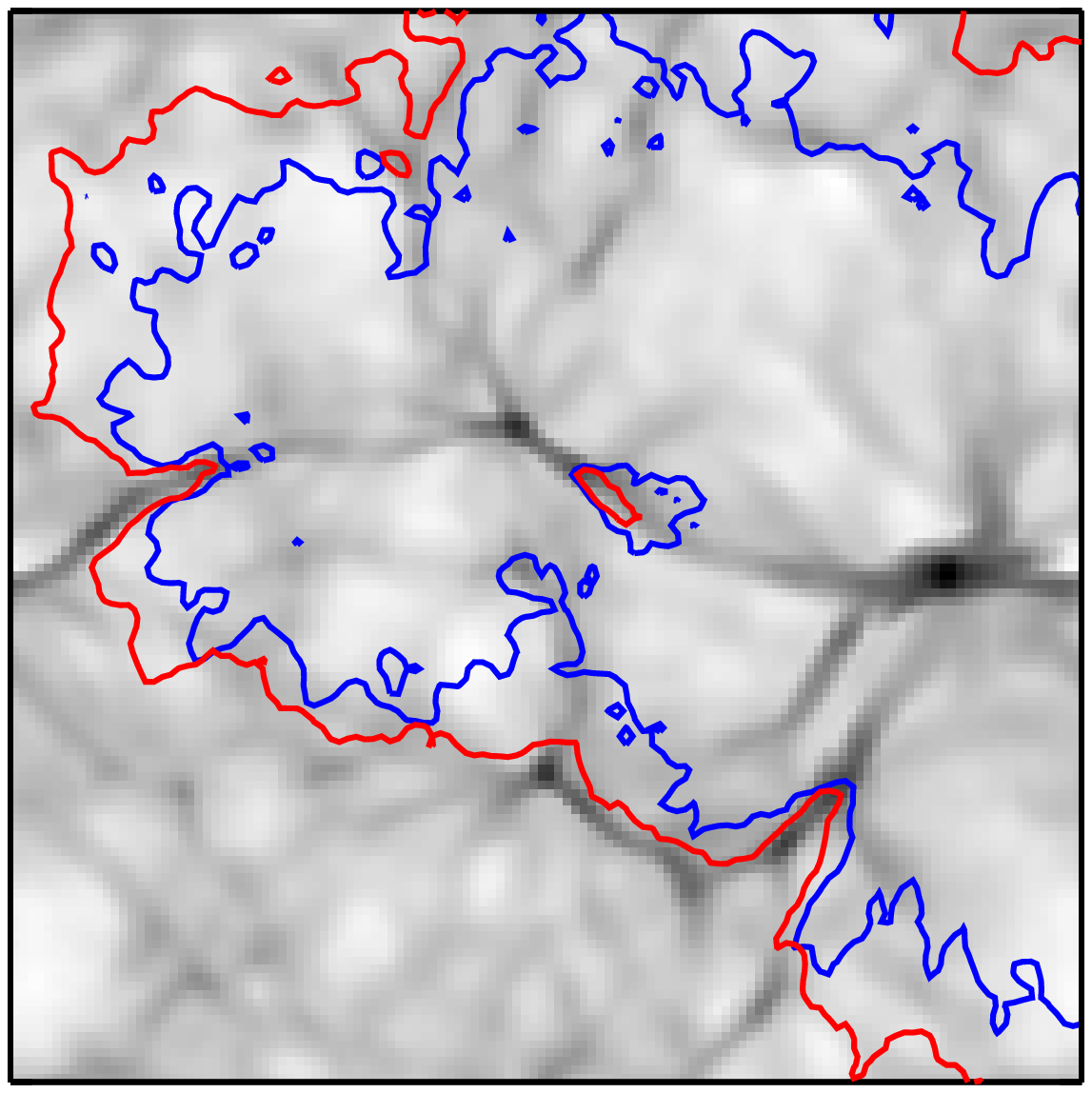} 
  \includegraphics[trim = 31mm 5mm 31mm 5mm, width=43mm, clip = true]{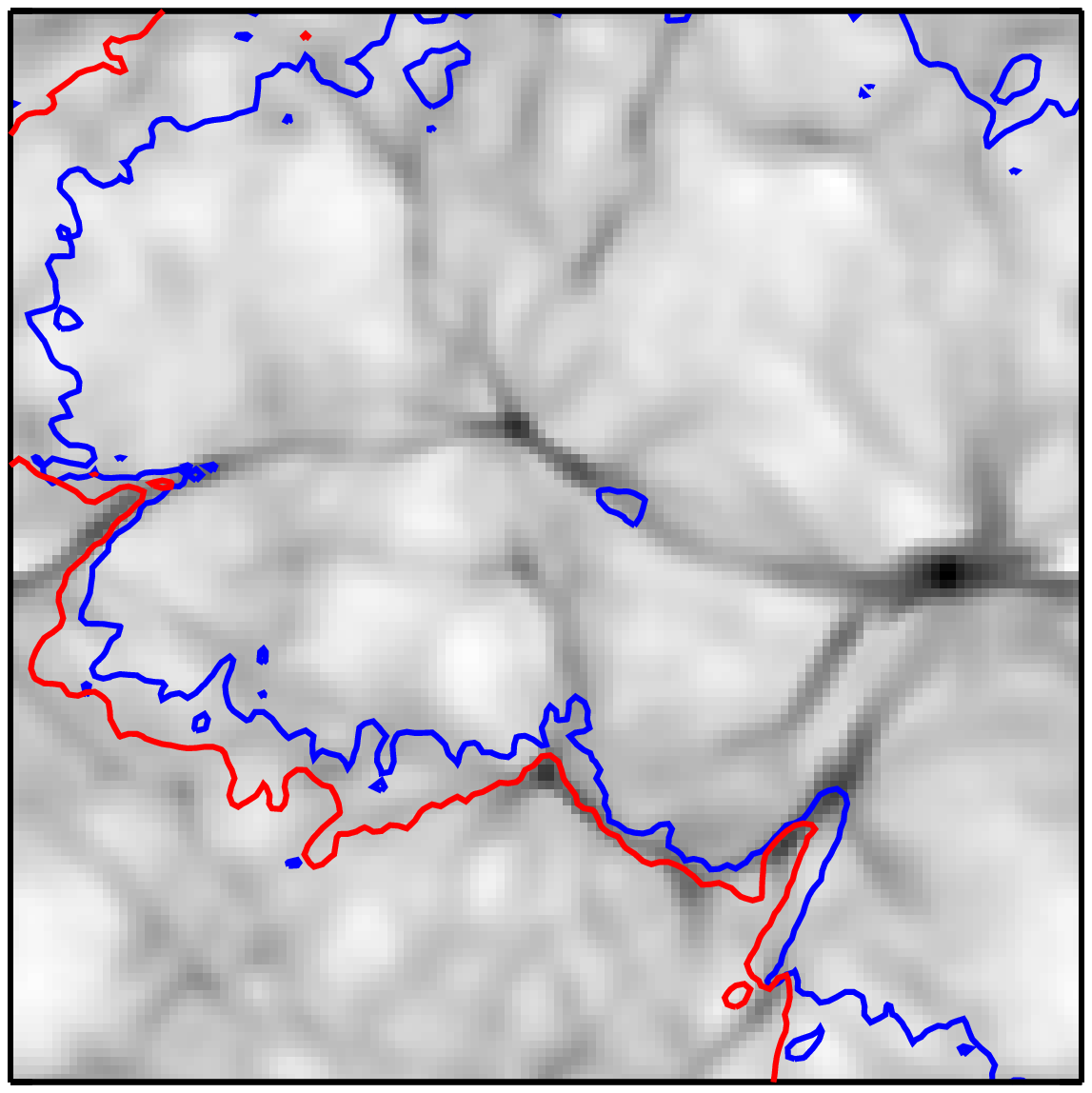} 

  \caption{Test 3: Slice through the density field at $z = L_{box}/2$.
    Contours show a neutral
    fraction of $\eta = 0.5$ at times $t =
    0.05, 0.1, 0.2$ and $0.3 \Myr$ (from left to right). 
    Red contours show the results of our fiducial $(N_c = 32,  \tilde{N}_{ngb}
    = 32)$ simulation (at times $0.05$ and $0.2 \Myr$ these are identical to the 
    red contours shown in the top-left resp. bottom-left panel of Fig.~\ref{Test3:Fig:Contours}). 
    Blue contours show the result of a simulation employing the same
    resolution $(N_c = 32,  \tilde{N}_{ngb} = 32)$, but using the clock of the
    photon packets to solve the time-dependent radiative transfer
    equation. Note that in the simulation solving the time-independent radiative transfer equation the
    ionized regions are too large, since in this simulation the ionization fronts are initially
    propagating at speeds larger than the speed of light.}
    \label{Test3:Fig:Clock} 
\end{figure*} 

\par
Finally, we show that for the present radiative transfer 
problem simulations solving the time-dependent radiative transfer equation 
give a significantly different result than the simulations solving the time-independent radiative transfer equation
discussed above. We carried out a simulation using $N_c = 32,  \tilde{N}_{ngb} = 32$
and additionally employed the photon packet clocks to limit the distance over which 
photon packets can propagate during each time step. 
The size of the radiative transfer time step was set to $\Delta t_r = 10^{-3}\Myr$. 
This time step is a factor 10 times larger than the time step used for solving the time-independent radiative transfer equation 
in the simulations presented above, which required a smaller time step because of our particular treatment of the time-independent 
radiative transfer equation (see Section~\ref{Sec:Timestep}). 
The locations of the ionization fronts (i.e. $\eta = 0.5$) obtained in this simulation are shown in 
Fig.~\ref{Test3:Fig:Clock} (blue curves), together with those obtained in the corresponding simulation 
solving the time-independent radiative transfer equation (red curves, cp. the
left-hand panels of Fig.~\ref{Test3:Fig:Contours}), 
at four different simulation times $t = 0.05, 0.1, 0.2$ and $0.3 \Myr$.
\par
It is clear from Fig.~\ref{Test3:Fig:Clock} that the simulation solving the time-independent 
radiative transfer equation produces ionized spheres
that are unphysically large. This is due to the well-known fact (see, e.g., the discussion in \citealp{Abel:1999})
that ionization fronts may propagate at speeds larger than the speed of 
light, if the time-independent radiative transfer equation is solved. The difference between the two 
simulations is larger at early times than at late times, which is expected, 
since in equilibrium, i.e. $t_r \to \infty$, the results of both simulations
must agree.
\par
In summary, in this section we studied the propagation of ionization fronts around multiple sources
in a static cosmological density field. We demonstrated the importance of the concept
of cones which underlies the photon transport in \traphic. Without the confinement by transmission cones of
sufficiently small solid angle, particle-to-neighbour transport is governed in
part by the spatial distribution of the particles, resulting in the preferential transport
of photons into high (particle) density regions. Thanks to the fact that
\traphic\ is adaptive in angle, a relatively modest formal angular resolution
of $N_c = 32$ is already sufficient to obtain a converged solution. Since the setup of our simulations followed
the description of the corresponding test in the cosmological radiative
transfer code comparison project (\citealp{Iliev:2006b}), we were able to benchmark our radiative transfer
scheme by comparing with the results obtained by the ray-tracing scheme \ctworay\ (\citealp{Mellema:2006}) and
the Monte Carlo code \crash\ (\citealp{Maselli:2003}; \citealp{Ciardi:2001}). We found excellent 
agreement in the positions of neutral fraction contours as well as the mass and volume-weighted mean ionized fractions. 
\par
We have furthermore seen that for the test problem presented in this section, simulations solving the time-independent 
radiative transfer equation lead to ionized regions that are unphysically
large during their early evolution. This illustrates the importance of correctly
accounting for the finite speed of light when performing radiative transfer
simulations to study the morphology of ionized regions that are strongly out of equilibrium.

\section{Conclusions} 
In this paper we have presented \traphic\ (TRAnsport of PHotons In Cones), a novel radiative transfer scheme
for SPH simulations. \traphic\ has been designed 
for use in radiation-hydrodynamical simulations exhibiting a wide dynamic range
of physical length scales and containing a large number of light sources, as for instance
encountered in cosmological simulations.
The requirements for this are high: Due to the limits on the 
computational power available today, radiative transfer simulations 
at the resolution of state-of-the-art hydrodynamical simulations need to be 
adaptive both in space and in angle and parallel on distributed memory machines. 
\par
\traphic\ meets these 
requirements. The radiative transfer equation is solved by transporting photons in an explicitly
photon-conserving manner directly on the discrete set of SPH particles.
Photons are transported globally by propagating them locally, between a particle and its neighbours. 
For a fixed number of SPH particles, the number of neighbours employed for the transport, the parameter
$\tilde{N}_{ngb}$, sets the adaptive spatial resolution. The conceptual similarity
between the photon transport and the calculation of
particle properties in SPH simulations allows a straightforward numerical
implementation of \traphic\ in SPH simulations for application on distributed memory machines.
\par
In \traphic\ photon packets are transported inside cones. Because source
particles emit into a set of cones that tessellates the simulation box, the
angular dependence of the emission can be modelled independently of the spatial distribution of the neighbouring SPH
particles. The number of cones, $N_c$, is the second 
parameter employed in \traphic\ and determines the formal angular resolution of
the radiative transfer simulation. After their emission, the photon packets are
transported downstream from SPH particle to SPH particle. They are confined
to the solid angle they were originally emitted into by regular cones of solid angle
$4\pi / N_c$. Mimicking the ray-splitting technique used in ray-tracing
schemes, the photon transport is adaptive in angle. The
effective angular resolution is therefore much higher than the formal one. Radiative
transfer simulations employing \traphic\ can thus be performed using only a
relatively small number of cones without loss of accuracy.
\par
The propagation of photons
inside cones that do not contain a neighbour requires the introduction of so-called virtual particles (ViPs). It is
the concept of cones combined with ViPs that enables the directed
transport of photons on the unstructured grid defined by the discrete set of irregularly distributed SPH
particles. In addition to the number of cones $N_c$ and
the number of SPH neighbours $\tilde{N}_{ngb}$, it is the ratio 
$N_c / \tilde{N}_{ngb}$ that directly influences the performance of
\traphic. It controls the amount of noise introduced by the representation of
the underlying continuum physics with a discrete set of particles. This
particle noise is small for both $N_c < \tilde{N}_{ngb}$ and $N_c >
\tilde{N}_{ngb}$  due to the large number of neighbours per cone and the large
number of ViPs, resp. For the choice of parameters $N_c = \tilde{N}_{ngb}$
the particle noise can be substantial. 
It can, however, be efficiently suppressed by employing a density field resampling strategy.
\par
A practical problem often encountered when solving the radiative transfer equation is
the linear scaling of the computational effort with the number of
sources. Such a scaling limits
the application of most of the existing radiative transfer schemes to simulations containing only a
few sources. In \traphic\ this problem is circumvented by introducing a source merging
procedure. Photon packets emitted by different sources are merged
in accordance with the angular resolution employed, such that at any point in
space at most $N_c$ photon packets need to be propagated. This renders the
photon transport effectively independent of the number of sources in the
simulation and makes it feasible to include a diffuse radiation
component emitted by the SPH particles. In the most extreme case of radiation
from an arbitrary number of sources completely filling the simulation box, the computational effort required by
\traphic\ merely scales as the product of spatial and angular resolution,
$N\times \tilde{N}_{ngb} \times N_c$, where $N$ is the total number of particles
(SPH + stars).
\par
Applications of \traphic\ include the solution of both the time-independent
and time-dependent radiative transfer equation in large-scale simulations
of cosmological reionization. Here we have specified \traphic\ for the transport of 
mono-chromatic hydrogen-ionizing radiation and described its implementation
in the parallel SPH code \gadget\ (\citealp{Springel:2005}). We showed 
that \traphic\ is able to accurately reproduce the expected growth
of the ionized sphere around a single point source in a homogeneous medium and 
to cast sharp shadows behind opaque obstacles. Furthermore, we tested our
scheme in a physical setting of complex geometry: The growth of ionized regions around
multiple point sources in a cosmological density field. Detailed comparisons with the
results obtained by other codes (\citealp{Iliev:2006b}) for identical problems
show excellent agreement.
\par
In this paper we have limited ourselves to the description of the radiative
transfer scheme \traphic\ and its implementation for use on static density
fields. However, 
\traphic\ can by design readily be coupled to the thermo- and hydrodynamic evolution of 
matter in SPH simulations. We will report on such a coupling and its extension to the transport of 
multi-frequency radiation in future work. 

\section*{Acknowledgments} 
We thank 
Claudio Dalla Vecchia for stimulating discussions,
enduring encouragement as well as technical help and 
Huub R\"ottgering for his valuable advice. We are grateful to Craig Booth and Benedetta
Ciardi for a thorough reading of the draft.
A.H.P. thanks the University of Potsdam and the Max Planck Institut
f\"ur Astrophysik for their hospitality.
Some of the simulations presented here were run on the Cosmology Machine
at the Institute for Computational Cosmology in Durham
as part of the Virgo Consortium research programme and
on Stella, the LOFAR BlueGene/L system in Groningen.
This work was supported by Marie Curie Excellence Grant
MEXT-CT-2004-014112.

\appendix 
\section[]{The Anisotropy of Particle-to-Neighbor Transport} 
\label{Appendix:A}
In Section~\ref{Section:EmissionBySourceParticles} we introduced an emission cone tessellation to accomplish
the emission of photons by a source particle to its neighboring gas
particles. Then we also mentioned that the cones are required to
achieve an emission in agreement with the angular dependence of the emissivity of the
source, independently of the spatial distribution of the neighbors. 
In this appendix we explain why this is the case. In particular, assuming that
particles have equal mass, we
show that when transporting photons (or more generally, any extensive quantity) in
particle simulations from a particle to its neighbors, the net transport direction
is generally correlated with the direction towards the centre of mass of the
neighbouring particles. As a result, the transport is partly governed by the geometry of the
simulation, in addition to the intrinsic emissivity of the source.
\par
Consider a particle located at the origin $O$ of a 3-dimensional  
coordinate system (the emitter frame). The particle emits photons to its
$\tilde{N}_{ngb}$ nearest neighbours, 
residing in the sphere of radius $\tilde{h}$ centred on the emitting
particle. Throughout this section we assume that all particles have  equal
mass, $m = 1$. The 
emission is performed by transferring the fraction $w_i/\sum^{\tilde{N}_{ngb}}_{j=1} w_j$ of the total number 
of photons to neighbour $i$ ($i = 1,2,...,\tilde{N}_{ngb})$, where the $w_i \ge 0$ are
weights to be specified. The emission properties will depend both on the 
weights and on the spatial distribution of the neighbours.
\par
To study the angular dependence of the emission process, we introduce the
vector sum
$\vect{s} = \sum_i w_i \vect{u}_i / \sum_i w_i$ over all unit vectors 
$\vect{u}_i \equiv \vect{r}_i/|\vect{r}_i|$, where $\vect{r}_i$  is the position 
of neighbour $i$. This vector sum can be interpreted as the net direction of the 
emission.  Let us denote the angle enclosed by $\vect{s}$ and the vector to the centre of 
mass $\vect{r}_{cm}$ of the neighbours, $\vect{r}_{cm} = \sum_i \vect{r}_i / \tilde{N}_{ngb}$, with $\alpha$.
We ask for the probability density function (pdf) $p(\cos\alpha)$  of the 
cosine of $\alpha$, where
\begin{equation}
\cos \alpha = \frac{\vect{s} \cdot \vect{r}_{cm}}{\left|\vect{s}\right|
  \left|\vect{r}_{cm}\right|}
= \frac{\sum_i w_i \vect{u}_i \cdot \sum_j \vect{r}_j}{
  |\sum_i w_i \vect{u}_i| |\sum_j \vect{r}_j |}.
\end{equation}
\par
The meaning of the pdf can be understood by
looking at its extremes. If $p(\cos\alpha)$ is strongly peaked around 
$\cos\alpha \approx 1$, the emission has a net direction biased towards
the centre of mass direction, and photons will be preferentially transported
into the high (particle) density regions. On the other hand, if 
$p(\cos\alpha)$ is strongly peaked around 
$\cos\alpha \approx -1$, photons will be preferentially transported into the
direction opposite to the centre of mass. Finally, a flat 
pdf $p(\cos\alpha) = 1/2$ indicates that the net emission 
direction and the direction towards the centre of mass are uncorrelated. 
\par
As an illustration, assume that all neighbours have identical weights $w_i = 1$ and that their positions $\vect{r}_i$ 
are drawn from a probability distribution that uniformly samples the 
surface of the sphere of radius $\tilde{h}$ surrounding the emitter at $O$. To obtain 
$p( \cos \alpha)$, we note that our
assumptions imply that the centre of mass vector $\vect{r}_{cm}$ and 
the vector sum $\vect{s}$ point in the same direction, so that 
$p(\cos\alpha)  =  2 \delta_D(\cos\alpha - 1)$, where $\delta_D(x)$ is the 
Dirac $\delta$~-function. Hence, there is a perfect correlation between  
$\vect{s}$ and $\vect{r}_{cm}$.
\par
\begin{figure} 
 \includegraphics[trim = 15mm 0mm 35mm 10mm, width=0.49\textwidth, clip=true]{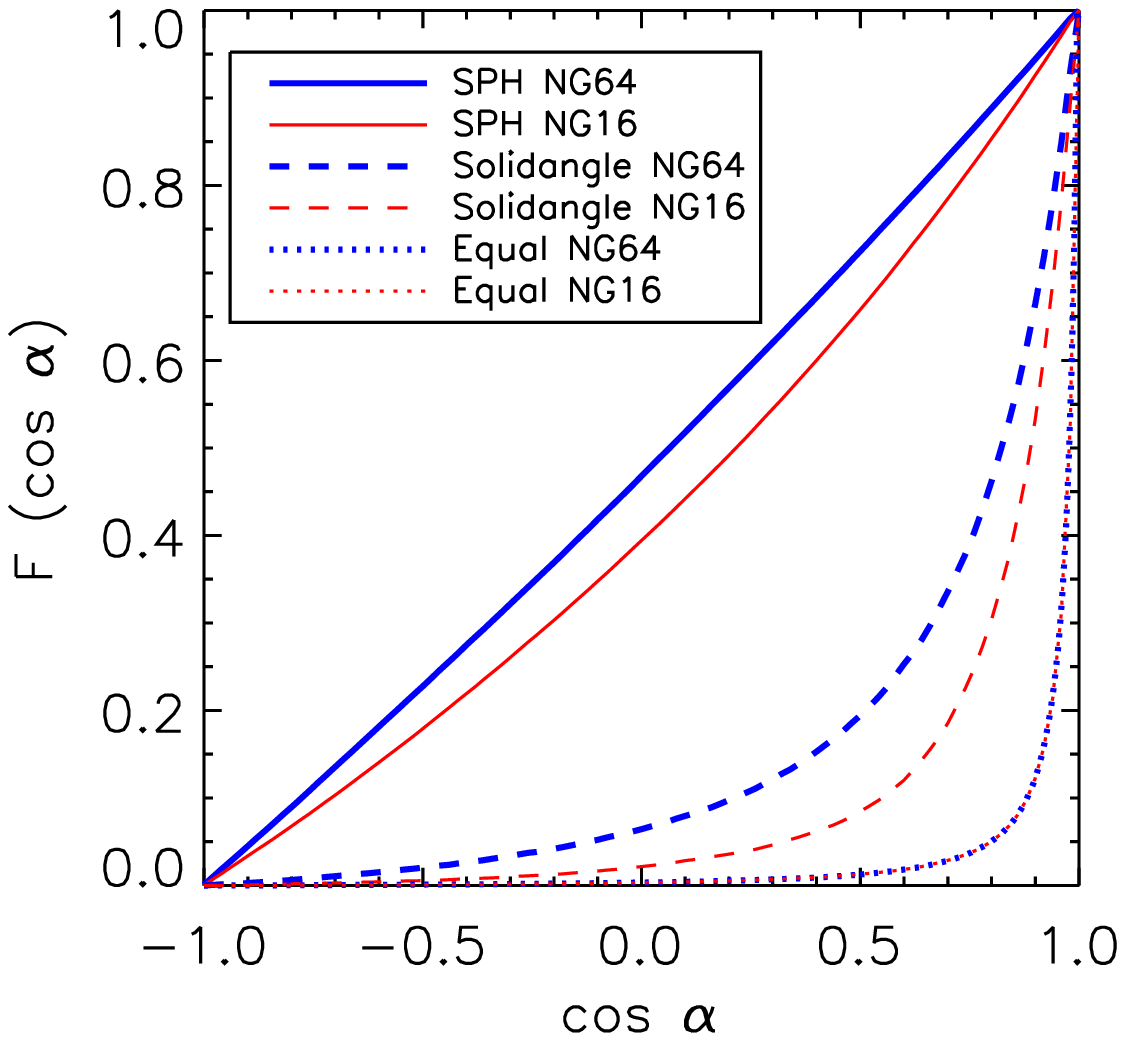} 
  \caption{Monte Carlo simulation of particle-to-neighbour transport. We show the cumulative
    probability distribution of $\cos \alpha$, where $\alpha$ is the
    angle between the net emission direction $\vect{s}$ and the direction towards the
    centre of mass $\vect{r}_{cm}$ of the neighbours. The curves correspond to the following weighting schemes, with the
    number of neighbours indicated in brackets (from top to bottom): 
    SPH (64), SPH (16), Solidangle (64), Solidangle (16), Equal (64),
    Equal (16). The last two curves are nearly on top of each other. } 
  \label{Appendix:A:PhotonDistribution} 
\end{figure}

Another example is given by assuming that the particles of unit mass are distributed
randomly within the sphere of radius $\tilde{h}$ around the emitter. The
resulting pdf $p(\cos \alpha)$ as obtained through 
a Monte Carlo simulation is shown in its cumulative 
form $F(\cos \alpha) = \int_{-1}^{\cos \alpha} dx\ p( x)$ in 
Fig.~\ref{Appendix:A:PhotonDistribution}  for two choices of the number of
neighbours, $\tilde{N}_{ngb} = 16$ (thin dotted curve) and $\tilde{N}_{ngb} = 64$ (thick
dotted curve; falling nearly on top of the thin dotted curve). Clearly, 
there is a very strong correlation between the net direction of emission and the 
direction towards the centre of mass of the neighbour distribution. 
\par
For the important case of isotropic sources of photons  the net emission direction should not be 
correlated with the direction towards the centre of mass, 
i.e. $p(\cos \alpha) = 1/2$, independent of $\cos \alpha$, and thus $F(\cos \alpha) =
(1+ \cos \alpha) / 2$. From here on, 
we will refer to this case as isotropic emission of radiation. We emphasize the statistical character of this
statement. An individual particle may still emit
anisotropically, which could be revealed by studying the amplitude
$|\vect{s}|$. For emission that is isotropic for individual particles, $|\vect{s}| = 0$.
Consider the neighbour distribution of the last example, for which we showed
the emission pattern to be far from isotropic. 
Can the emission be isotropized by tuning the weights? 
\par
For instance, consider solid angle weighting, defined as follows. 
We project all neighbours radially onto the unit sphere. To each neighbour we
attach a weight proportional to its area of influence, which we define by the collection of 
all points on the unit sphere which are closer to the projected neighbour 
for which we calculate the weight than to all other projected neighbours. Hereby,
the distance of two points on the sphere is given by the arclength of
the connecting geodesic (great circle). This definition of the area of
influence results in a specific tessellation of the surface of the sphere, 
the so-called Voronoi tessellation (e.g. \citealp{Okabe:2000}), and provides 
us with well-defined non-overlapping solid angles. 
\par
The resulting cumulative distribution $F(\cos \alpha)$ is shown in 
Fig.~\ref{Appendix:A:PhotonDistribution} (dashed curves). Although the weighted emission has 
weakened the correlation between  $\vect{s}$ and $\vect{r}_{cm}$ as compared to the
case of equal weights, the net emission direction is still markedly peaked 
towards the direction to the centre of mass. We note that weighting by solid
angle is the most natural choice and its inability to
isotropize the transport might be surprising. Alternatively, one
could introduce {\it ad hoc} weights $\hat{w}_i = w_i^r$, where $r>1$ is 
some exponent and the $w_i$ are solid angle weights, to re-shape the
distribution $p( \cos\alpha)$ to become less and less peaked around
$\cos\alpha \sim 1$. 
\par
However, the reason why the solid angle weights
fail to isotropize the transport is not that the weights are inappropriate, 
but that the directions to the neighbours are fixed and thus cannot be freely
chosen along with the weights, as mentioned in
Section~\ref{Section:EmissionBySourceParticles}. 
This strongly limits one's ability to influence the shape of $p(\cos\alpha)$.
For illustration, consider an emitter having all its neighbours located in 
one hemisphere. The net emission direction $\vect{s}$
will then necessarily lie in that hemisphere, too, {\it regardless} of the chosen
weights. For a statistical ensemble of emitters, $\vect{s}$ will then  
necessarily be correlated with the direction towards the centre of mass.
\par
This also explains  the dependence of
$p(\cos\alpha)$ on the number of neighbours. The larger $\tilde{N}_{ngb}$, the
greater the probability of finding a neighbour in any chosen solid angle. Through this
increase in directional sampling, $p(\cos\alpha)$ can be expected to
approach the uncorrelated case. This behaviour can indeed be observed in  
Fig.~\ref{Appendix:A:PhotonDistribution}. 
\par
We also compute the statistic $p(\cos \alpha)$ for the case of 
neighbours clustered in space, which is more typical for cosmological simulations of
structure formation. We did this both using a toy clustering model
(\citealp{Soneira:1978}) and for a cosmological density field obtained from
a $\Lambda$CDM hydrodynamical simulation (the same field that is used in Section~\ref{Sec:Test3}). 
For the weighting schemes we investigated 
the dependence of the shape of $p(\cos \alpha)$ 
on the chosen weights is qualitatively very similar to that in the case of the random
distribution of neighbours discussed earlier.
\par
We emphasize that our discussion is not limited to the transport of
photons. Indeed, particle-to-neighbour transport is for example routinely employed in SPH
simulations of galaxy formation, where a star particle distributes its metals and
energy amongst its neighbours. Different recipes for distributing 
metals and energy can be found in the literature. Most of these apply 
a weighting related to the SPH formalism. As an example, to each neighbour $i$ we associate the weight
\begin{equation}
  w_i = \frac{ \frac{m_i}{\rho_i} W(r_i, \tilde{h})}{\sum_j \frac{m_j}{\rho_j} W(r_j, \tilde{h})}.
\end{equation} 
Here, $\tilde{h}$ is the radius of the neighbourhood of the emitting particle (at $O$), $r_i = |\vect{r}_i|$ 
and $W$ is the spline kernel Eq.~\ref{Eq:Kernel:Spline}. Performing the transport on the same random neighbour
distribution as before, we obtain the distribution  $p(\cos \alpha)$. As
can be seen in Fig.~\ref{Appendix:A:PhotonDistribution} (solid curves), the direction of
net transport and the direction to the centre of mass are found to be almost 
uncorrelated. We note that this result, the reason of which at the moment
is not clear to us, is only a statistical statement. We calculated
$|\vect{s}|$ and found that the emission of individual particles is still not
close to isotropic. We arrive at qualitatively similar results for different definitions of the SPH
weights, both for  random and clustered neighbour distributions. A more
quantitative statement, however, must depend on both the specific properties of the spatial distribution of the
neighbours, and the precise form of the adopted SPH weights. 
\par
In summary, in this appendix we have shown that when transporting a quantity
from a particle to its neighbours, the net transport direction is generally
governed by the spatial distribution of the neighbours, in addition to the
the intrinsic properties of the emitting particle. We emphasize that our observations apply to the
particle-to-neighbour transport of arbitrary quantities in {\it any} particle simulation.
We would like to remind the reader that in Section~\ref{Sec:Method} we 
presented a specifically designed particle-based transport scheme, \traphic, that overcomes the problems of
particle-to-neighbour based transport discussed in this appendix, not just
statistically, but also on a particle by particle basis. We achieved this by
employing cones to confine photons to the solid angle into which they are
emitted, making the transport independent of the geometry of the SPH
simulation, on the scale of the chosen angular resolution. 
\par

\section[]{Numerical Implementation} 
\label{Appendix:B}
\subsection{Cones}
\label{Appendix:B:Cones}
In this section we describe the numerical implementation of the cones employed for the
emission and reception of photon packets in \traphic. For the emission
(Section~\ref{Section:EmissionBySourceParticles}), 
each source particle divides its neighbourhood using a set of tessellating emission cones. The same
tessellation is also employed for the reception of photon packets by gas
particles (Section~\ref{Sec:TRAPHIC:Merging}). In the
following, we employ spherical coordinates $(r, \phi, \theta)$, which are related to the
Cartesian components $(r_x, r_y, r_z)$ of an arbitrary vector $\vect{r}$ through
\begin{eqnarray}
r_x &=& r \cos\phi \sin\theta\\
r_y &=& r \sin\phi \sin\theta\\
r_z &=& r \cos\theta.
\end{eqnarray}
In our implementation, the emission (reception) cones sample the volume around each 
particle  isotropically. Since the surface element of the unit sphere is given by
$ds = d(\cos\theta) d\phi$, this is achieved  by distributing
the cone boundaries\footnote{Strictly speaking, one should distribute the cone
{\it axes} uniformly in $(\cos\theta,\phi)$, but this implies asymmetric
cones.} uniformly (i.e. on a regular lattice with indices $i, j$)  
in $(\cos\theta,\phi)$. Thus, the boundaries of cone $(i,j)$ are described
by the four arcs of constant 
\begin{eqnarray}
  \phi^{ij}_1 &=& i\frac{2\pi}{N_\phi}, \quad 0 \le i < N_\phi, \label{Cones:Boundaries:1}\\ 
  \phi^{ij}_2 &=& (i + 1)\frac{2\pi}{N_\phi}, \quad 0 \le i < N_\phi, \label{Cones:Boundaries:2}\\ 
  \theta^{ij}_1 &=& \arccos (1 - 2 \frac{j}{N_\theta}), \quad 0 
  \le j < N_\theta , \label{Cones:Boundaries:3}\\
  \theta^{ij}_2 &=& \arccos (1 - 2 \frac{j+1}{N_\theta}), \quad 0 
  \le j < N_\theta. \label{Cones:Boundaries:4}
\end{eqnarray} 
Correspondingly, we define the emission (reception) cone axes by
\begin{eqnarray}
  \phi^{ij} &=& \frac{\phi^{ij}_1 + \phi^{ij}_2}{2}  \label{Appendix:A:Eq:Method3:Phi},\\
  \theta^{ij} &=& \frac{\theta^{ij}_1 + \theta^{ij}_2}{2} \label{Appendix:A:Eq:Method3:Theta},
\end{eqnarray} 
Note that each of the $N_c = N_\phi \times N_\theta$ emission (reception) cones 
has the same solid angle $\Omega = 4\pi / N_c$, as can be seen from
integrating over the surface element of the unit sphere within the boundaries 
(Eqs.~\ref{Cones:Boundaries:1}-\ref{Cones:Boundaries:4}) of a single cone. 
We also implemented the tessellation used in 
\cite{Abel:1999}, which leads to cones that are more similar in shape. 
We could not find any systematic differences in the test problems described in
Sections~\ref{Section:Test1} - \ref{Sec:Test3} using either of the
two types of tessellations. This is not surprising because any artefacts due
to the shape of the cones will be suppressed by the random rotations of the
emission (reception) cones we perform before each emission (reception), as we
will describe in the following section.
\subsection{Random rotations}
\label{Appendix:B:Rotation}
Recall from Sections \ref{Section:Transmission} and
\ref{Sec:TRAPHIC:Merging} 
that we apply a random rotation to each cone
tessellation. Consequently, each cone tessellation has a random orientation. 
The primary motivation for randomly rotating cones is to 
increase the angular sampling. Furthermore, randomly rotating the emission and
reception cones leads to a reduction of artefacts 
arising from the particular definition we employ to construct the cone
tessellation, as noted in the last section. 
Here we describe our numerical implementation of the random rotations.
\par
We can think of the set of cones that comprises a particular cone tessellation as a rigid body, 
to which we can attach a local Cartesian coordinate system with axes 
$\{e_x^{\prime}, e_y^{\prime}, e_z^{\prime}\}$. The orientation of this coordinate system with
respect to the canonical Cartesian coordinate system, e.g.~the simulation box axes $\{e_x, e_y, e_z\}$, is
fully described by three variables, the Eulerian angles (e.g.~\citealp{Goldstein:1980}).
\par 
Eulerian angles are defined as the three successive angles of rotations
that map the axes  $\{e_x, e_y, e_z\}$ onto the axes 
$\{e_x^{\prime}, e_y^{\prime}, e_z^{\prime}\}$. There exist several conventions. 
In the $zxz$ convention we employ here, the initial system of axes $\{e_x, e_y, e_z\}$ is 
first rotated counter-clockwise by an angle $\phi$ around the $z$-axis, with the 
resulting coordinate system labelled $\{e_\xi, e_\eta, e_\zeta\}$. Second, 
the coordinate system $\{e_\xi, e_\eta, e_\zeta\}$ is rotated
by an angle $\theta$ counter-clockwise about the $\xi$-axis, leaving the new 
coordinate system $\{e_\xi^{\prime}, e_\eta^{\prime}, e_\zeta^{\prime}\}$. The
third and last rotation is carried out counter-clockwise by an angle $\psi$ around 
the $\zeta^{\prime}$-axis, giving the desired 
$\{e_x^{\prime}, e_y^{\prime}, e_z^{\prime}\}$ coordinate system.
\par
To obtain random Eulerian angles, we note that the invariant measure $\mu$ (the
``volume element'') on $\mathrm{SO(3)}$, the group of proper rotations 
in $\mathbb{R}^3$, in the $zxz$ Eulerian angle parametrization reads 
(e.g.~\citealp{Miles:1965}), 
\begin{equation}
  \mu(\phi, \theta,\psi) d\phi d\theta d\psi 
  = \frac{1}{8\pi}\sin\theta d\phi d\theta d\psi.
\end{equation}
Random Eulerian angles are therefore obtained by 
drawing random variables $u_1, u_2, u_3$ from a uniform distribution on the 
interval $[0,1]$ and applying the usual transformation (cp.~\citealp{NumericalRecipes:1992}),
\begin{eqnarray}
  \phi &=& 2\pi u_1 \\
  \theta &=& \arccos(1 - 2 u_2)\\
  \psi &=& 2 \pi u_3.
\end{eqnarray}
We implement random rotations using rotation
matrices, which are obtained from the random Eulerian angles. 
The matrix elements of a matrix representing a rotation associated with a
given set of Eulerian angles can be readily calculated. For details, we
refer the reader to \cite{Goldstein:1980}.
\par
In principle, random rotation matrices can be obtained 
without drawing random Eulerian angles (e.g.~\citealp{Stewart:1980}). 
However, we find that it is faster to generate random Eulerian angles,
and then calculate the corresponding rotation matrices. Moreover, 
storing the three Eulerian angles instead of the nine rotation matrix 
elements reduces the memory cost.
\subsection{Reduction of particle noise}
\label{Sec:Resampling:Implementation}
For some of the simulations presented in
Sections~\ref{Section:Test1} - \ref{Sec:Test3} we employ the resampling approach described
in Section~\ref{Sec:Method:Regularization} to reduce numerical artefacts arising
from the representation of the continuous density field with a discrete set of
particles. The resampling requires the sampling of the SPH 
kernel used in \gadget, which is the spline given by Eq.~\ref{Eq:Kernel:Spline}.
We approximate this kernel by a Gaussian,
\begin{equation}
  W_\sigma(r) = \frac{1}{(2 \pi)^{3/2} \sigma^3} \exp(-r^2/2\sigma^2).
  \label{Kernel:Gaussian}
\end{equation}
We find that with a standard deviation of $\sigma \sim 0.29\times h$, the 
Gaussian provides a reasonable fit to the spline. A similar relation was
employed in \cite{Alvarez:2006}. When we resample the 
SPH density field, all SPH particles are redistributed by randomly displacing them from
the positions given by the hydrodynamical simulation, with the probability to
find a given particle displaced by the distance $r$ given by 
Eq.~\ref{Kernel:Gaussian}. The (rare) displacements larger than $h$ are discarded; 
in this case the original particle positions as determined by the hydrodynamical simulation are used.
\par

\bsp 
 
\label{lastpage} 
 
\end{document}